\documentclass[onecolumn,floatfix,showpacs,showkeys,nofootinbib,preprint]{revtex4}
\usepackage{epsfig}
\usepackage{amssymb,latexsym,amsmath}
\usepackage{color}
 
\begin{document}

\title{Statistical Ensembles With Finite Bath:\\
 A Description for an Event Generator}
\author{M. Hauer}
\affiliation{Helmholtz Research School, University of Frankfurt, Frankfurt,
  Germany}
\affiliation{UCT-CERN Research Centre and Department of Physics,University of   
  Cape Town, Rondebosch 7701, South Africa}
\author{S. Wheaton}   
\affiliation{UCT-CERN Research Centre and Department of Physics,University of   
  Cape Town, Rondebosch 7701, South Africa}  

\begin{abstract}
A Monte Carlo event generator has been developed assuming thermal 
production of hadrons. The system under consideration is
sampled grand canonically in the Boltzmann approximation. A
re-weighting scheme is then introduced to account for conservation of charges
(baryon number, strangeness, electric charge) and energy and momentum, 
effectively allowing for extrapolation
of grand canonical results to the microcanonical limit. This method
has two strong advantages compared to analytical approaches and
standard microcanonical Monte Carlo techniques, in that it is capable
of handling resonance decays as well as (very) large system sizes. 
\end{abstract}



\maketitle

\section{Introduction}
\label{Sec_Intro}

The statistical hadronization model, first introduced by Fermi~\cite{Fermi} 
and Hagedorn~\cite{Hagedorn}, has been remarkably successful in the
description of experimentally measured average hadron production yields 
in heavy ion collisions ranging from SIS \cite{GSIfits}, and AGS \cite{AGSfits}, 
over SPS \cite{SPSfits} to RHIC \cite{RHICfits} energies. 
Over time this has led to the establishment of the 
`chemical freeze-out line`~\cite{FreezeOut}, which is now a vital part of our understanding 
of the phase diagram of strongly interacting matter. 
Model predictions for the upcoming LHC and future 
FAIR~\cite{SHM_predictions_LHC,SHM_predictions_FAIR} experiments largely follow these trends.

Somewhere above this freeze-out line in the phase diagram we expect, in general, 
a phase transition from hadronic degrees of freedom to a phase of deconfined quarks 
and gluons, generally termed the quark gluon plasma; and more specifically, a first 
order phase transition at low temperature and high baryon chemical potential, 
and a cross-over at high temperature and low baryon chemical potential. In between,
a second order endpoint or a critical point might emerge. 
For recent reviews see \cite{QCD_pd,Model_pd}.

Fluctuation and correlation observables are amongst the most promising candidates 
suggested to be suitable for signaling the formation
of new states of matter, and transitions between them. For recent reviews here 
see \cite{OnsetOfDecon,PhaseTrans,CriticalPoint,Koch}.  

The statistical properties of a sample of events are, however, certainly not solely 
determined by critical phenomena. More broadly 
speaking, they depend strongly on the way events are chosen for the analysis, 
and on the information available about the system.

The ideal gas approximation of the statistical hadronization model will 
again serve as our testbed. 
Its strong advantage is that it is simple, and to 
some extent intuitive. 
Given its success in describing experimentally 
measured average hadron yields, and its ability to reproduce 
low temperature lattice susceptibilities \cite{Karsch_susc}, 
the question arises as to whether fluctuation and correlation observables 
also follow its main line. 
Critical phenomena (and many more), however, remain beyond the present study.

Conventionally in statistical mechanics three standard ensembles are   
discussed; the microcanonical ensemble (MCE), the canonical ensemble (CE),  
and the grand canonical ensemble (GCE). In the MCE\footnote{The term MCE is also often   
applied to ensembles with energy but not momentum conservation.} one considers an ensemble   
of microstates with exactly fixed values of extensive conserved quantities   
(energy, momentum, electric charge, etc.), with `a priori equal probabilities` of   
all microstates (see e.g. \cite{Patriha}). The CE introduces the concept of temperature by   
introduction of an infinite thermal bath, which can exchange energy (and momentum)
 with the system.  
The GCE introduces further chemical potentials by attaching the system under consideration   
to an infinite charge bath\footnote{Note that a system with many charges can have some   
charges described via the CE and others via the GCE.}.  
Only if the experimentally accessible system is just a small   
fraction of the total, and all parts have had the opportunity   
to mutually equilibrate, can the appropriate ensemble be the grand canonical ensemble.  

A statistical hadronization model Monte Carlo event generator affords us with the possibility 
of studying fluctuation and correlation observables in equilibrium systems.
Data analysis can be done in close relation to experimental analysis techniques.
Imposing global constraints on a sample is always technically a bit more challenging.
Direct sampling of MCE events (or microstates) has only been done in 
the non-relativistic limit~\cite{Randrup}. 
Sample and reject procedures, suitable for relativistic systems, become rapidly 
inefficient with large system size. However, they have the advantage of being very 
successful for small system sizes \cite{Bec_MC,Bec_MCE}. 

In this article we try a different approach: we sample the GCE, then re-weight events  
according to their values of extensive quantities, and approach the sample-reject limit 
(MCE) in a controlled manner. In this way one can study the statistical properties 
of a global equilibrium system in their dependence on the size of their thermodynamic bath.
As any of the three standard ensembles remain idealizations of physical systems,
one might find intermediate ensembles to be of phenomenological interest too.

We study the first and, in particular, second moments 
of joint distributions of extensive quantities. We concentrate
mainly on particle number distributions and distributions of `conserved' charges,
and discuss the influence of acceptance cuts in momentum space, conservations laws, and
resonance decay on the statistical properties of a sample of 
hadron resonance gas model events. 
We extend our previous studies of ideal particle and anti-particle 
gases~\cite{acc,baseline} and of gases of altogether massless particles \cite{feq}.


The numerical code has been written for inclusion into the already 
existing THERMUS package~\cite{THERMUS}. We make frequent use of the
functionality provided by the ROOT framework~\cite{ROOT}.

The paper is organized as follows:
In Section~\ref{Sec_SEfB} the basic ideas of this article are formulated. 
The GCE Monte Carlo sampling 
procedure is described in Section~\ref{Sec_GCEsampling}. 
The first and second moments of the distributions of fully phase space
integrated extensive quantities are then extrapolated to the microcanonical limit 
in Section~\ref{Sec_ExtraMCE}. 
Section~\ref{Sec_MomSpect} contains an analysis of GCE momentum spectra. 
The momentum space dependence of correlations between conserved charges 
is studied in Section~\ref{Sec_LCfluc}.
Section~\ref{Sec_MultFluc} then deals with
multiplicity fluctuations and correlations in limited acceptance 
and their extrapolation to the MCE limit.
A summary is given in Section~\ref{Sec_Summary}.

\section{Statistical Ensembles With Finite Bath}
\label{Sec_SEfB}
We start out as Patriha \cite{Patriha}, and Challa and Hetherington \cite{extgauss}, 
but quickly take a different route.\\
 
Let us define two microcanonical partition functions, i.e. the number of microstates, 
for two separate systems.
The first system is assumed to be enclosed in a volume $V_1$ and to have fixed values 
of extensive quantities $P_1^{\mu}=(E_1,P_{x,1},P_{y,1},P_{z,1})$, and $Q_1^j=(B_1,S_1,Q_1)$, 
while the second system is enclosed in a volume $V_2$ and has fixed values of extensive 
quantities $P_2^{\mu}=(E_2,P_{x,2},P_{y,2},P_{z,2})$, and $Q_2^j=(B_2,S_2,Q_2)$, 
where $E$ is the energy of the system, $P_{x,y,z}$ are the components of its 
three-momentum, and $B$, $S$, and $Q$, are baryon number, strangeness and 
electric charge, respectively. Thus we have:
\begin{equation}\label{eq_one}
Z(V_1,P_1^{\mu},Q_1^j) ~=~ \sum_{\{N_1^i \}} ~ Z_{N_1^i}(V_1,P_1^{\mu},Q_1^j)~, ~~
~~\textrm{and}~~~~Z(V_2,P_2^{\mu},Q_2^j)~, 
\end{equation}
where $Z_{N_1^i}(V_1,P_1^{\mu},Q_1^j)$ denotes the number of microstates of system 1 with 
additionally fixed multiplicities $N_1^i$ of particles of species $i$.
Suppose that system 1 and system 2 are subject to the following constraints:
\begin{eqnarray}
V_g &=& V_1 ~+~ V_2 ~,\label{constraint_V}\\
P_g^{\mu} &=& P_1^{\mu} ~+~ P_2^{\mu}~,\label{constraint_P} \\
Q_g^{j} &=& Q_1^{j} ~+~ Q_2^{j}~. \label{constraint_Q}
\end{eqnarray}
We can then construct the partition function $Z(V_g,P_g^{\mu},Q_g^j)$ of the joint 
system as the sums over all possible charge and energy-momentum split-ups:
\begin{equation} \label{PF_combined}
Z(V_g,P_g^{\mu},Q_g^j) = \sum \limits_{\{P_1^{\mu}\}} \sum \limits_{\{Q_1^{j}\}}  
Z(V_g-V_1,P_g^{\mu}-P_1^{\mu},Q_g^j-Q_1^j)~ Z(V_1,P_1^{\mu},Q_1^j)~.
\end{equation}
Next we construct the distribution of extensive quantities in the subsystem $V_1$. This is
given by the ratio of the number of all microstates consistent with a given charge 
and energy-momentum split-up and a given set of particle multiplicities to the number of 
all possible configurations:
\begin{equation}
P(P_1^{\mu},Q_1^j,N_1^i) ~=~ 
\frac{Z(V_g-V_1,P_g^{\mu}-P_1^{\mu},Q_g^j-Q_1^j)}{Z(V_g,P_g^{\mu},Q_g^j)} ~
~Z_{N_1^i}(V_1,P_1^{\mu},Q_1^j)~.
\end{equation}
We then define the weight factor $W(V_1,P_1^{\mu},Q_1^j;V_g,P_g^{\mu},Q_g^j) $ such that:
\begin{equation}\label{basic}
P(P_1^{\mu},Q_1^j,N_1^i) ~=~ W(V_1,P_1^{\mu},Q_1^j;V_g,P_g^{\mu},Q_g^j) ~
~Z_{N_1^i}(V_1,P_1^{\mu},Q_1^j)~.
\end{equation}
By construction, the first moment of the weight factor is equal to unity:
\begin{equation}
\langle W \rangle ~=~ \sum_{\{P_1^{\mu}\}} \sum_{\{Q_1^j \}}  \sum_{\{N_1^i \}} ~  
W(V_1,P_1^{\mu},Q_1^j;V_g,P_g^{\mu},Q_g^j) ~
~Z_{N_1^i}(V_1,P_1^{\mu},Q_1^j) = 1~,
\end{equation}
as the distribution is properly normalized.

The weight factor $W(V_1,P_1^{\mu},Q_1^j;V_g,P_g^{\mu},Q_g^j) $ generates 
an ensemble with statistical properties different from the limiting cases 
$V_g \rightarrow V_1$ (MCE), and $V_g \rightarrow \infty$ (GCE).
This effectively allows for extrapolation of GCE results to the MCE limit. 
In the thermodynamic limit ($V_1$ sufficiently large) a family of
thermodynamically equivalent (same densities) ensembles is generated. 
In principle any other (arbitrary) choice of 
$W(V_1,P_1^{\mu},Q_1^j;V_g,P_g^{\mu},Q_g^j) $ could be taken.
In this work we confine ourselves, however, to the situation discussed above. Please note
that all microstates consistent with the same set of extensive quantities 
$(P_1^{\mu},Q_1^j)$ have `a priori equal probabilities`. 

In the large volume limit, ensembles are equivalent in the sense that 
densities are the same. The ensembles defined by  Eq.(\ref{basic}) and later on by Eq.(\ref{thetrick}) 
are no exception. If both $V_1$ and $V_g$ are sufficiently large, then the average 
densities in both systems will be the same, $Q^j_g / V_g$ and $P^{\mu}_g / V_g$ respectively. 
The system in $V_1$ will hence carry on average a certain fraction:
\begin{equation}\label{lambda_def}
\lambda ~\equiv~ V_1/V_g~,
\end{equation}
 of the total charge $Q^j_g$ and four-momentum $P^{\mu}_g$, i.e.: 
\begin{equation}
\langle Q^j_1 \rangle ~=~ \lambda~ Q^j_g ~,\qquad \textrm{and} 
\qquad \langle P_1^{\mu} \rangle ~=~ \lambda~ P^{\mu}_g~.
\end{equation} 
By varying the ratio $\lambda = V_1 /V_g$, while keeping $\langle Q^j_1 \rangle$ and 
$\langle P_1^{\mu} \rangle$ constant, we can thus study a class of systems with the same 
average charge content and four-momentum, but 
different statistical properties.

\subsection{Introducing the Monte Carlo Weight $\mathcal{W}$}
Since Eq.(\ref{basic}) poses a formidable challenge, both mathematically and numerically,
 we write instead:
\begin{equation}\label{thetrick}
P(P_1^{\mu},Q_1^j,N_1^i) ~=~ \mathcal{W}^{P_1^{\mu},Q_1^j;P_g^{\mu},Q_g^j}(V_1;V_g|\beta,u_{\mu},\mu_j) ~
~P_{gce}(P_1^{\mu},Q_1^j,N_1^i|\beta,u_{\mu},\mu_j)~,
\end{equation}
where the distribution of extensive quantities $P_1^{\mu}$, $Q_1^j$ and particle multiplicities 
$N_1^i$ of a GCE system with temperature $T=\beta^{-1}$, volume $V_1$, chemical potentials $\mu_j$ and 
collective four-velocity $u_{\mu}$ is given by:
\begin{equation}\label{Pgce}
P_{gce}(P_1^{\mu},Q_1^j,N_1^i|\beta,u_{\mu},\mu_j) ~\equiv~ \frac{e^{-P_1^{\mu} u_{\mu} \beta }~ 
e^{Q_1^j \mu_j \beta}}{Z(V_1,\beta,u_{\mu},\mu_j)} ~ Z_{N_1^i}(V_1,P_1^{\mu},Q_1^j)~,
\end{equation}
where $\mu_j = (\mu_B,\mu_S,\mu_Q)$, summarizes the chemical potentials associated with 
baryon number, strangeness and electric charge in a vector.
The normalization in Eq.(\ref{Pgce}) is given by the GCE partition function 
$Z(V_1,\beta,u_{\mu},\mu_j)$, i.e. the number of all microstates averaged over 
the Boltzmann weights $e^{-P_1^{\mu} u_{\mu} \beta }$ and $e^{Q_1^j \mu_j \beta}$:
\begin{equation}
Z(V_1,\beta,u_{\mu},\mu_j)~=~ \sum_{\{P_1^{\mu}\}} \sum_{\{Q_1^{j} \}} \sum_{\{N_1^i \}}
~ e^{-P_1^{\mu} u_{\mu} \beta }~e^{Q_1^j \mu_j \beta}~  Z_{N_1^i}(V_1,P_1^{\mu},Q_1^j)~.
\end{equation}
The new  weight factor $ \mathcal{W}^{P_1^{\mu},Q_1^j;P_g^{\mu},Q_g^j}(V_1;V_g|\beta,u_{\mu},\mu_j)$ 
now reads:
\begin{eqnarray}\label{newW}
 \mathcal{W}^{P_1^{\mu},Q_1^j;P_g^{\mu},Q_g^j}(V_1;V_g|\beta,u_{\mu},\mu_j) &=& 
Z(V_1,\beta,u_{\mu},\mu_j)~
\frac{e^{-(P_g^{\mu}-P_1^{\mu}) u_{\mu} \beta }~ 
e^{(Q_g^j-Q_1^j) \mu_j \beta} }{e^{-P_g^{\mu} u_{\mu} \beta}~ e^{Q_g^j \mu_j \beta} }  \nonumber \\
&\times&\frac{Z(V_g-V_1,P_g^{\mu}-P_1^{\mu},Q_g^j-Q_1^j)}{Z(V_g,P_g^{\mu},Q_g^j)}~.
\end{eqnarray}
In the case of an ideal (non-interacting) gas, Eq.(\ref{newW}) can be 
written \cite{clt,baseline} as:
\begin{eqnarray}\label{simpleW}
 \mathcal{W}^{P_1^{\mu},Q_1^j;P_g^{\mu},Q_g^j}(V_1;V_g|\beta,u_{\mu},\mu_j) &=& 
Z(V_1,\beta,u_{\mu},\mu_j)~
\frac{\mathcal{Z}^{P_g^{\mu}-P_1^{\mu},Q_g^j-Q_1^j}(V_g-V_1,\beta,u_{\mu},\mu_j)}{
\mathcal{Z}^{P_g^{\mu},Q_g^j} (V_g,\beta,u_{\mu},\mu_j)} ~.
\end{eqnarray}
The advantage of Eq.(\ref{thetrick}), compared to Eq.(\ref{basic}), is that the 
distribution $P_{gce}(P_1^{\mu},Q_1^j,N_1^i|\beta,u_{\mu},\mu_j)$ can easily be sampled for Boltzmann particles, 
while a suitable approximation for the weight 
$\mathcal{W}^{P_1^{\mu},Q_1^j;P_g^{\mu},Q_g^j}(V_1;V_g|\beta,u_{\mu},\mu_j)$ is available.

Again, by construction, the first moment of the new weight factor is equal to unity:
\begin{equation}
\langle \mathcal{W} \rangle ~=~ \sum_{\{P_1^{\mu}\}} \sum_{\{Q_1^j \}} \sum_{\{N_1^i \}} ~  
\mathcal{W}^{P_1^{\mu},Q_1^j;P_g^{\mu},Q_g^j}(V_1;V_g|\beta,u_{\mu},\mu_j) 
~P_{gce}(P_1^{\mu},Q_1^j,N_1^i|\beta,u_{\mu},\mu_j)  = 1~.
\end{equation}
In principle, Eq.(\ref{basic}) and Eq.(\ref{thetrick}) 
are equivalent. In fact, Eq.(\ref{basic}) can be obtained by taking the limit 
$(\mu_B,\mu_S,\mu_Q) = (0,0,0)$, $u_{\mu}=(1,0,0,0)$, and $\beta \rightarrow 0$ of 
Eq.(\ref{thetrick}). However, as one can already see, $\langle \mathcal{W}^n \rangle  
\not= \langle W^n \rangle $. Higher, and in particular the second, moments 
of the weight factors $W$ and $\mathcal{W}$ are a 
measure of the statistical error to be expected for a finite sample of events. 
The larger the higher moments of the weight factor, the larger the statistical error, 
and the slower the convergence with sample size. 
Please see also Appendices \ref{App_ConvStudy} and \ref{App_CBG}.

As GCE and MCE densities are the same in the 
system $V_g$, these values are effectively regulated by intensive parameters 
$\beta$, $\mu_j$ and $u_{\mu}$. In essence, if you want to study a system 
with average $\langle Q^j_1 \rangle$, then sample the GCE with $\langle Q^j_1 \rangle$ and calculate 
the weight according to  Eq.(\ref{simpleW}). This will result in a low statistical 
error for finite samples (as shown in later sections), and allow for extrapolation to the MCE limit.

We will now first calculate the weight factor Eq.(\ref{simpleW}) and then 
take the appropriate limits. With the appropriate choice of $\beta$, $\mu_j$ 
and $u_{\mu}$ the calculation of Eq.(\ref{simpleW}) is particularly easy in the 
large volume limit \cite{clt}.

\subsection{Calculating the Monte Carlo Weight $\mathcal{W}$}
In this article, the total number of (potentially) conserved extensive quantities in
a hadron resonance gas is $L=J+4 = 3+4=7$, where $J=3$ is the number of charges $(B,S,Q)$ 
and there are four components of the four-momentum. Including  all extensive quantities 
into a single vector:
\begin{equation}
\mathcal{Q}^l = (Q^j,P^{\mu}) = (B,S,Q,E,P_x,P_y,P_z)~,
\end{equation}
the weight Eq.(\ref{simpleW}) can be expressed as:
\begin{eqnarray}\label{curly_W}
 \mathcal{W}^{\mathcal{Q}_1^l;\mathcal{Q}_g^l}(V_1;V_g|\beta,u_{\mu},\mu_j) &=& 
Z(V_1,\beta,u_{\mu},\mu_j)~
\frac{\mathcal{Z}^{\mathcal{Q}_g^l-\mathcal{Q}_1^l}(V_g-V_1,\beta,u_{\mu},\mu_j)}{
\mathcal{Z}^{\mathcal{Q}_g^l} (V_g,\beta,u_{\mu},\mu_j)} ~.
\end{eqnarray}
The general expression for the partition function 
$\mathcal{Z}^{\mathcal{Q}^l}(V,\beta,u_{\mu},\mu_j)$ 
in the large volume limit reads~\cite{clt}:
\begin{equation}\label{clt_approx}
\mathcal{Z}^{\mathcal{Q}^l}(V,\beta,u_{\mu},\mu_j) ~\simeq~Z(V,\beta,u_{\mu},\mu_j)~ 
\frac{1}{(2 \pi V)^{L/2} \det \sigma}~ 
\exp \left( - \frac{1}{2}~ \frac{1}{V}~ \xi^l \xi_l \right)~,
\end{equation}
where:
\begin{equation}\label{xi}
\xi^l ~=~ \left(\mathcal{Q}^k - V \kappa_1^k \right) ~ \left( \sigma^{-1} \right)_k^l~,
\end{equation}
and:
\begin{equation}
\sigma_k^l ~=~ \left( \kappa_2^{1/2} \right)_k^l ~.
\end{equation}
Here $\kappa_1$ and $\kappa_2$ are the GCE vector of mean values and the
GCE covariance matrix respectively. The values of $\beta$, $\mu_j$ and $u_{\mu}$ are chosen 
such that:
\begin{equation}
\frac{\partial \mathcal{Z}^{\mathcal{Q}^l}}{\partial \mathcal{Q}^l}
\Big|_{\mathcal{Q}^l=\mathcal{Q}^l_{eq}} ~=~ 0_l.
\end{equation}
The approximation (\ref{clt_approx}) gives then a reliable description of 
$\mathcal{Z}^{\mathcal{Q}_g^l}$ around the equilibrium value 
$\mathcal{Q}_g^l = V_g \kappa_1^l$, provided $V_g$ is sufficiently large.
The charge vector, Eq.(\ref{xi}), is then equal 
to the null-vector $\xi_l = 0_l$ ($\mathcal{Q}_g^l = V_g \kappa_1^l$).

For the normalization in Eq.(\ref{curly_W}) we then find:
\begin{equation} \label{W_norm}
\mathcal{Z}^{\mathcal{Q}_g^l}(V_g,\beta,u_{\mu},\mu_j)\Big|_{\mathcal{Q}_g^l=\mathcal{Q}^l_{g,eq}} 
~\simeq~ \frac{Z(V_g,\beta,u_{\mu},\mu_j)}{(2 \pi V_g)^{L/2}\det \sigma}~ 
\exp \left( 0 \right)~.
\end{equation}
For the numerator we obtain:
\begin{equation}  \label{W_numm}
\mathcal{Z}^{\mathcal{Q}_g^l-\mathcal{Q}_1^l}(V_g-V_1,\beta,u_{\mu},\mu_j)
\Big|_{\mathcal{Q}_g^l=\mathcal{Q}^l_{g,eq}} ~\simeq~
\frac{Z(V_g-V_1,\beta,u_{\mu},\mu_j) }{(2 \pi \left(V_g -V_1 \right))^{L/2} \det \sigma}~
 \exp \left( -\frac{1}{2}~ \frac{1}{(V_g-V_1) }~ \xi^l \xi_l\right)~,
\end{equation}
where in Eq.(\ref{W_numm}) we write for the charge vector Eq.(\ref{xi}):
\begin{equation}
\xi^l = \left( \Delta \mathcal{Q}_2 \right)^k \left( \sigma^{-1} \right)_k^l~.
\end{equation}
Then, using $\mathcal{Q}^k_g = \mathcal{Q}^k_{g,eq} = V_g \kappa_1^k$, we find:
\begin{equation}
\left( \Delta \mathcal{Q}_2 \right)^k = \left(\mathcal{Q}_g - \mathcal{Q}_1 \right)^k - 
\left( V_g-V_1\right) \kappa_1^k ~=~  -\left( \mathcal{Q}_1 -V_1\kappa_1 \right)^k~.
\end{equation}
Substituting Eq.(\ref{W_norm}) and Eq.(\ref{W_numm}) into Eq.(\ref{curly_W}) yields:
\begin{eqnarray}\label{almost_solved_curly_W}
 \mathcal{W}^{\mathcal{Q}_1^l;\mathcal{Q}_g^l}(V_1;V_g|\beta,u_{\mu},\mu_j)
\Big|_{\mathcal{Q}_g^l=\mathcal{Q}^l_{g,eq}}
 &\simeq& \frac{Z(V_1,\beta,u_{\mu},\mu_j)~Z(V_g-V_1,\beta,u_{\mu},\mu_j)}{
Z(V_g,\beta,u_{\mu},\mu_j)} \nonumber \\
 &&\times ~\frac{(2 \pi V_g )^{L/2} \det \sigma}{(2 \pi \left(V_g -V_1 \right))^{L/2} \det \sigma} 
\exp \left( -\frac{1}{2}~ \frac{1}{(V_g-V_1) }~ \xi^l \xi_l\right)~.
\end{eqnarray}
The GCE partition functions are multiplicative in the sense that 
$Z(V_1,\beta,u_{\mu},\mu_j)~Z(V_g-V_1,\beta,u_{\mu},\mu_j) = Z(V_g,\beta,u_{\mu},\mu_j)$,
 and thus the first term in Eq.(\ref{almost_solved_curly_W}) is equal to unity. 
Now using Eq.(\ref{lambda_def}),  $\lambda = V_1 / V_g$, we can re-write 
Eq.(\ref{almost_solved_curly_W}) as:
\begin{eqnarray}\label{solved_curly_W}
 \mathcal{W}^{\mathcal{Q}_1^l;\mathcal{Q}_g^l}(V_1;V_g|\beta,u_{\mu},\mu_j)
\Big|_{\mathcal{Q}_g^l=\mathcal{Q}^l_{g,eq}}
 &\simeq& ~\frac{1}{(1-\lambda)^{L/2} } 
\exp \left( -\frac{1}{2} \left( \frac{\lambda}{1-\lambda}\right) 
\frac{1}{V_1}~ \xi^l \xi_l\right)~.
\end{eqnarray}
Model parameters are hence the intensive variables inverse temperature $\beta$, four-velocity
$u^{\mu}$ and chemical potentials $\mu^j$, which regulate energy and charge densities, 
and collective motion.
Provided $V_1$ is sufficiently large, we have defined a family of thermodynamically equivalent 
ensembles, which can now be studied in their dependence of fluctuation and correlation 
observables on the size of the bath $V_2 = V_g - V_1$. Hence, we can test the sensitivity of
such observables, for example, to globally applied conservation laws. The expectation 
values $\langle \dots \rangle$ are then identical to GCE expectation values, 
while higher moments will depend  crucially on the choice of $\lambda$.

\subsection{The Limits of $\mathcal{W}$}
The largest weight is given to states for which $\xi^l \xi_l = 0$, i.e. with extensive 
quantities $\mathcal{Q}_1^l = \mathcal{Q}_{1,eq.}^l$. Hence, the maximal weight a microstate 
 (or event) at a given value of $\lambda = V_1/V_g$ can assume is 
$\mathcal{W}_{max}^{\mathcal{Q}_1^l;\mathcal{Q}_g^l}(V_1;V_g|\beta,u_{\mu},\mu_j) 
= (1-\lambda)^{-L/2}$. Taking the limits of Eq.(\ref{solved_curly_W}), it is easy to see that:
\begin{equation}
\lim_{\lambda \rightarrow 0} ~ \mathcal{W}^{\mathcal{Q}_1^l;\mathcal{Q}_g^l}
(V_1;V_g|\beta,u_{\mu},\mu_j) ~=~ 1~.
\end{equation}
I.e. for $\lambda = 0$ we sample the GCE, and all events have a weight equal to unity. 
Hence, we also find $\langle \mathcal{W}^2 \rangle = 1$ and therefore 
$\langle (\Delta \mathcal{W})^2 \rangle = 0$, implying a low statistical error.
For $\lambda \rightarrow 1$, we effectively approach a ''sample-reject'' 
procedure, as (for instance) used in  \cite{Bec_MCE,Bec_MC}, and:
\begin{equation}
\lim_{\lambda \rightarrow 1} ~ \mathcal{W}^{\mathcal{Q}_1^l;\mathcal{Q}_g^l}
(V_1;V_g|\beta,u_{\mu},\mu_j) ~\propto~ \delta(\mathcal{Q}_1^l - V_1 \kappa_1^l)~.
\end{equation}
However, as now not all events have equal weight, $\langle (\Delta \mathcal{W})^2 \rangle$ grows 
and so too the statistical error of finite samples. Also, the larger the number
$L$ of extensive quantities considered for re-weighting, the larger 
will be the statistical uncertainty.

\section{The GCE sampling procedure}
\label{Sec_GCEsampling}
The Monte Carlo sampling procedure for a GCE system in the Boltzmann 
approximation is now explained.
The system to be sampled is assumed to be in an equilibrium state enclosed in a volume 
$V_1$ with temperature $T = \beta^{-1}$ and chemical potentials $\mu_j = (\mu_B,\mu_S,\mu_Q)$. 
Additionally, the system is assumed to
be at rest. The four-velocity is then $u^{\mu} = (1,0,0,0)$ and the four-temperature 
is $\beta^{\mu}= (\beta,0,0,0)$. In this case, multiplicity distributions are Poissonian, 
while momentum spectra are of Boltzmann type. \\

The GCE sampling process is composed of four steps, each discussed below.

\subsubsection{Multiplicity Generation}
\label{none}
In the first step, we randomly sample multiplicities $N_1^i$
of all particle species $i$ considered in the model. The expectation value of 
the multiplicity of thermal Boltzmann particles in the GCE is given by:
\begin{equation}\label{psi_meanN}
\langle N_1^i \rangle ~=~ \frac{g_i V_1}{2 \pi^2} ~m_i^2 ~ T~ K_2\left(\frac{m_i}{T}\right)~ 
e^{\mu_i/T}~.
\end{equation}
Multiplicities $\lbrace N_1^i \rbrace_n$ are randomly generated for each event $n$ 
according to Poissonians with mean values $\langle N_1^i \rangle$:
\begin{equation}
P(N_1^i) ~=~ \frac{\langle N_1^i \rangle^{N_1^i}}{N_1^i!} ~ e^{-\langle N_1^i \rangle}~.
\end{equation}
In the above, $m_i$ and $g_i$ are the mass and degeneracy factor of a particle of 
species $i$ respectively.
The chemical potential $\mu_i = \mu_j q_i^j = \mu_B b_i + \mu_S s_i + \mu_Q q_i $, where 
$q_i^j = (b_i,s_i,q_i)$ represents the quantum number content of a particle of species $i$.

\subsubsection{Momentum Spectra}
\label{none}
In the second step, we generate momenta for each particle according to 
a Boltzmann spectrum. For a static thermal source spherical coordinates are convenient:
\begin{equation}
\frac{dN_i}{d|p|} ~=~ \frac{g_i V_1}{2 \pi^2}~ T^3 ~|p|^2~ e^{-\varepsilon/T}~.
\end{equation}
These momenta are then isotropically distributed in momentum space. Hence:
\begin{eqnarray}
p_x &=& |p| ~ \sin \theta ~\cos \phi ~,\\
p_y &=& |p| ~ \sin \theta ~\sin \phi ~,\\
p_z &=& |p| ~ \cos \theta  ~,\\
\varepsilon &=& \sqrt{|p|^2 + m^2}~,
\end{eqnarray}
where $p_x$, $p_y$, and $p_z$ are the components of the three-momentum, $\varepsilon$
is the energy, and $|p| = \sqrt{p_x^2+p_y^2+p_z^2}$ is the total momentum.
The polar and azimuthal angles are sampled according to:
\begin{eqnarray}
\theta &=& \cos^{-1} \left[ 2 \left(x-0.5 \right) \right] ~,\\
\phi &=& 2~\pi \left(x-0.5 \right)~,
\end{eqnarray}
where $x$ is uniformly distributed between $0$ and $1$. Additionally, we calculate 
the transverse momentum $p_T$ and rapidity $y$ for each particle:
\begin{eqnarray}
p_T &=& \sqrt{p_x^2 + p_y^2}~,\\
y &=& \frac{1}{2} \ln \left(\frac{\varepsilon+p_z}{\varepsilon-p_z} \right)~.
\end{eqnarray}
Finally, we distribute particles homogeneously in a sphere of radius $r_1$ and 
calculate decay times based on the Breit-Wigner width of the resonances.

\subsubsection{Resonance Decay}
\label{none}
The third step (if applicable) is resonance decay. We follow the prescription 
used by the authors of the THERMINATOR package~\cite{THERMINATOR},
and perform only 2 and 3 body decays, 
while allowing for successive decay of unstable daughter particles. Only strong decays are 
considered, while weak and electromagnetic decays are omitted.
Particle decay is first calculated in the parent's rest frame, with daughter momenta 
then boosted into the lab frame. Finally, decay positions are generated based on 
the parent's production point, momentum and life time.

Throughout this article, always only the lightest states of the following baryons:
\begin{equation}
\textrm{p} \qquad \textrm{n} \qquad \Lambda \qquad \Sigma^+ \qquad \Sigma^- \qquad \Xi^- \qquad 
\Xi^0 \qquad \Omega^-
 \end{equation}
and mesons:
\begin{equation}
\pi^+ \qquad \pi^- \qquad \pi^0 \qquad K^+ \qquad K^- \qquad K^0 
 \end{equation}
are considered as stable. The system could now be given collective velocity $u^{\mu}$.

\subsubsection{Re-weighting}
\label{none}
In the fourth step, we calculate the values of extensive quantities for the 
events generated by iterating over the particle list of each event. 
For the values of extensive quantities $\mathcal{Q}^l_{1,n}~=~(B_{1,n},S_{1,n},Q_{1,n},E_{1,n},
P_{x,1,n},P_{y,1,n},P_{z,1,n})$ in subsystem $V_1$ of event $n$ we write: 
\begin{equation}
\mathcal{Q}^l_{1,n} ~=~ \sum_{\textrm{particles } i_n } \mathfrak{q}^l_{i_n}~,
\end{equation}
where $\mathfrak{q}^l_{i_n} = \left( b_{i_n}, s_{i_n}, q_{i_n}, \varepsilon_{i_n}, p_{x,i_n}, 
p_{y,i_n}, p_{z,i_n}\right)$ is the `charge vector' of particle $i$  in event $n$.
Based on $\mathcal{Q}^l_{1,n}$ we calculate the weight $w_n$ for the event:
\begin{equation}
w_n = \mathcal{W}^{\mathcal{Q}_{1,n}^l;\mathcal{Q}_g^l}(V_1;V_g|\beta,u_{\mu},\mu_j)~,
\end{equation}
according to Eq.(\ref{solved_curly_W}). 
Please note that all microstates with the same set of extensive 
quantities $\mathcal{Q}^l_{1,n}$ are still counted equally.

\section{Extrapolating Fully Phase Space Integrated Quantities to the MCE}
\label{Sec_ExtraMCE}

We now attempt to extrapolate fully phase space integrated grand canonical results to 
the microcanonical limit. For this we iteratively generate, re-weight, and 
analyze samples of events for various values of $\lambda = V_1/V_g$.
By construction of the weight factor $\mathcal{W}$, Eq.(\ref{solved_curly_W}), we
extrapolate in a systematic fashion such that, for instance, particle momentum spectra 
as well as mean values of extensive quantities remain unchanged. On the other hand, 
all variances and covariances of extensive quantities subject to re-weighting 
converge linearly to their microcanonical values.

This can be seen from the form of the analytical approximation to the grand canonical 
distribution of (fully phase space integrated) extensive quantities 
$P_{gce}(\mathcal{Q}^l_1)$ (from Eq.(\ref{clt_approx})):
\begin{equation}
P_{gce}(\mathcal{Q}^l_1)~\simeq~\frac{1}{(2 \pi V_1)^{L/2} \det \sigma}~ 
\exp \left( - \frac{1}{2}~ \frac{1}{V_1}~ \xi^l \xi_l \right)~,
\end{equation}      
where the variable $\xi^l$ is given by Eq.(\ref{xi}). Now taking the weight factor 
$\mathcal{W}_{\lambda}$, Eq.(\ref{solved_curly_W}), 
($\sigma$ and $\xi_l$ are the same in both equations) we obtain for the distribution
$P_{\lambda}(\mathcal{Q}^l_1)$ of extensive quantities $\mathcal{Q}^l_1$ in subsystem $1$:
\begin{eqnarray}
P_{\lambda}(\mathcal{Q}^l_1) &\simeq&  \mathcal{W}_{\lambda}^{\mathcal{Q}_1^l;\mathcal{Q}_g^l}
~P_{gce}(\mathcal{Q}^l_1)\\
\label{Pbath}
&\simeq& \frac{1}{(2 \pi(1-\lambda) V_1)^{L/2} \det \sigma}~ 
\exp \left( - \frac{1}{2}~ \frac{1}{(1-\lambda)~V_1}~ \xi^l \xi_l \right)~.
\end{eqnarray}
This is essentially the same multivariate normal distribution as the 
grand canonical version $P_{gce}(\mathcal{Q}^l_1)$, 
however linearly contracted. We will compare Monte Carlo results to Eq.(\ref{Pbath}).

The Monte Carlo output is essentially a distribution $P_{MC}(X_1,X_2,X_3,...)$ of a set
of observables $X_1$, $X_2$, $X_3$, etc. For all practical purposes this distribution
is obtained by histograming all events $n$ according to their values of 
$X_{1,n}$, $X_{2,n}$, $X_{3,n}$, etc. and their weight $w_n$. 
One can then define moments of two observables $X_i$ and $X_j$ through:
\begin{equation}\label{MCmoment}
  \langle X_i^n X_j^m \rangle ~\equiv~ \sum_{X_i,X_j} X_i^n X_j^m P_{MC}(X_i,X_j)~.
\end{equation}
Additionally, we define the variance $\langle \left( \Delta X_i \right)^2 \rangle$
and the covariance $\langle \Delta X_i \Delta X_j \rangle$ respectively as:
\begin{eqnarray}\label{variance}
\langle \left( \Delta X_i \right)^2 \rangle &\equiv&  \langle X_i^2\rangle ~-~ 
\langle X_i  \rangle^2~, \qquad \textrm{and} \\
\label{covariance}
\langle \Delta X_i \Delta X_j \rangle &\equiv&  \langle X_i  X_j \rangle 
~-~ \langle X_i \rangle  \langle X_j \rangle~.
\end{eqnarray}
In the following, we use the scaled variance $\omega_i$ and the correlation 
coefficient $\rho_{ij}$ defined as:
\begin{eqnarray}\label{omega}
\omega_i &\equiv& \frac{\langle \left( \Delta X_i \right)^2 \rangle}{\langle X_i \rangle}~, 
\qquad \textrm{and} \\
\label{rho}
\rho_{ij} &\equiv& \frac{\langle \Delta X_i \Delta X_j \rangle}{
\sqrt{\langle \left( \Delta X_i \right)^2 \rangle 
\langle \left( \Delta X_j \right)^2 \rangle }}~.
\end{eqnarray}

Let us consider a static and neutral system with four-velocity 
$u_{\mu} = (1,0,0,0)$, chemical potentials $\mu_j = (0,0,0)$, local temperature 
$T~=~\beta^{-1}=0.160 GeV$, and volume ${V_1=2000fm^3}$. 
This is a system large enough\footnote{Generally it is not easy to say when a system is 
`large enough` for the large volume approximation to be valid. Here we find good 
agreement with asymptotic analytic solutions. Charged systems, or Bose-Einstein/Fermi-Dirac
systems, usually converge more slowly to their asymptotic solution.
}
 for using the large volume 
approximation worked out in Section~\ref{Sec_SEfB}.

\begin{figure}[ht!]
  \epsfig{file=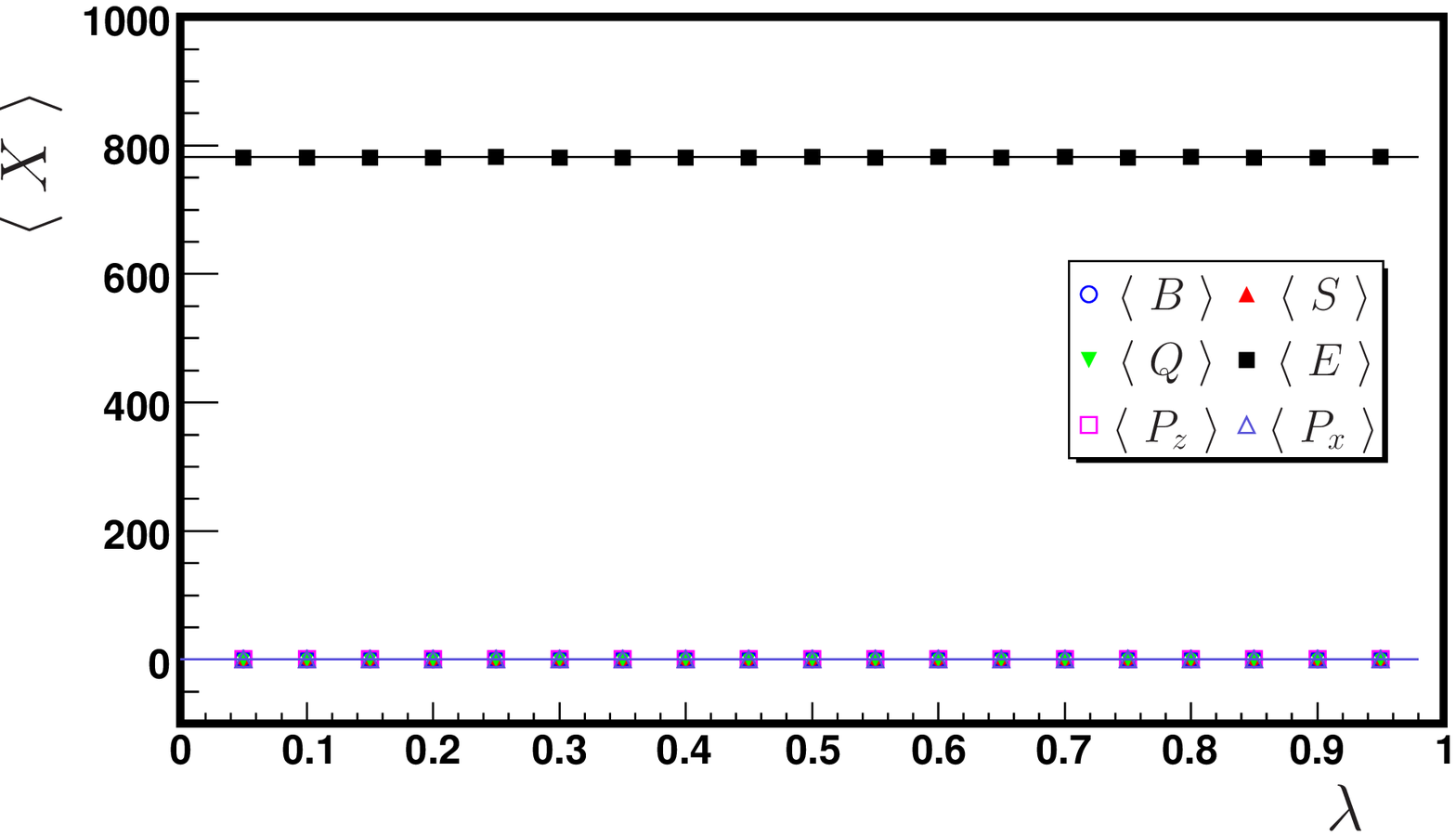,width=8.4cm,height=6.5cm}
  \epsfig{file=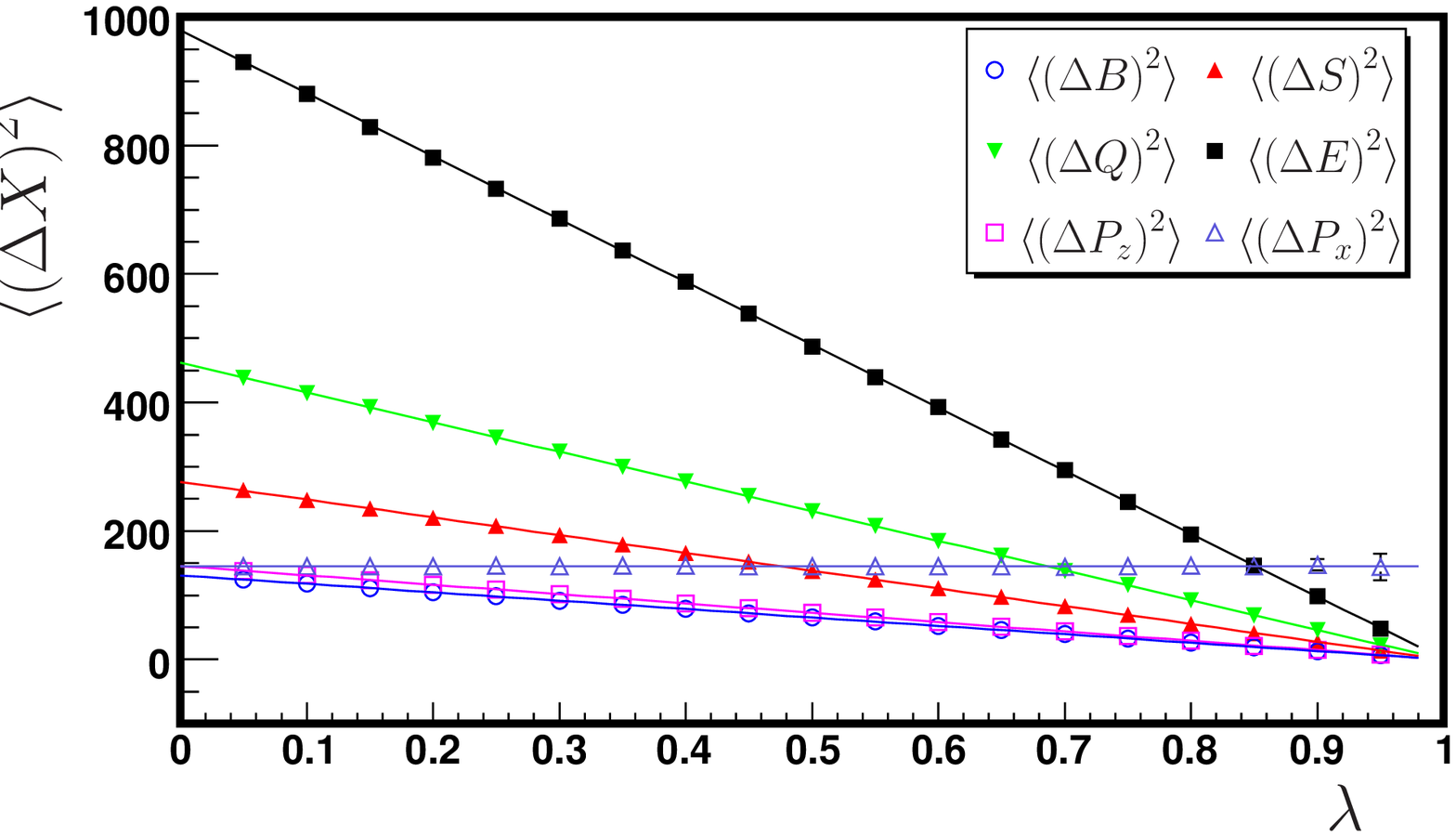,width=8.4cm,height=6.5cm}
  \caption{Mean values ({\it left}) and variances ({\it right}) of various extensive quantities, 
    as listed in the legends, as a function of $\lambda$. 
    Each marker and its error bar represents the result of 20 Monte Carlo 
    runs of $2.5 \cdot 10^4$ events each. 
    19 different equally spaced values of $\lambda$ have been investigated.
    Solid lines indicate GCE values ({\it left}), or linear extrapolations from the GCE value 
    to the MCE limit ({\it right}).}  
  \label{CwL_plot_fluc}
 \end{figure}

\begin{figure}[ht!]
  \epsfig{file=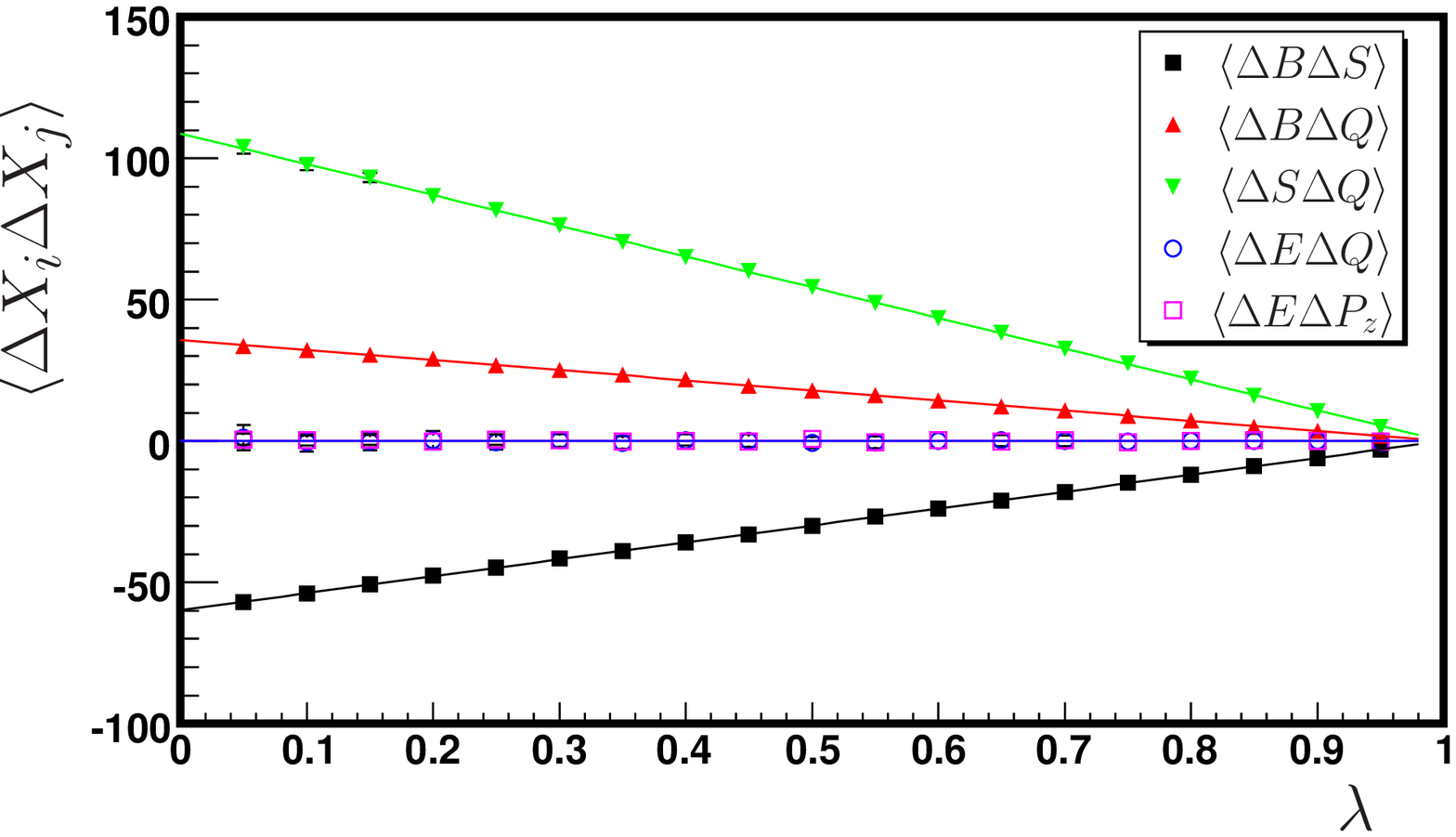,width=8.4cm,height=6.5cm}
  \epsfig{file=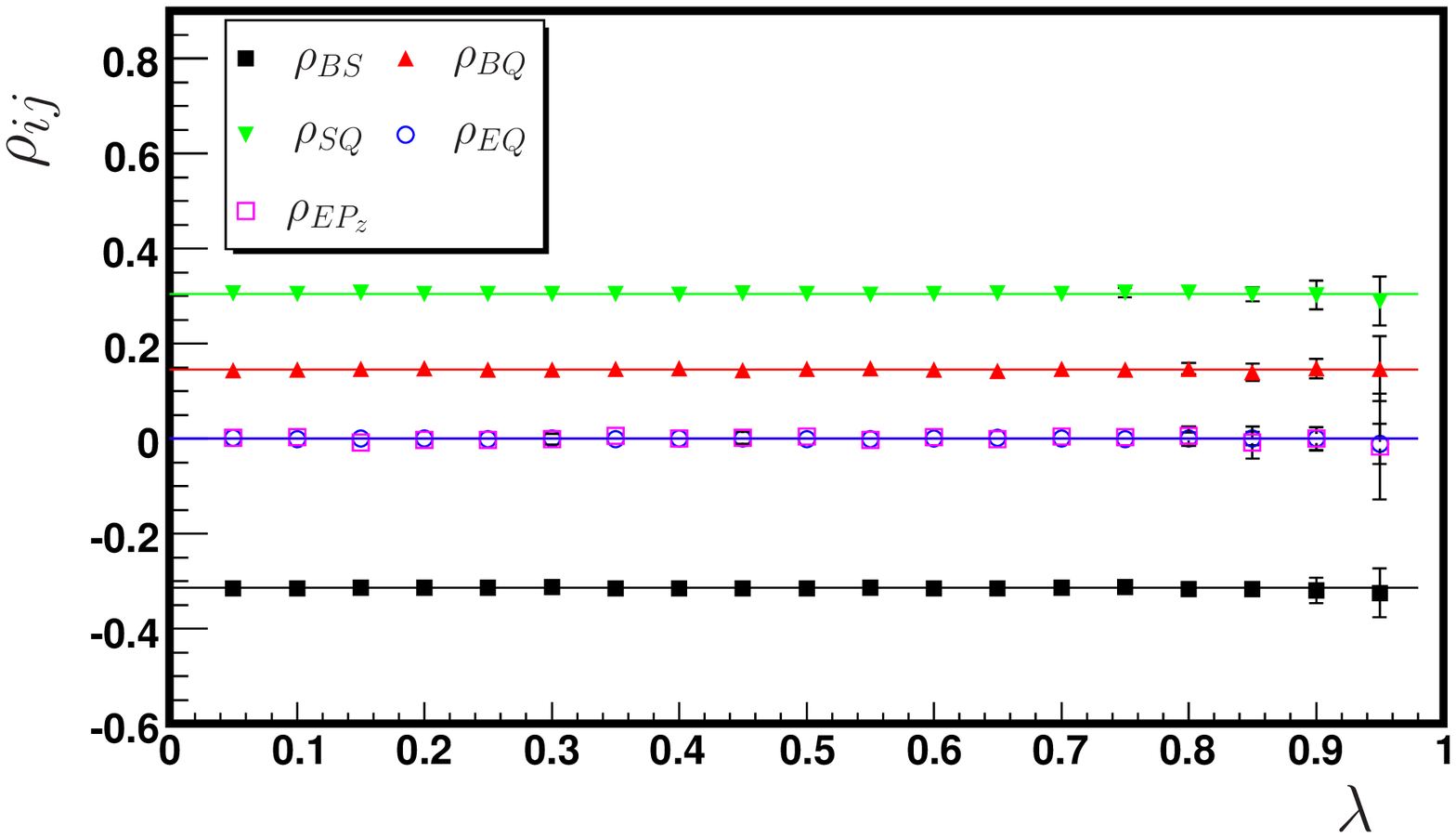,width=8.4cm,height=6.5cm}
  \caption{Covariances ({\it left}) and correlation coefficients ({\it right}) between 
    various extensive quantities, as listed in the legends, as a function of $\lambda$. 
    Solid lines indicate  linear extrapolations from the GCE value 
    to the MCE limit ({\it left}), or GCE values ({\it right}).
    The rest as in Fig.(\ref{CwL_plot_fluc}).}  
  \label{CwL_plot_corr}
  \end{figure}

In Figs.(\ref{CwL_plot_fluc}) and (\ref{CwL_plot_corr}) we show the results of Monte Carlo
runs of $2.5 \cdot 10^4$ events each. Each value of $\lambda$ has been sampled $20$ times 
to allow for calculation of a statistical uncertainty estimate. 19 different values 
of $\lambda$ have been studied. In this case study, the extensive quantities baryon 
number $B$, strangeness $S$, electric charge $Q$, energy $E$, and longitudinal 
 momentum~$P_z$ are considered for re-weighting.
Conservation of transverse momenta $P_x$ and $P_y$ can be shown not to affect the 
$\Delta p_{T,i}$ and $\Delta y_i$ dependence of multiplicity fluctuations and correlations 
studied in the following sections.
Their $\Delta y_i$ dependence is, however, rather sensitive to $P_z$ conservation. 
Angular correlations (not studied in this article), on the other hand, are strongly 
sensitive to joint $P_x$ and $P_y$ conservation~\cite{acc,baseline}.

In Fig.(\ref{CwL_plot_fluc}) ({\it left}) we show the results for mean values 
of baryon number $\langle B \rangle$, strangeness $\langle S \rangle$, electric charge 
$\langle Q\rangle$, energy $\langle E \rangle$, and the
momenta $\langle P_x \rangle$ and $\langle P_z \rangle$.
The solid lines represent GCE values. Only the expectation value of energy is not equal to 0, 
as the system sampled is assumed to be static and neutral with $T \not= 0$. 
The evolution of the respective variances is shown in Fig.(\ref{CwL_plot_fluc}) ({\it right}). 
Variances of extensive quantities subject to re-weighting converge linearly 
to~$0$ as~$\lambda$ goes to~$1$. One notes that 
$\langle \left( \Delta P_x \right)^2 \rangle$ remains constant (within error bars), 
as this quantity is not re-weighted in this case study.
Please note that on many data points the error bars are smaller than the symbol used.

In Fig.(\ref{CwL_plot_corr}) ({\it left}) we show the evolution of covariances 
$\langle \Delta B \Delta S \rangle$, $\langle \Delta B \Delta Q \rangle$, 
$\langle \Delta S \Delta Q \rangle$, and $\langle \Delta E \Delta Q \rangle$
with the `size of the bath'. As seen, the covariances between quantities 
considered for re-weighting also converge linearly to 0. In a neutral system,  
covariances between energy and charge are equal to 0. As an example, we show 
$\langle \Delta E \Delta Q \rangle$. In a static system, also the covariances
between momenta and any other extensive quantity are equal to 0. As
an example, we show $\langle \Delta E \Delta P_z \rangle$. The correlation 
coefficients, Eq.(\ref{rho}), on the other hand, remain constant as a function 
of $\lambda$, as shown in Fig.(\ref{CwL_plot_corr}) ({\it right}). The values 
of fully phase space integrated correlation coefficients $\rho_{BS}$, $\rho_{BQ}$,
and $\rho_{SQ}$ can be compared to the GCE results denoted by the 
solid lines shown in Figs.(\ref{lc_bs_0000} - \ref{lc_sq_0000}) in Section \ref{Sec_LCfluc}.

The variances and covariances converge linearly from their GCE values to their respective 
MCE limits in the large volume limit. 
The dependence of $\langle (\Delta X_i)^2\rangle$, Eq.(\ref{variance}), and 
$\langle \Delta X_i \Delta X_j \rangle$, Eq.(\ref{covariance}), on the size of the bath 
$\lambda$ is given by:
\begin{eqnarray}\label{variance_lambda}
\langle (\Delta X_i)^2\rangle_{\lambda} &=& 
(1-\lambda) ~\langle (\Delta X_i)^2\rangle_{gce}
~+~ \lambda ~\langle (\Delta X_i)^2\rangle_{mce} \\
\label{covariance_lambda}
\langle \Delta X_i \Delta X_j \rangle_{\lambda}&=& 
(1-\lambda) ~\langle \Delta X_i \Delta X_j \rangle_{gce}~
~+~ \lambda ~\langle \Delta X_i \Delta X_j \rangle_{mce}~.
\end{eqnarray}
Mean values $\langle  X_i \rangle_{\lambda}$ remain constant. This implies that the scaled
variance $\omega$ of multiplicity fluctuations, Eq.(\ref{omega}), also converges linearly:
\begin{equation}\label{acc_scaling}
\omega_{\lambda}~\equiv~ 
\frac{\langle (\Delta N_i)^2\rangle_{\lambda}}{\langle  N_i \rangle_{\lambda}}
~=~ (1-\lambda) ~\omega_{gce} ~+~ \lambda ~\omega_{mce}~,
\end{equation}
from its GCE value $\omega_{gce}$ to the MCE limit $\omega_{mce}$. Please note that 
Eqs.(\ref{variance_lambda},\ref{covariance_lambda},\ref{acc_scaling}) are equivalent to the `acceptance 
scaling` approximation\footnote{For the situation discussed here one could 
  equivalently say that particles 
  are randomly drawn from coordinate space of the total volume~$V_g$.
  For the derivation of the acceptance scaling formula \cite{CEfirst} it was, however,
 assumed that particles are randomly drawn from a sample in momentum space.}
 used in \cite{MCEvsData,Res,CEfirst}. For the correlation 
coefficient, Eq.(\ref{rho}), 
\begin{equation}\label{lambda_rho}
\rho_{\lambda} ~\equiv~ \frac{\langle \Delta X_i \Delta X_j \rangle_{\lambda}}{
\sqrt{\langle (\Delta X_i)^2\rangle_{\lambda}\langle (\Delta X_j)^2\rangle_{\lambda}}}~,
\end{equation}
the story is more complicated. In case both $X_i$ and $X_j$ are re-weighted and 
measured in full phase space, we find:
\begin{equation}
\langle (\Delta X_i)^2 \rangle_{mce}~=~
\langle (\Delta X_j)^2 \rangle_{mce}~=~
\langle \Delta X_i  \Delta X_j \rangle_{mce}~=~0~,
\end{equation}
and the correlation coefficient $\rho_{\lambda}$, Eq.(\ref{lambda_rho}), 
is independent of the value of $\lambda$, see Fig.(\ref{CwL_plot_corr}).
In all other cases, one needs to extrapolate 
Eqs.(\ref{variance_lambda},\ref{covariance_lambda})
separately, and then calculate the correlation coefficient.

We have therefore successively transformed our Monte Carlo sample. As $\lambda \rightarrow 1$,
we give larger and larger weight to events in the immediate vicinity of the equilibrium
expectation value, and smaller and smaller weight to events away from it. 
The distribution of extensive quantities considered for re-weighting 
(a multivariate normal distribution in the GCE in the large volume limit) 
hence gets contracted to a $\delta$-function with 
vanishing variances and covariances. I.e., we successively
highlight the properties of events which have very similar values of extensive quantities. 
This will have a bearing on charge correlations and, in particular, 
multiplicity fluctuations and correlations discussed in the following sections.

\begin{figure}[ht!]
  \epsfig{file=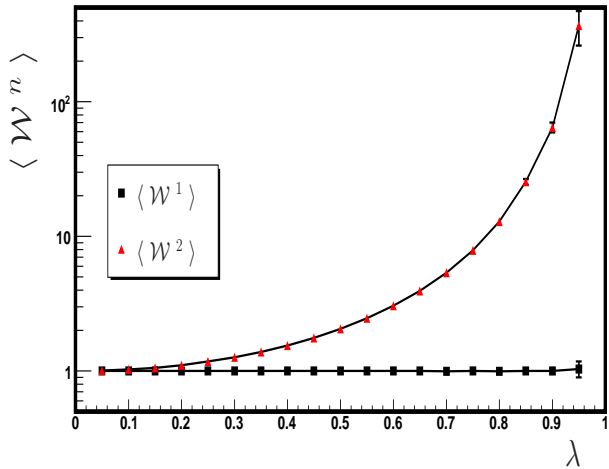,width=8.4cm,height=6.5cm}
  \caption{First and second moment of the weight factor Eq.(\ref{solved_curly_W}) as a 
    function of $\lambda$.
    The rest as in Fig.(\ref{CwL_plot_fluc}).
  }  
  \label{CwL_plot_weight_2nd_mom}
\end{figure} 

The price we pay is that, as $\lambda$ grows, so too does the statistical uncertainty. In the 
limit $\lambda \rightarrow 1$, we approach a sample-reject type of formalism. We
cannot, therefore, directly obtain the microcanonical limit for the large system size 
studied here, as this is prohibited by available computing power.
On the bright side, however, we can extrapolate to this limit. In Fig.(\ref{CwL_plot_weight_2nd_mom}) we 
show the second moment of the weight factor, Eq.(\ref{solved_curly_W}), as a function 
of $\lambda$. A large second moment $\langle \mathcal{W}^2 \rangle$ implies a large 
statistical uncertainty and, hence, usually requires a larger sample. We mention in 
this context that the intermediate ensembles, between the limits of GCE and MCE, may also be 
of phenomenological interest.

\section{Momentum Spectra}
\label{Sec_MomSpect}
\begin{figure}[ht!]
  \epsfig{file=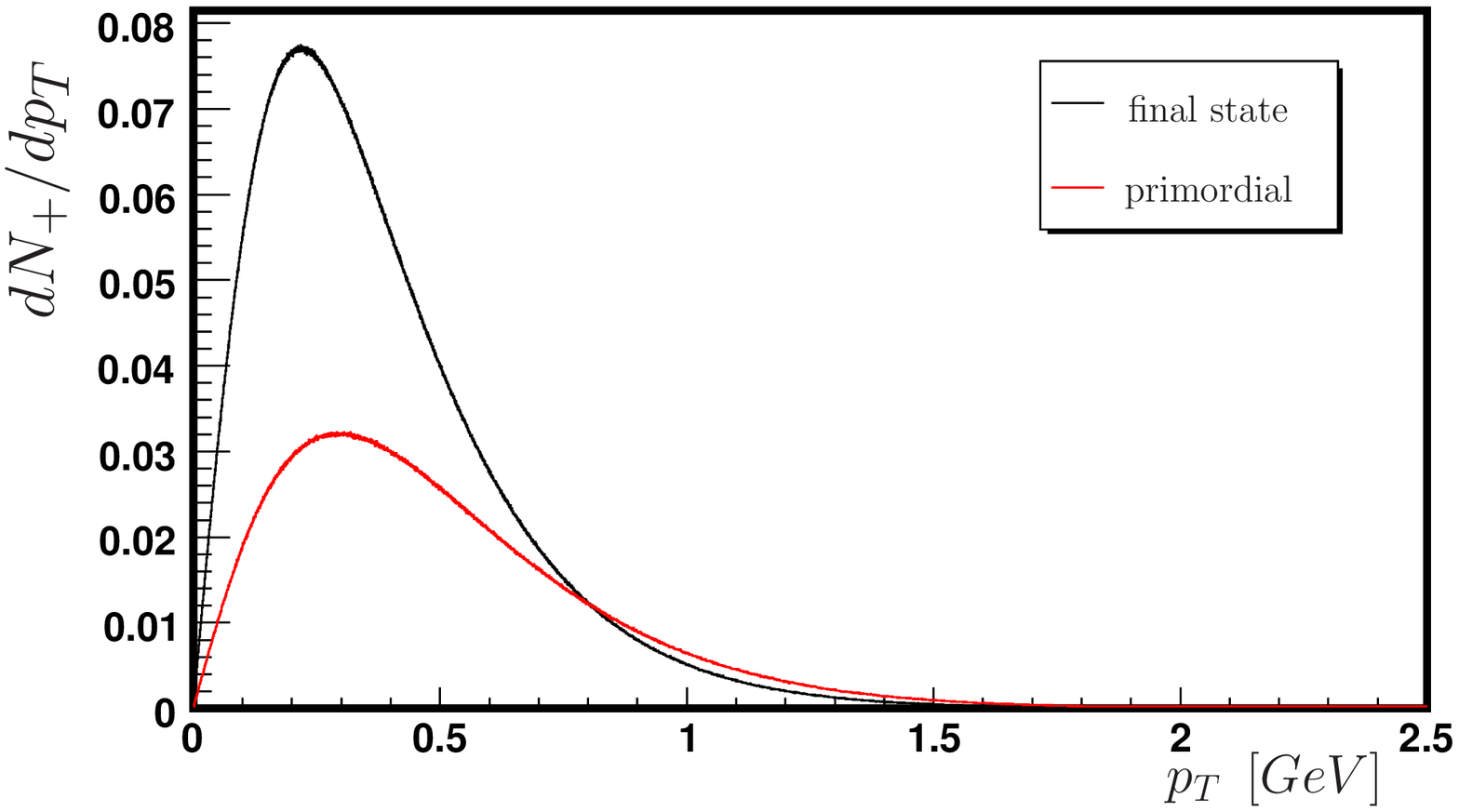,width=8.4cm,height=6.5cm}
  \epsfig{file=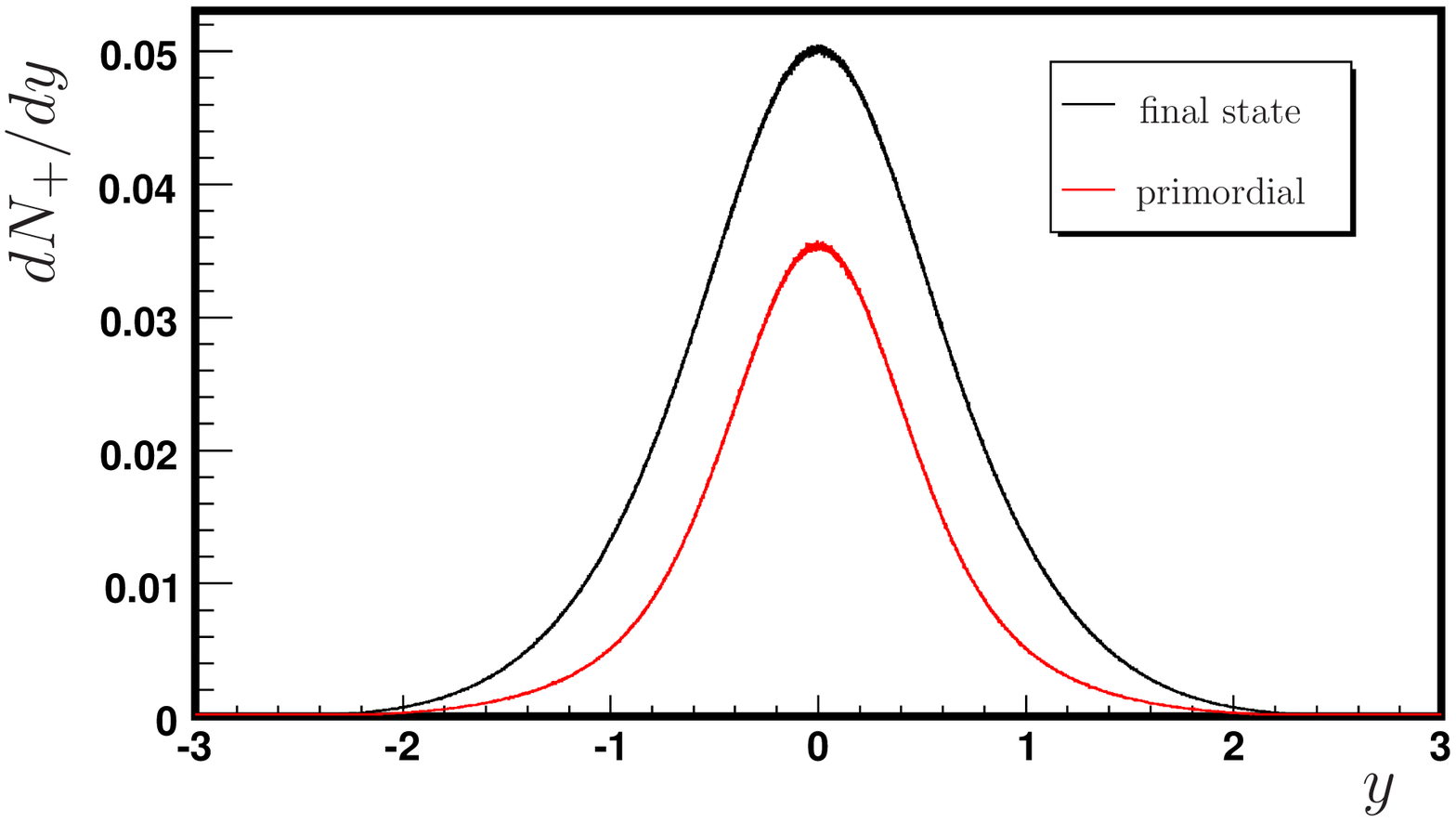,width=8.4cm,height=6.5cm}
  \caption{({\it Left:}) Transverse momentum spectrum of positively charged hadrons,
    both primordial and final state. ({\it Right:}) Rapidity spectrum of 
    positively charged hadrons, both primordial and final state.
    $2 \cdot 10^6$ events have been sampled.}  
  \label{mom_spect}
 \end{figure}

We next consider momentum spectra. 
In Fig.(\ref{mom_spect}) we show transverse momentum and rapidity spectra 
of positively charged hadrons, both primordial and final state, 
for a static thermal system. 

Based on these momentum spectra we construct acceptance bins $\Delta p_{T,i}$ 
and $\Delta y_{i}$, as in~\cite{acc,feq,baseline} and \cite{beni_urqmd,beni_data}. 
Momentum bins are constructed such that each 
of the five bins constructed contains on average one fifth of the total 
yield of positively charged particles. The values defining the bounds of the momentum 
space bins $\Delta p_{T,i}$ and $\Delta y_i$ are summarized in Table~\ref{accbins}.

Resonance decay shifts the transverse momentum distribution to lower average 
transverse momentum $\langle p_T \rangle$ and widens the rapidity distribution of 
thermal `fireballs` \cite{ResDecay}. Final state transverse momentum bins are, 
hence, slightly `contracted`, while final state rapidity bins get slightly `wider`, when 
compared to their respective primordial counterparts.

\begin{table}[h!]
  \begin{center}
    \begin{tabular}{||c||c|c|c|c|c|c||}\hline
      & ~~$p_{T,1}$ [GeV] & $p_{T,2}$ [GeV]  & $p_{T,3}$ [GeV] & $p_{T,4}$ [GeV]  
      & $p_{T,5}$ [GeV]  & ~~$p_{T,6}$ [GeV]  \\
      \hline
      ~primordial~ & 0.0 & 0.22795 & 0.36475 & 0.51825 & 0.73995 & 5.0 \\ 
      ~final state & 0.0 & 0.17105 & 0.27215 & 0.38785 & 0.56245 & 5.0 \\ 
      \hline \hline
      & $y_1$  & $y_2$  & $y_3$  & $y_4$  & $y_5$  & $y_6$  \\
      \hline 
      ~primordial~ & -5.0 & -0.4275 & -0.1241 & 0.1241 & 0.4273 & 5.0 \\ 
      ~final state & -5.0 & -0.5289 & -0.1553 & 0.1551 & 0.5289 & 5.0 \\
      \hline
    \end{tabular}
    \caption{Transverse momentum and rapidity bins 
      $\Delta p_{T,i} = \left[p_{T,i},p_{T,i+1} \right]$
      and $\Delta y_{i} = \left[y_{i},y_{i+1} \right]$, both primordial and final state,
      for a static neutral Boltzmann system with temperature $T=0.160GeV$.
    } \label{accbins}
  \end{center}
\end{table}

Resonance decay combined with transverse as well as longitudinal flow is believed to 
provide a rather good description of experimentally observed momentum spectra 
in relativistic heavy ion collisions at SPS and RHIC energies 
\cite{SolfrankHeinz,THERMINATOR,BecCley}. Our spectra, on the other hand, contain no flow and 
our results thus cannot be directly compared to experimental data
or transport simulations.
However, qualitatively one might observe effects of the kind discussed in the following.

\section{The Momentum Space Dependence of Correlations between conserved charges}
\label{Sec_LCfluc}

An interesting example of quantities for which the measured value depends on the 
observed part of the momentum spectrum are the correlation coefficients between the 
charges baryon number $B$, strangeness $S$ and electric charge~$Q$. 
Please note that also the variances and covariances of the baryon number, strangeness, 
and electric charge distribution are sensitive to the acceptance cuts applied. Their 
values are additionally rather sensitive to the effects of globally 
enforced conservation laws. If the size of the `bath` is reduced, a change in one 
interval of phase space will have to be balanced (preferably) by a change in another 
interval, and not by the `bath`.

\subsection{Grand Canonical Ensemble}
\label{none}

We will now consider the correlation coefficients $\rho_{BS}$, $\rho_{BQ}$, and $\rho_{SQ}$
in limited acceptance bins $\Delta p_{T,i}$ and $\Delta y_i$, as defined in Table~\ref{accbins},
in the grand canonical ensemble. Particles in one momentum bin are then essentially sampled
independently from particles in any other momentum space segment, due to the 
`infinite bath` assumption.  Nevertheless, the 
way in which quantum numbers are correlated is different in different momentum bins, as 
different particle species have, due to their different masses, different momentum spectra.

Let us first make some basic observations about the hadron resonance gas and the way
in which quantum numbers are correlated in a GCE. Charge fluctuations directly probe
 the degrees of freedom of a system, i.e. they are sensitive to its particle mass spectrum (and 
its quantum number configurations). We first consider the contribution of different particle
species to the covariance $\langle \Delta X_i \Delta X_j \rangle$, Eq.(\ref{covariance}), 
and hence to the correlation coefficient $\rho_{ij}$, Eq.(\ref{rho}).

\begin{figure}[ht!]
  \epsfig{file=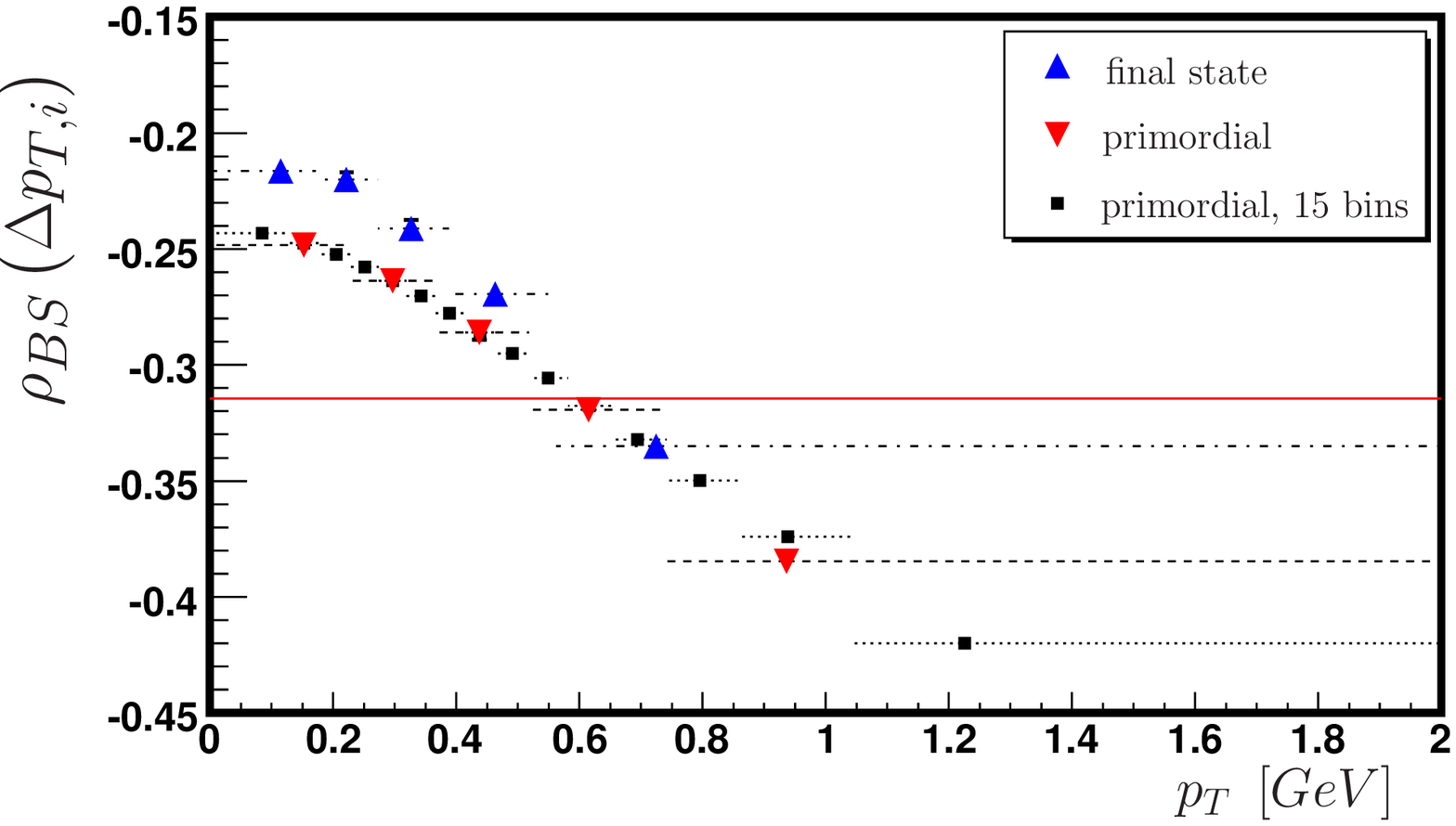,width=8.4cm,height=6.5cm}
  \epsfig{file=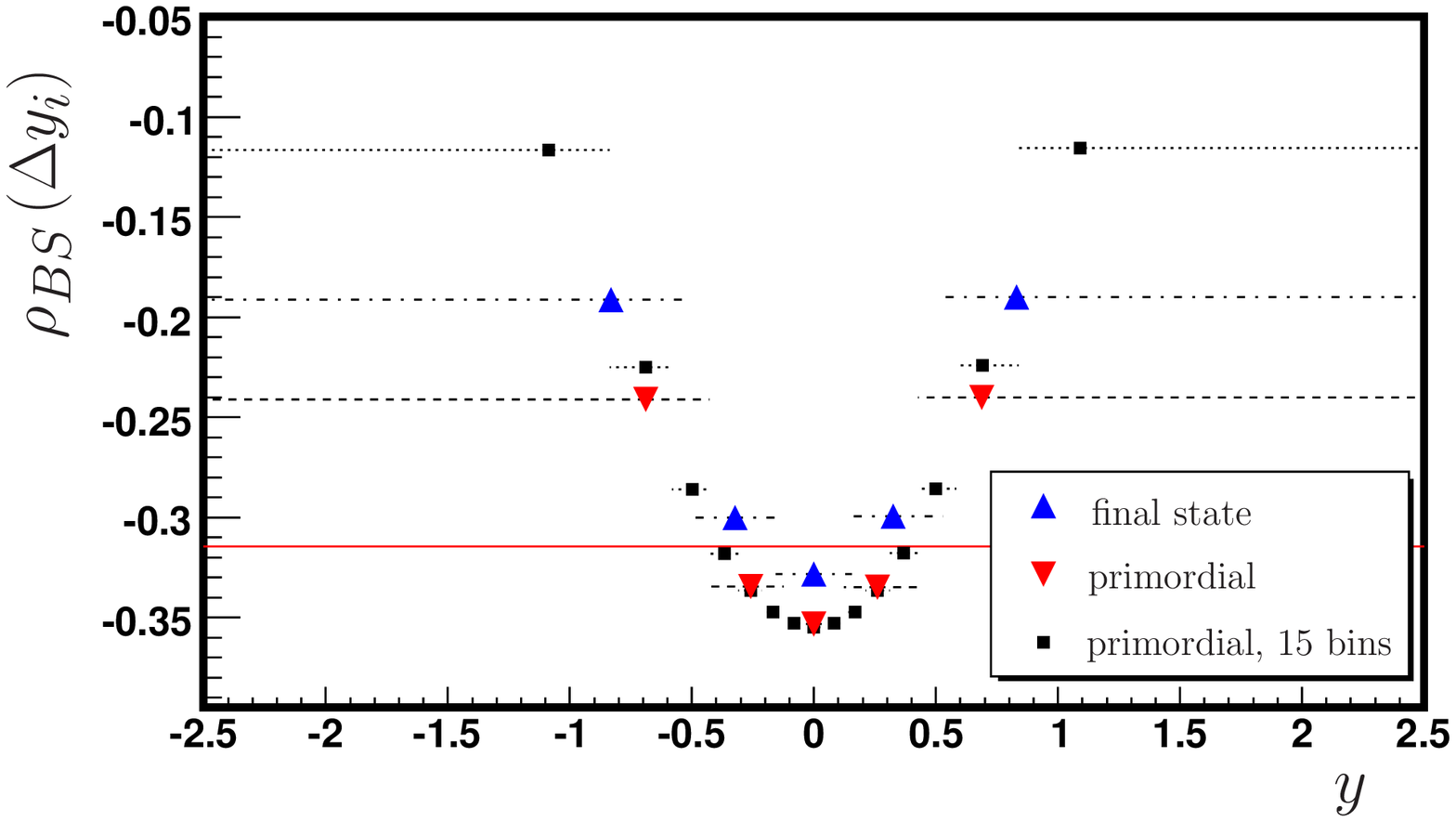,width=8.4cm,height=6.5cm}
  \caption{Baryon-strangeness correlation coefficient $\rho_{BS}$ in the GCE in limited acceptance windows,
    both primordial and final state.
    ({\it Left:}) transverse momentum bins  $\Delta p_{T,i}$. 
    ({\it Right:}) rapidity bins $\Delta y_i$.  
    Horizontal error bars indicate the width and position of the momentum bins 
    (And not an uncertainty!).
    Vertical error bars indicate the statistical uncertainty of $20$ Monte Carlo 
    runs of $10^5$ events each.
    The marker indicates the center of gravity of the corresponding bin.
    The solid lines show the fully phase space integrated GCE result.}  
  \label{lc_bs_0000}
 \end{figure}

All baryons have baryon number $b=+1$. Baryons can only carry strange quarks, i.e. 
their strangeness is always $s\le0$. Anti-baryons have $b=-1$, and $s\ge0$. Hence, both 
groups contribute negatively to the baryon-strangeness covariance, and so 
$\langle \Delta B \Delta S \rangle < 0$, and therefore ${\rho_{BS}<0}$, as indicated 
by the solid lines in Fig.(\ref{lc_bs_0000}).

\begin{figure}[ht!]
  \epsfig{file=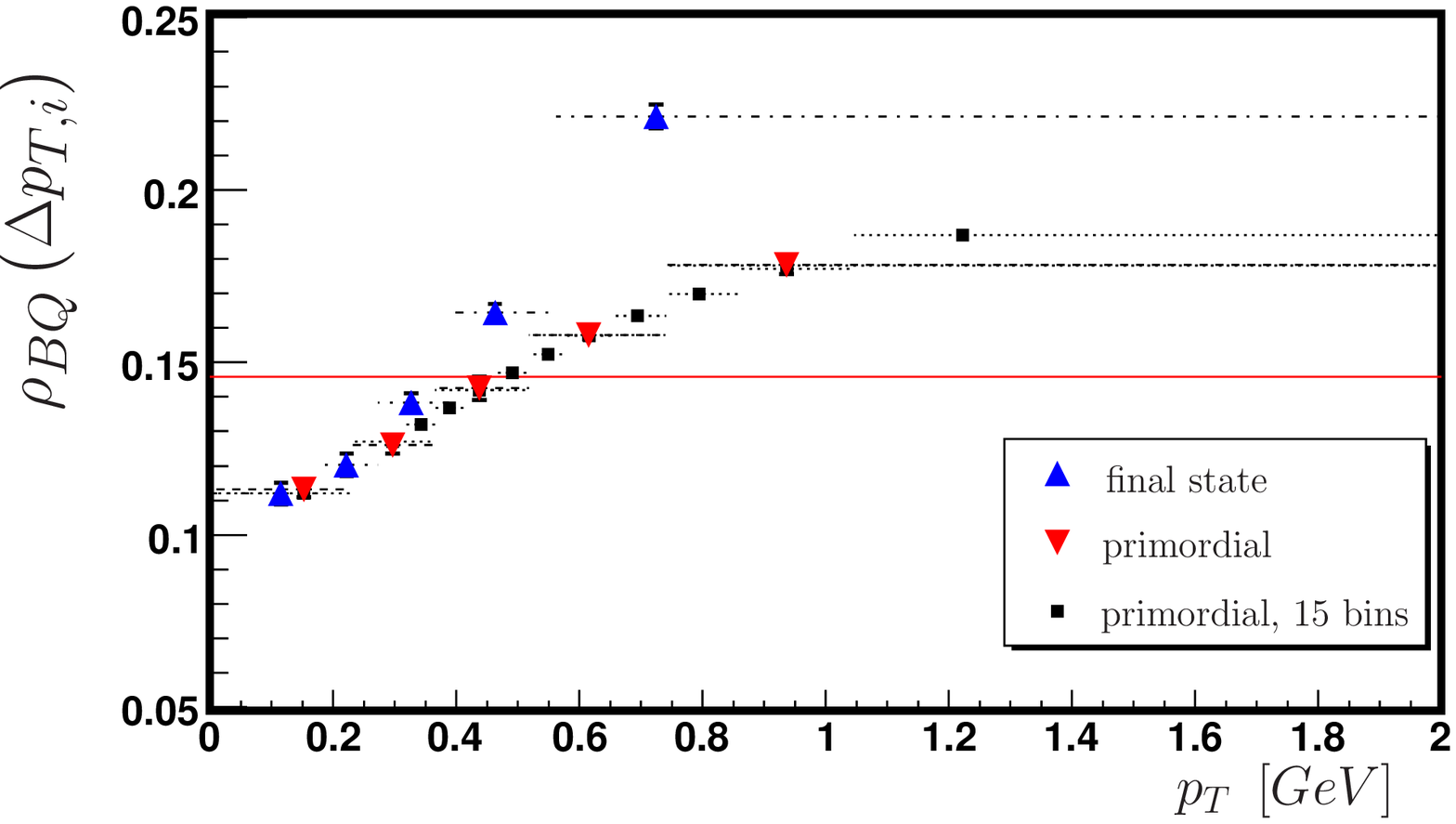,width=8.4cm,height=6.5cm}
  \epsfig{file=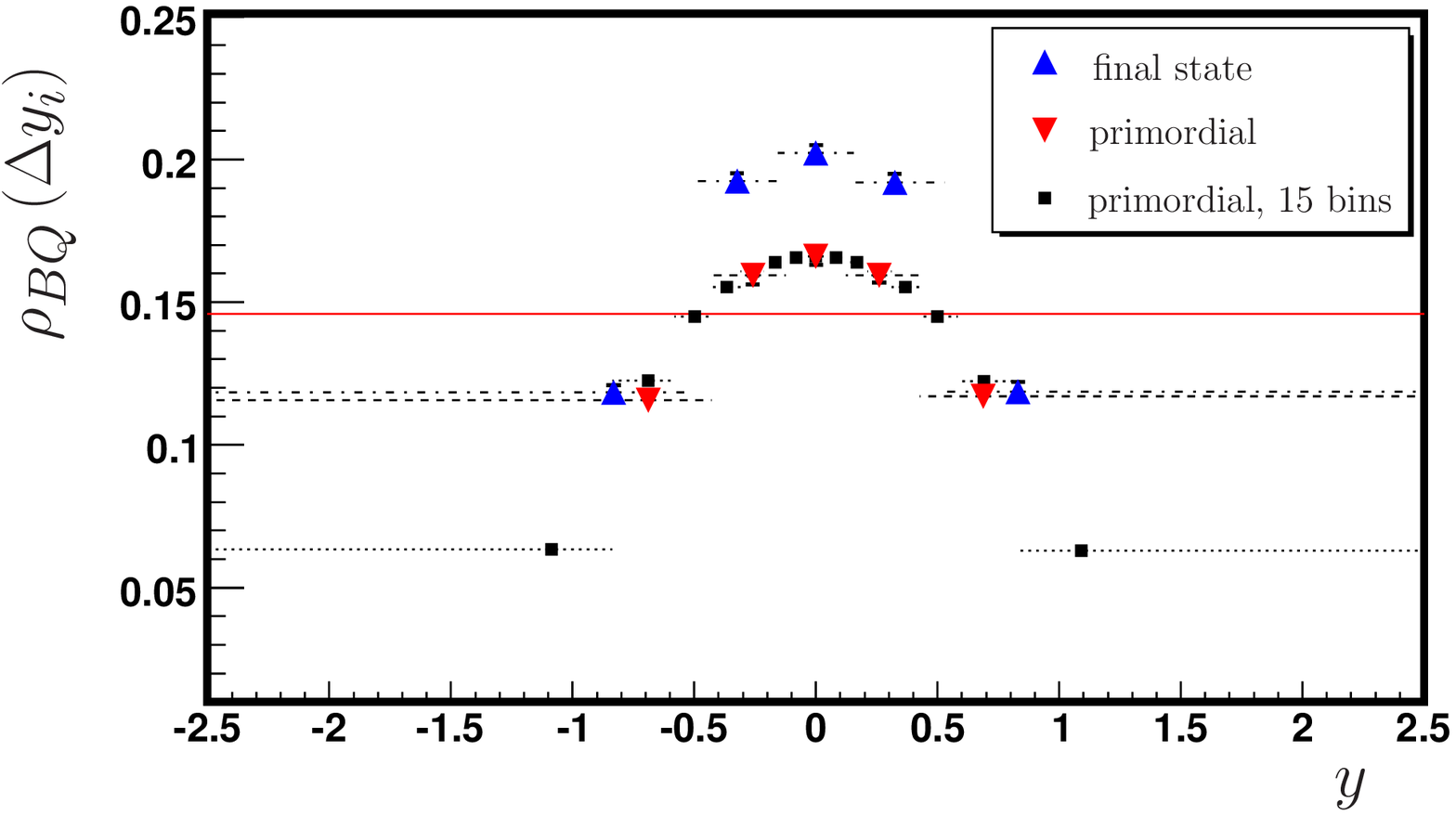,width=8.4cm,height=6.5cm}
  \caption{Baryon-electric charge correlation coefficient $\rho_{BQ}$ in the GCE in limited 
    acceptance windows, both primordial and final state.
    ({\it Left:}) transverse momentum bins $\Delta p_{T,i}$. 
    ({\it Right:}) rapidity bins  $\Delta y_i$.
    The rest as in Fig.(\ref{lc_bs_0000}).  }  
  \label{lc_bq_0000}
 \end{figure}

Positively charged baryons and their anti-particles contribute positively to
the baryon-electric charge covariance $\langle \Delta B \Delta Q \rangle$, while negatively 
charged baryons (and their anti-particles) contribute negatively. Two observations can
be made on the hadron resonance gas mass spectrum: there are more positively charged baryons
than negatively charged ones, and their average mass is lower. I.e., in a neutral gas 
($\mu_B=\mu_Q=\mu_S = 0$) the contribution of positively charged baryons dominates and 
therefore $\langle \Delta B \Delta Q \rangle >0$ and ${\rho_{BQ}>0}$, as indicated 
by the solid lines in Fig.(\ref{lc_bq_0000}).

\begin{figure}[ht!]
  \epsfig{file=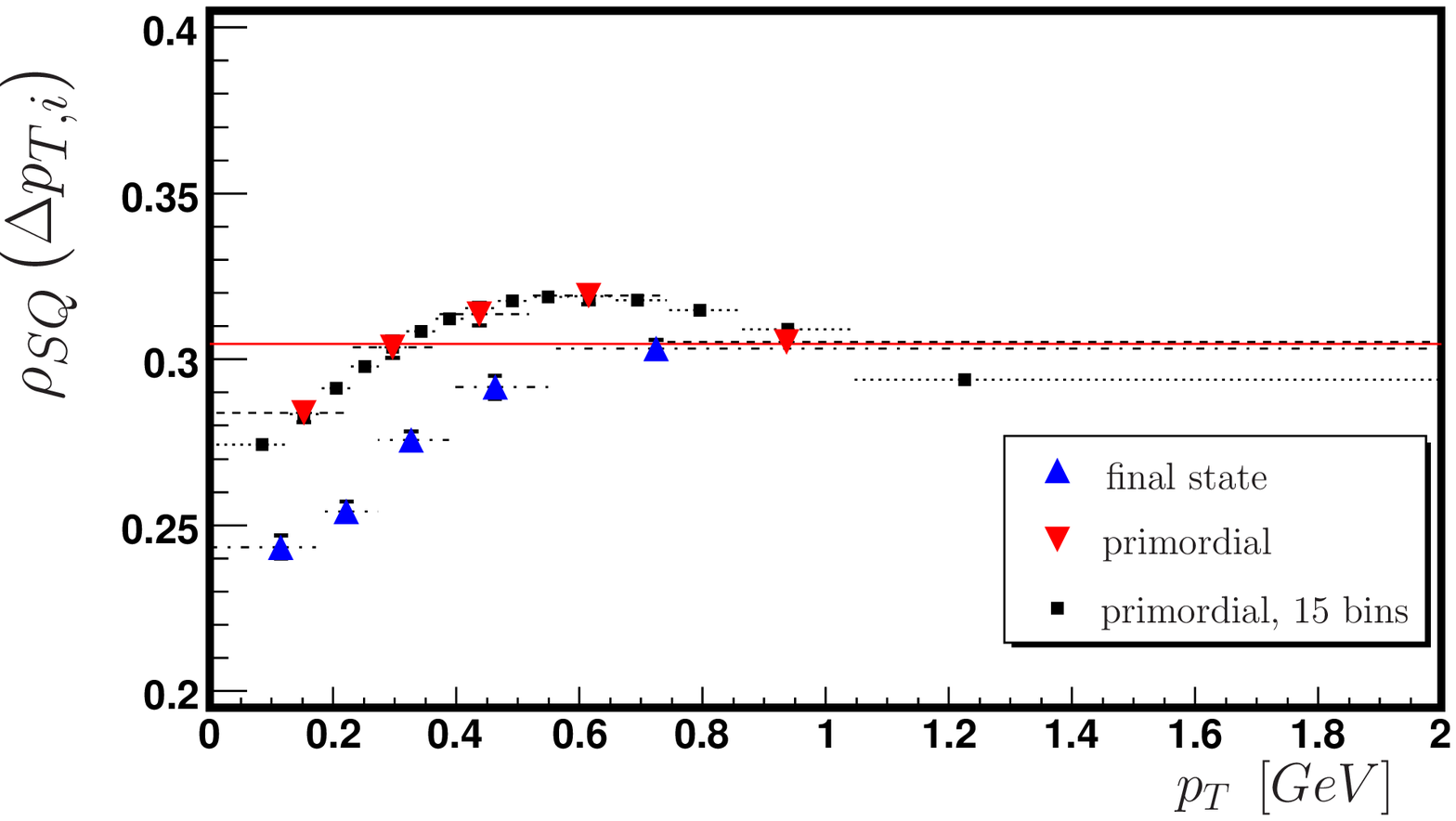,width=8.4cm,height=6.5cm}
  \epsfig{file=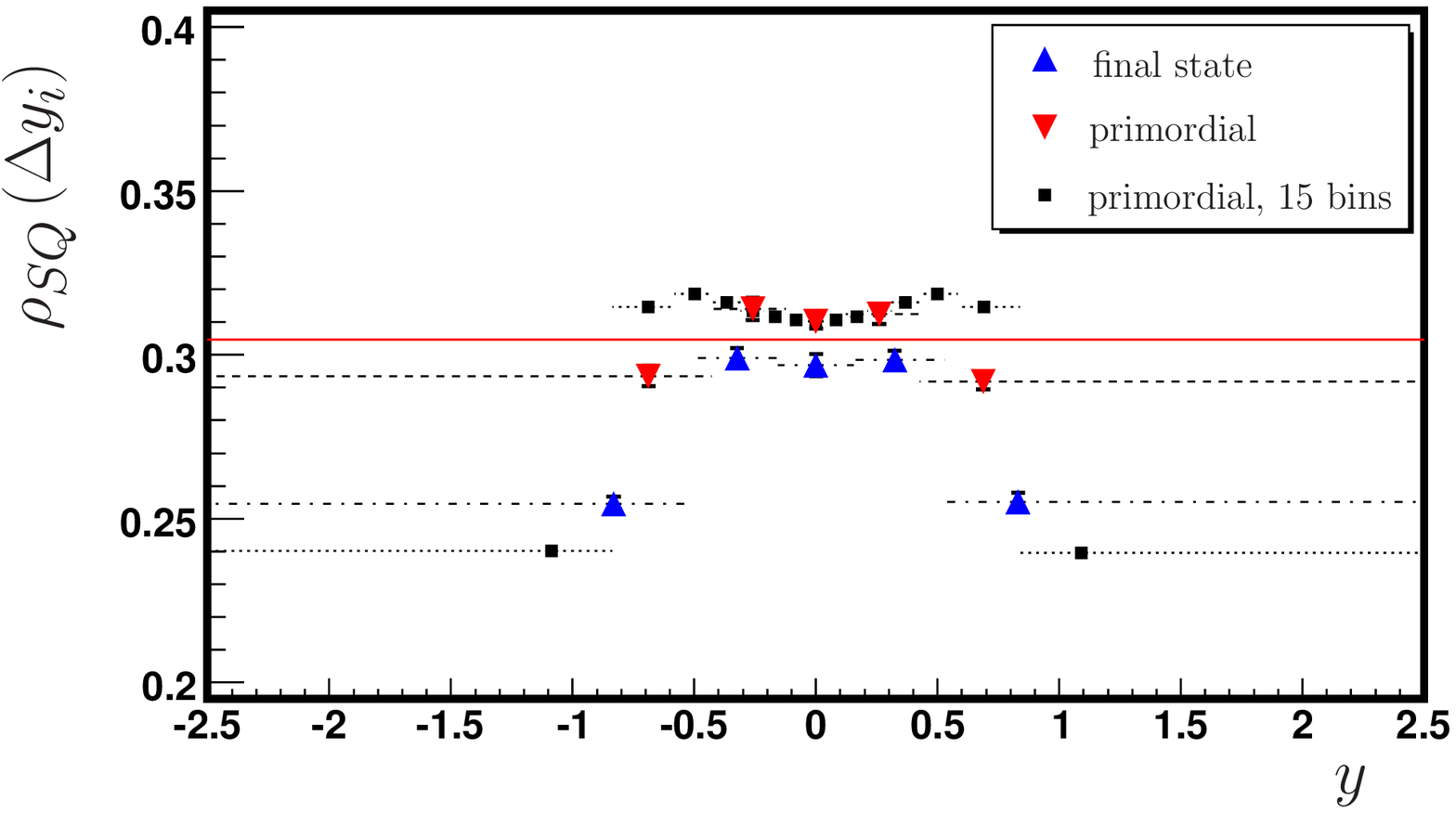,width=8.4cm,height=6.5cm}
  \caption{Strangeness-electric charge correlation coefficient $\rho_{SQ}$ in the GCE in limited 
    acceptance windows, both primordial and final state.
    ({\it Left:}) transverse momentum bins $\Delta p_{T,i}$. 
    ({\it Right:}) rapidity bins  $\Delta y_i$.
    The rest as in Fig.(\ref{lc_bs_0000}).}
  \label{lc_sq_0000}
 \end{figure}

Mesons and their anti-particles always contribute positively to the 
strangeness-electric charge correlation coefficient $\rho_{SQ}$. Electrically 
charged strange mesons are either composed of a $u$-quark and an {$\bar{s}$-quark}, 
or of an $\bar{u}$-quark and a $s$-quark (and superpositions thereof). 
Their contribution to 
$\langle \Delta S \Delta Q \rangle $ is in either case positive. On the baryonic side, only the
$\Sigma^+$ (as well as its degenerate states and their respective anti-particles) has a negative 
contribution to $\langle \Delta S \Delta Q \rangle $, while all other strangeness carrying
baryons have either electric charge $q=-1$, or $q=0$. Therefore, we find ${\rho_{SQ} >0}$, 
as indicated by the solid lines in Fig.(\ref{lc_sq_0000}).

In Figs.(\ref{lc_bs_0000}-\ref{lc_sq_0000}) we show the correlation coefficients 
$\rho_{BS}$ (baryon number - strangeness), $\rho_{BQ}$ (baryon number - electric charge), 
and $\rho_{SQ}$ (strangeness - electric charge) as measured in the acceptance 
bins $\Delta p_{T,i}$ and $\Delta y_i$ defined in Table~\ref{accbins}, 
both primordial and final state. 
The average baryon number, strangeness, and electric charge in each
bin is equal to zero, as the system is assumed to be neutral.
The analytical primordial values  (15 bins) shown in Figs.(\ref{lc_bs_0000}-\ref{lc_sq_0000}) 
are calculated using analytical spectra. 
Please note that, again, on many data points the error bars are smaller than the symbol used.

In Tables~\ref{accbins_LC_BS}~to~\ref{accbins_LC_SQ} we summarize the transverse 
momentum and rapidity  dependence of the correlation coefficients 
$\rho_{BS}$, $\rho_{BQ}$, and $\rho_{SQ}$. The statistical error quoted corresponds to 
20 Monte Carlo runs of $10^5$ events each. The analytical values (5 bins) listed 
in the tables are calculated using the momentum bins defined in Table~\ref{accbins}.
 Mild differences between Monte Carlo and analytical
results are unavoidable. The analytical values are also not exactly 
symmetric in $\Delta y_i$, as the exact size of the acceptance bins constructed is 
sensitive to the number of bins used for the calculation of the momentum spectra. 
The values of the correlation coefficient $\rho$ are also  
rather sensitive to exact bin size, and the fourth digit becomes somewhat unreliable.

\begin{table}[h!]
  \begin{center}
    \begin{tabular}{||c||c|c|c|c|c||}\hline
      $\rho_{BS}$ & $\Delta p_{T,1}$ & $\Delta p_{T,2}$ & $\Delta p_{T,3}$ & $\Delta p_{T,4}$ 
      & $\Delta p_{T,5}$   \\
      \hline
      ~$\rho_{prim}^{calc}$ ~&~ $-0.2479$ ~&~$-0.2641$ ~&~$-0.2864$ ~&~$-0.3188$ ~&~$-0.3839$ ~ \\
      ~$\rho_{prim}$ ~&~ $-0.248 \pm 0.003$ ~&~$-0.264 \pm 0.003$ ~&~$-0.286 \pm 0.003$ ~
      &~$-0.319 \pm 0.002$ ~&~$-0.385 \pm 0.002$ ~ \\
      ~$\rho_{final}$ ~&~ $-0.216 \pm 0.002$ ~&~$-0.220 \pm 0.003$ ~
      &~$-0.241 \pm 0.004$ ~&~$-0.269 \pm 0.003$ ~&~$-0.335 \pm 0.003$ ~ \\
      \hline \hline
      $\rho_{BS}$ & $\Delta y_1$  & $\Delta y_2$  & $\Delta y_3$  & $\Delta y_4$  
      & $\Delta y_5$ \\
      \hline 
      ~$\rho_{prim}^{calc}$ ~&~ $-0.2407$ ~&~$-0.3345$ ~&~$-0.3536$ ~&~$-0.3345$ ~&~$-0.2408$ ~ \\
      ~$\rho_{prim}$ ~&~ $-0.241 \pm 0.003$ ~&~$-0.334 \pm 0.003$ ~
      &~$-0.353 \pm 0.003$ ~&~$-0.335 \pm 0.003$ ~&~$-0.240 \pm 0.003$ ~ \\
      ~$\rho_{final}$ ~&~ $-0.191 \pm 0.002$ ~&~$-0.300 \pm 0.002$ ~
      &~$-0.328 \pm 0.002$ ~&~$-0.299 \pm 0.002$ ~&~$-0.190 \pm 0.002$ ~ \\
      \hline
    \end{tabular}
    \caption{Baryon-strangeness correlation coefficient $\rho_{BS}$ in the GCE in transverse momentum 
      bins $\Delta p_{T,i}$ and rapidity bins $\Delta y_i$, both primordial and final state. 
      For comparison, analytical values $\rho_{prim}^{calc}$ for primordial correlations 
      are included.
      The statistical uncertainty corresponds to $20$ Monte Carlo runs of $10^5$ events each.
    }
    \label{accbins_LC_BS}
  \end{center}
\end{table}

\begin{table}[h!]
  \begin{center}
    \begin{tabular}{||c||c|c|c|c|c||}\hline
      $\rho_{BQ}$ & $\Delta p_{T,1}$ & $\Delta p_{T,2}$ & $\Delta p_{T,3}$ & $\Delta p_{T,4}$ 
      & $\Delta p_{T,5}$   \\
      \hline
      ~$\rho_{prim}^{calc}$ ~&~ $0.1120$ ~&~$0.1271$ ~&~$0.1420$ ~&~$0.1579$ ~&~$0.1781$ ~ \\
      ~$\rho_{prim}$ ~&~ $0.113 \pm 0.002$ ~&~$0.126 \pm 0.002$ ~&~$0.143 \pm 0.003$ ~
      &~$0.158 \pm 0.002$ ~&~$0.178 \pm 0.003$ ~ \\
      ~$\rho_{final}$ ~&~ $0.112 \pm 0.003$ ~&~$0.120 \pm 0.003$ ~&~$0.138 \pm 0.003$ ~
      &~$0.164 \pm 0.003$ ~&~$0.221 \pm 0.003$ ~ \\
      \hline \hline
      $\rho_{BQ}$ & $\Delta y_1$  & $\Delta y_2$  & $\Delta y_3$  & $\Delta y_4$  
      & $\Delta y_5$ \\
      \hline
      ~$\rho_{prim}^{calc}$ ~&~ $0.1160$ ~&~$0.1601$ ~&~$0.1658$ ~&~$0.1601$ ~&~$0.1160$ ~ \\
      ~$\rho_{prim}$ ~&~ $0.116 \pm 0.002$ ~&~$0.160 \pm 0.003$ ~&~$0.166 \pm 0.003$ ~
      &~$0.159 \pm 0.003$ ~&~$0.117 \pm 0.002$ ~ \\
      ~$\rho_{final}$ ~&~ $0.118 \pm 0.003$ ~&~$0.192 \pm 0.003$ ~&~$0.202 \pm 0.003$ ~
      &~$0.192 \pm 0.003$ ~&~$0.119 \pm 0.003$ ~ \\
      \hline
    \end{tabular}
    \caption{
      Baryon-electric charge correlation coefficient $\rho_{BQ}$ in the GCE in transverse momentum 
      bins $\Delta p_{T,i}$ and rapidity bins $\Delta y_i$, both primordial and final state. 
    }
    \label{accbins_LC_BQ}
  \end{center}
\end{table}

\begin{table}[h!]
  \begin{center}
    \begin{tabular}{||c||c|c|c|c|c||}\hline
      $\rho_{SQ}$ & $\Delta p_{T,1}$ & $\Delta p_{T,2}$ & $\Delta p_{T,3}$ & $\Delta p_{T,4}$ 
      & $\Delta p_{T,5}$   \\
      \hline
      ~$\rho_{prim}^{calc}$ ~&~ $0.2831$ ~&~$0.3033$ ~&~$0.3150$ ~&~$0.3185$ ~&~$0.3055$ ~ \\
      ~$\rho_{prim}$ ~&~ $0.284 \pm 0.003$ ~&~$0.304 \pm 0.003$ ~&~$0.314 \pm 0.003$ ~
      &~$0.319 \pm 0.002$ ~&~$0.305 \pm 0.002$ ~ \\
      ~$\rho_{final}$ ~&~ $0.243 \pm 0.003$ ~&~$0.254 \pm 0.003$ ~&~$0.276 \pm 0.003$ ~
      &~$0.292 \pm 0.003$ ~&~$0.303 \pm 0.002$ ~ \\
      \hline \hline
      $\rho_{SQ}$ & $\Delta y_1$  & $\Delta y_2$  & $\Delta y_3$  & $\Delta y_4$  
      & $\Delta y_5$ \\
      \hline 
      ~$\rho_{prim}^{calc}$ ~&~ $0.2934$ ~&~$0.3137$ ~&~$0.3104$ ~&~$0.3137$ ~&~$0.2934$ ~ \\
      ~$\rho_{prim}$ ~&~ $0.294 \pm 0.003$ ~&~$0.314 \pm 0.003$ ~&~$0.310 \pm 0.002$ ~
      &~$0.312 \pm 0.003$ ~&~$0.292 \pm 0.002$ ~ \\
      ~$\rho_{final}$ ~&~ $0.255 \pm 0.002$ ~&~$0.299 \pm 0.003$ ~&~$0.297 \pm 0.003$ ~
      &~$0.298 \pm 0.003$ ~&~$0.255 \pm 0.003$ ~ \\
      \hline
    \end{tabular}
    \caption{
      Strangeness-electric charge correlation coefficient $\rho_{SQ}$ in the GCE in transverse momentum 
      bins $\Delta p_{T,i}$ and rapidity bins $\Delta y_i$, both primordial and final state. 
    }
    \label{accbins_LC_SQ}
  \end{center}
\end{table}

We next attempt to explain, in turn, the rapidity dependence of $\rho_{BS}$, $\rho_{BQ}$, 
and $\rho_{SQ}$. Strange baryons are, on average, heavier than non-strange baryons, so their 
rapidity distributions are narrower. The kaon rapidity distribution is then, compared
to baryons, again wider.
A change in baryon number (strangeness) at high $|y|$ is less likely to be accompanied 
by a change in strangeness (baryon number) than at low $|y|$. 
The value of $\rho_{BS}$, therefore, drops 
toward higher rapidity, as shown in Fig.(\ref{lc_bs_0000}), ({\it right}). 
By the same argument, we find 
a weakening of the baryon-electric charge correlation $\rho_{BQ}$ at higher rapidity 
(Fig.(\ref{lc_bq_0000}), ({\it right})) as the rapidity distribution of electrically charged particles 
is wider than that of baryons. For the strangeness-electric charge 
correlation coefficient we find first a mild rise, and then a somewhat stronger drop 
of $\rho_{SQ}$ towards higher rapidity. As one shifts ones acceptance window towards higher
values of $|y|$, first the contribution of baryons (in particular $\Sigma^+$) decreases and, as
the meson contribution grows, $\rho_{SQ}$ rises slightly. Towards the highest $|y|$, 
pions again dominate and de-correlate the quantum numbers.\\

The transverse momentum  dependence can be understood as follows:
heavier particles have higher average transverse momentum $\langle p_T \rangle$ and, 
hence, their influence increases towards higher~$p_T$. Heavy particles have a tendency 
to carry several charges, causing the correlation coefficients to grow.

The contribution of strange baryons compared to non-strange baryons grows towards higher 
transverse momentum, as strange baryons have on average larger mass than non-strange baryons.
The correlation coefficient $\rho_{BS}$ thus becomes strongly negative at high $p_T$. As the 
contribution of baryons compared to mesons grows stronger towards larger $p_T$, 
a change in baryon number (electric charge) is now more likely to be accompanied 
by a change in electric charge (baryon number) than at low $p_T$, and $\rho_{BQ}$ increases
with $p_T$ (The $\Delta$ resonances\footnote{Included in the THERMUS particle table 
up to the $\Delta(2420)$ .} ensure it keeps rising). For the $\Delta p_{T,i}$
dependence of $\rho_{SQ}$ we finally note that one of the strongest contributors at higher 
$p_T$ is the $\Omega^-$, with a relatively low mass of $m_{\Omega^-} = 1.672GeV$. So after a
rise, $\rho_{SQ}$ drops again towards highest $p_T$, due to an increasing $\Sigma^+$ 
contribution\footnote{Included in the THERMUS particle table up to the $\Sigma(2030)$.}. \\

Since resonance decay has the habit of dropping the lighter particles (mesons) 
at low $p_T$ and higher~$|y|$, while keeping heavier particles (baryons) 
at higher $p_T$ and at mid-rapidity, none of the above arguments about the transverse
momentum and rapidity dependence are essentially changed by resonance decay. The 
correlation coefficient $\rho_{BS}$ becomes more negative towards higher $p_T$, 
while becoming weaker towards higher $|y|$. Similarly, $\rho_{BQ}$ grows larger at high $p_T$
and drops towards higher~$y$. The larger contributions of baryons to the high $p_T$
tail of the transverse momentum spectrum, and their decreased contribution to the tails
of the rapidity distribution, compared to mesons, are to blame. 
The bump in the $p_T$ dependence of $\rho_{SQ}$, presumably caused by the $\Sigma^+$,
has vanished, as the $\Sigma^+$ is only considered as stable in its lightest version 
with mass $m_{\Sigma^+}=1.189GeV$. The small bump in the $y$ dependence of $\rho_{SQ}$, however, stays.
The correlation is  presumably first increased by a growing kaon contribution 
and then again decreased by a growing pion contribution at larger rapidities.\\

The values of $\rho$ after resonance decay are directly sensitive to how the data is 
analyzed. In the above study we analyzed final state particles (stable against strong decays)
only. One could, however, also reconstruct decay positions and momenta of parent resonances 
and could then count them as belonging to the acceptance bin the parent momentum 
would fall into. In the situation above, however, this would again yield the 
primordial scenario. If reconstruction of resonances is not done, one is 
sensitive to charge correlations carried by final state particles. 
As in the primordial case, a larger acceptance bin effectively averages over 
smaller bins. However, the smaller the acceptance bin, the more information is lost due to 
resonance decay. In full acceptance, final state and primordial correlation coefficients 
ought to be the same, since quantum numbers (and energy-momentum) are conserved
in the decays of resonances.

\subsection{Extrapolating to the MCE}
\label{none}

We next consider the extrapolation to the MCE limit of variances and covariances and, hence, correlation 
coefficients, of joint distributions of charges in limited acceptance. The primordial 
joint baryon number - strangeness distributions in different transverse momentum 
bins will serve as examples. In this subsection, we use an 
extended data set of $20 \cdot 8 \cdot 10^5$ events. 

\begin{figure}[ht!]
  \epsfig{file=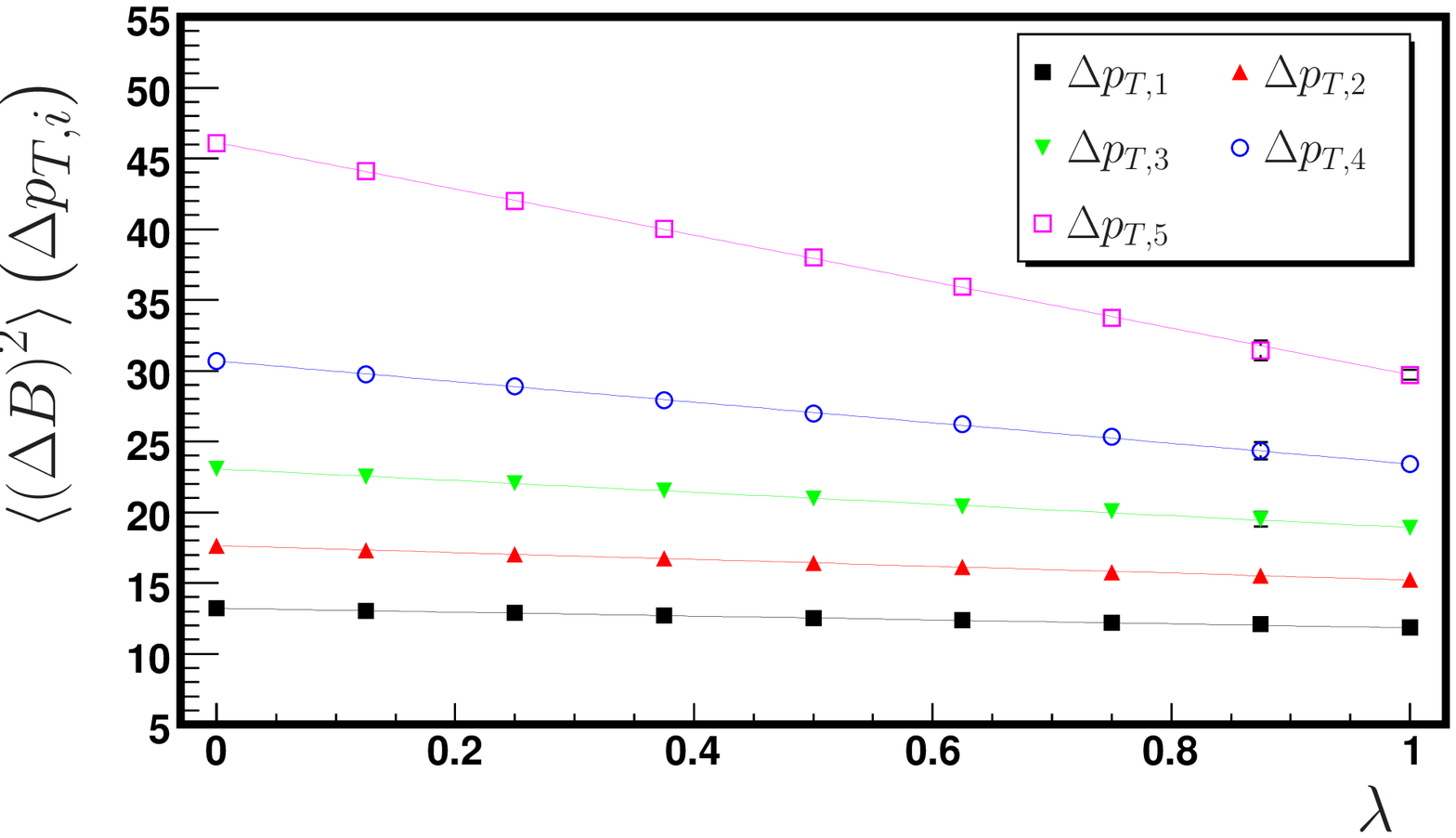,width=8.4cm,height=6.5cm}
  \epsfig{file=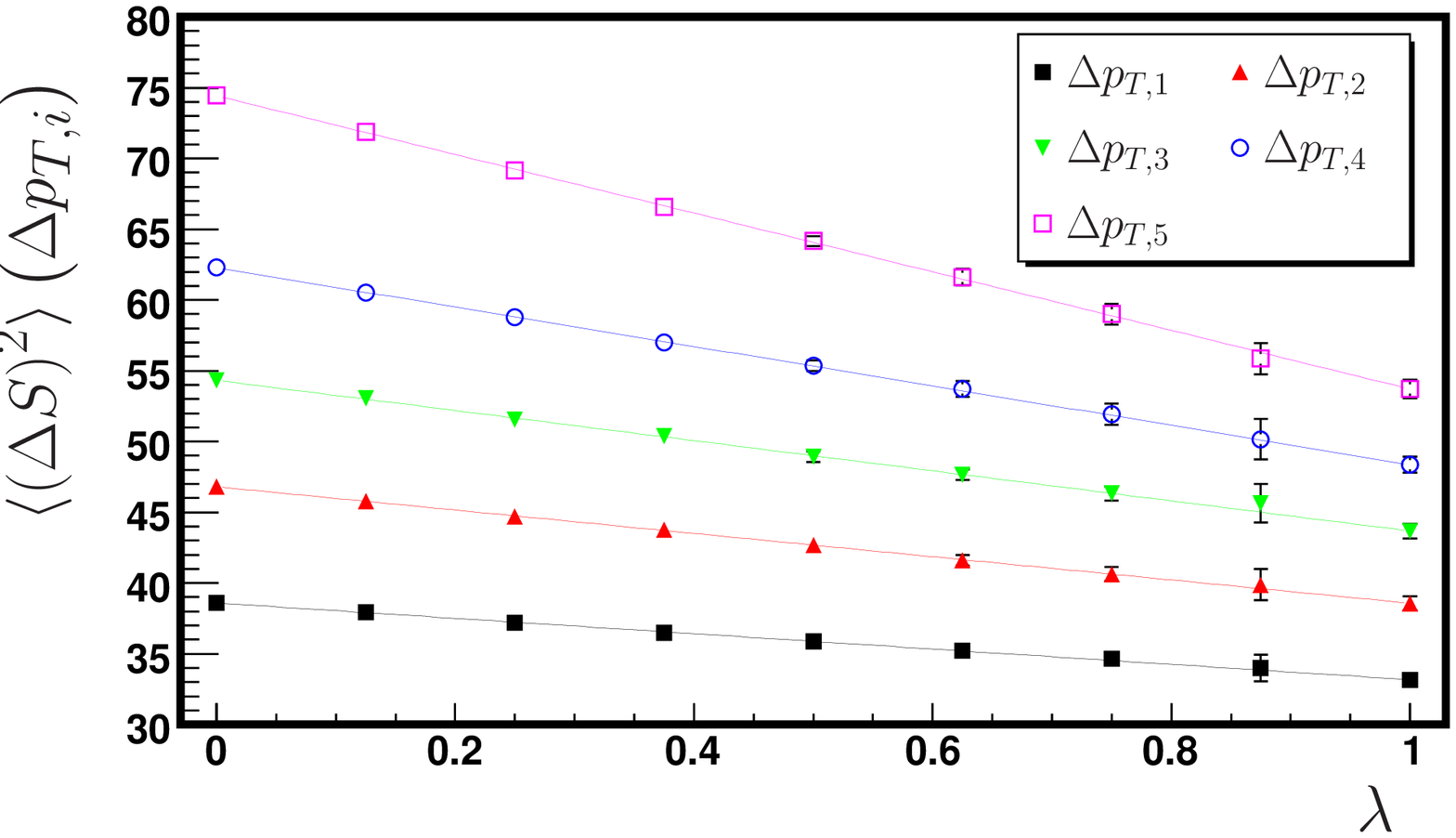,width=8.4cm,height=6.5cm}
  \caption{Evolution of the variance of the marginal baryon number distribution 
    $\langle (\Delta B)^2 \rangle$ ({\it left}) and the variance of the marginal strangeness distribution
    $\langle (\Delta S)^2 \rangle$ ({\it right}) with $\lambda$ for a primordial hadron 
    resonance gas in different $\Delta p_{T,i}$ bins. 
    Each marker and its error bar except the last 
    represents the result of $20$ Monte Carlo runs of $10^5$ events
    each. 8 different equally spaced values of $\lambda$ have been investigated.
    The last marker denotes the result of the extrapolation.
    Solid lines indicate  
    extrapolations from the GCE value to the MCE limit. 
  }  
  \label{CwL_varCC}
\end{figure}

\begin{figure}[ht!]
  \epsfig{file=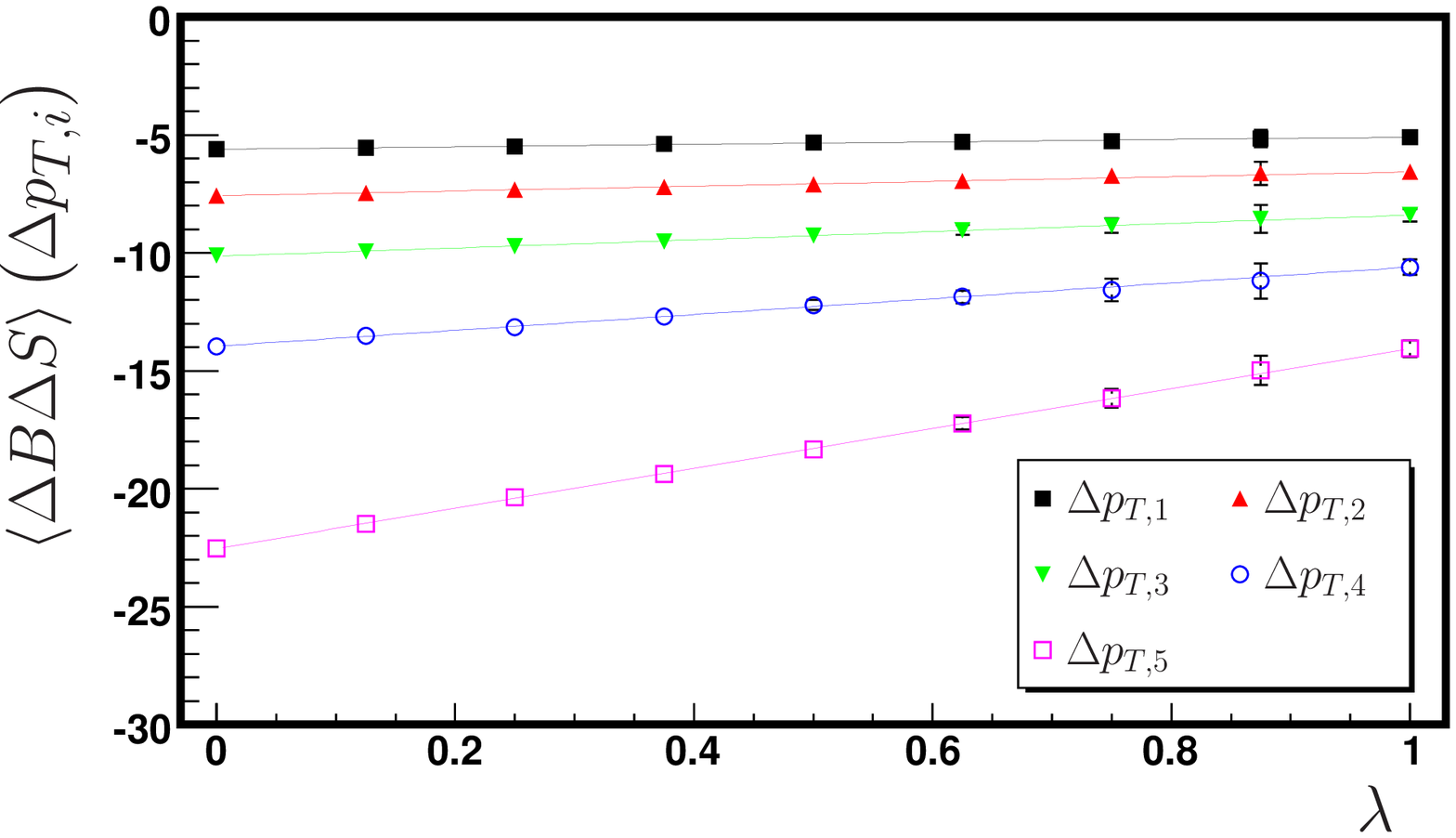,width=8.4cm,height=6.5cm}
  \epsfig{file=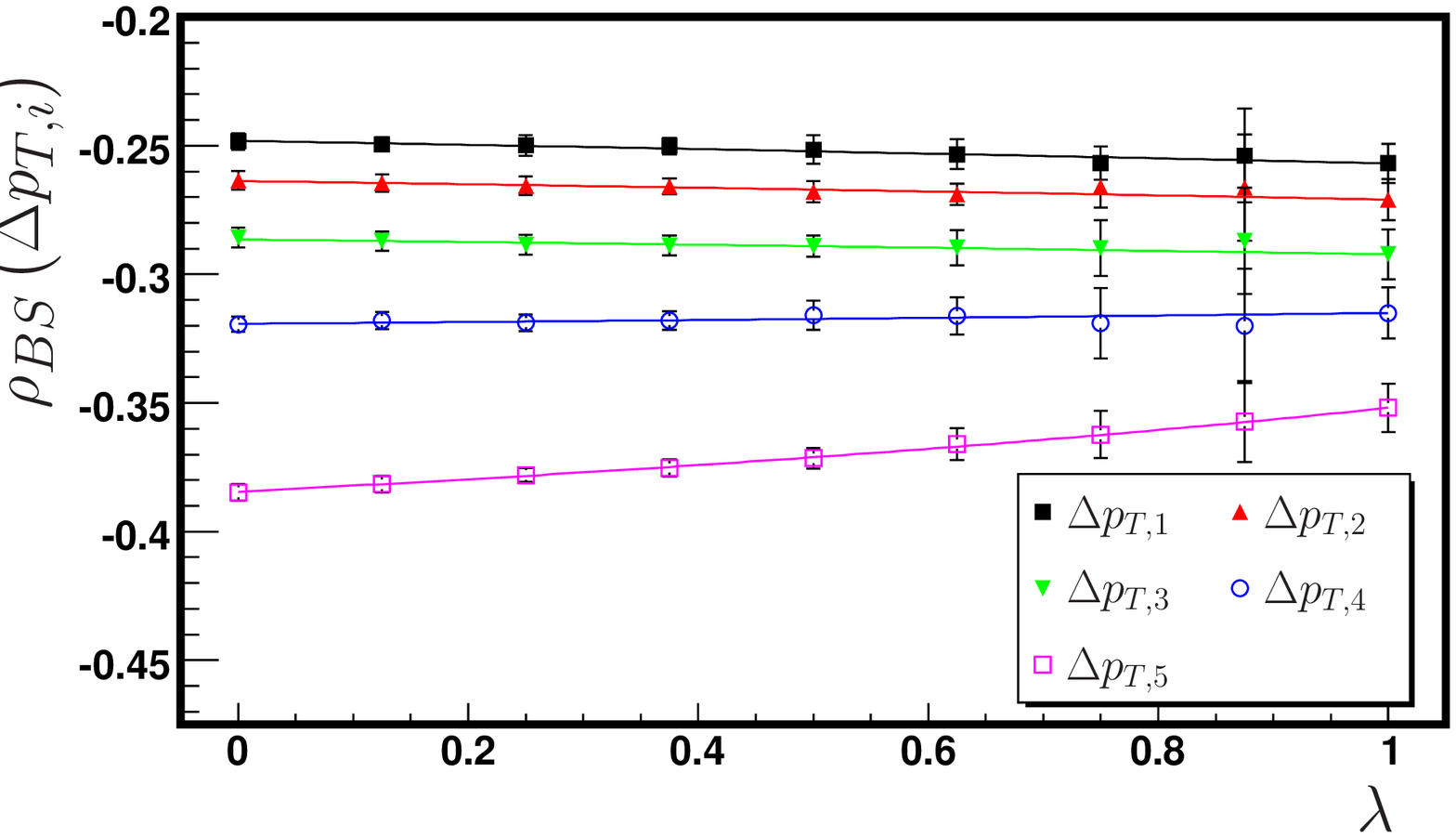,width=8.4cm,height=6.5cm}
  \caption{Evolution of the covariance $\langle \Delta B \Delta S\rangle$ ({\it left}) 
    and the correlation coefficient $\rho_{BS}$ ({\it right}) 
    of the baryon number - strangeness distribution 
    with $\lambda$ for a primordial hadron 
    resonance gas in different $\Delta p_{T,i}$ bins.
    The rest as in Fig.(\ref{CwL_varCC}).
  }  
  \label{CwL_corrCC}
\end{figure}

In Fig.(\ref{CwL_varCC}) we show the evolution of the variances of the marginal primordial
 baryon number distribution $\langle (\Delta B)^2 \rangle$ ({\it left}) and of the marginal 
primordial strangeness distribution $\langle (\Delta S)^2 \rangle$ ({\it right})
in the transverse momentum bins $ \Delta p_{T,i}$, defined in Table \ref{accbins}, as 
a function of the size of the bath $\lambda= V_1/V_g$. $8$ equally spaced values of $\lambda$
have been investigated. The last marker denotes the result of the extrapolation. In
Fig.(\ref{CwL_corrCC}) we show the dependence of the primordial covariance 
$\langle \Delta B \Delta S\rangle$ ({\it left}) and the primordial correlation coefficient 
$\rho_{BS}$ ({\it right}) of the joint baryon number - strangeness distribution on the size of the 
bath $\lambda$.

Let us first comment on the GCE values of variances (the left most markers  
in Fig.(\ref{CwL_varCC})).
As each of the 5 momentum bins holds one fifth of the 
charged particle yield and, hence, less than one fifth of the baryonic contribution
in the lowest bin $\Delta p_{T,1}$, and more than one fifth in the highest 
bin $\Delta p_{T,5}$, we find the baryon number variance $\langle (\Delta B)^2 \rangle$
largest in $\Delta p_{T,5}$, and smallest in $\Delta p_{T,1}$. 
If binned in rapidity: $\Delta y_{3}$ has the strongest baryon contribution, and, hence,  
$\langle (\Delta B)^2 \rangle$ is largest there. The same goes for 
the variance $\langle (\Delta S)^2 \rangle$ of the marginal strangeness distribution.
Strangeness carrying particles are on average heavier than electrically charged 
particles and, hence, the strangeness contribution is strongest around mid-rapidity 
and towards larger transverse momentum (i.e. $\langle (\Delta S)^2 \rangle$ is largest 
in $\Delta y_{3}$ and $\Delta p_{T,5}$,
while being smallest in $\Delta y_{1}$, $\Delta y_{5}$, and $\Delta p_{T,1}$).

\begin{figure}[ht!]
  \epsfig{file=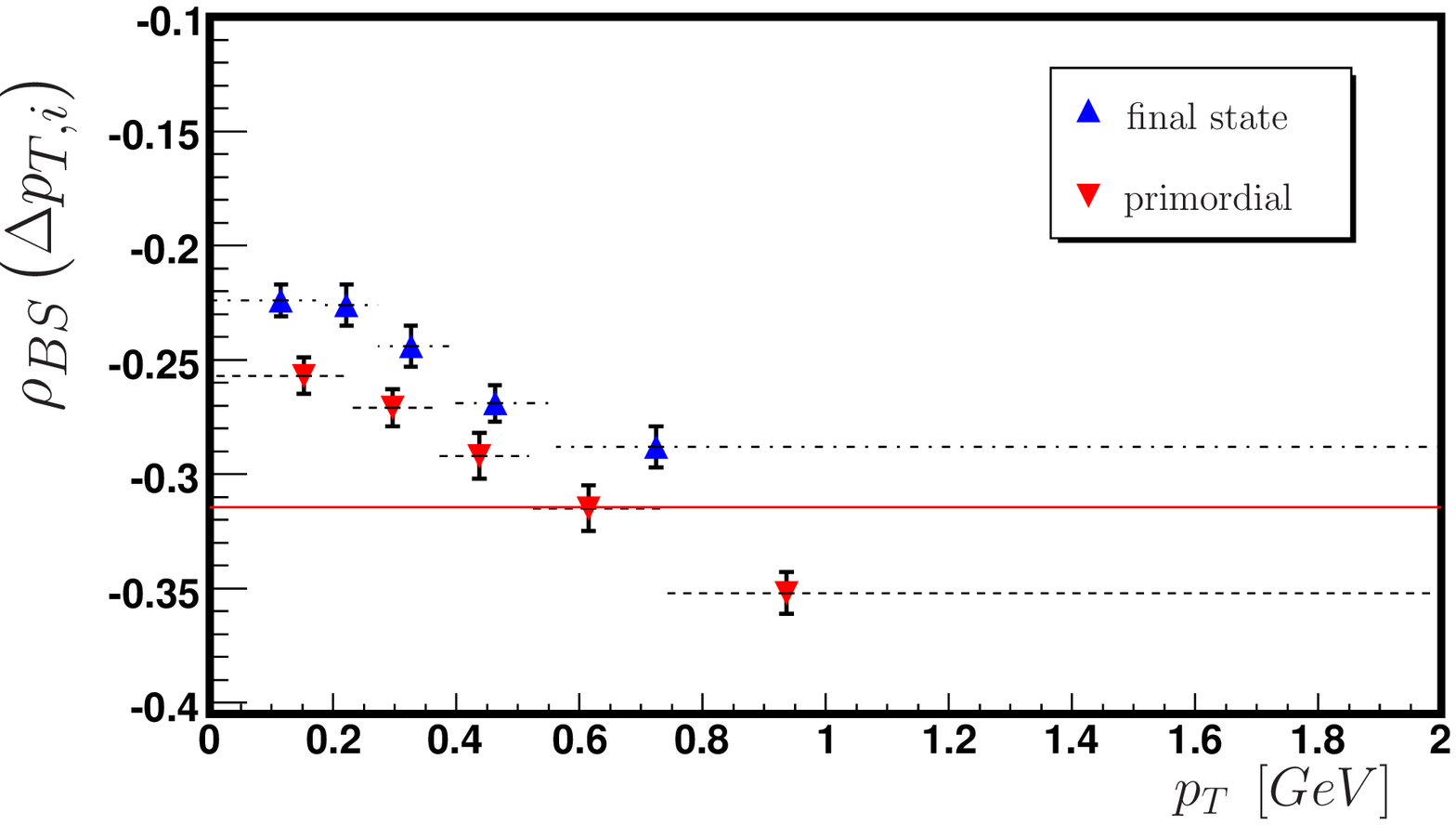,width=8.4cm,height=6.5cm}
  \epsfig{file=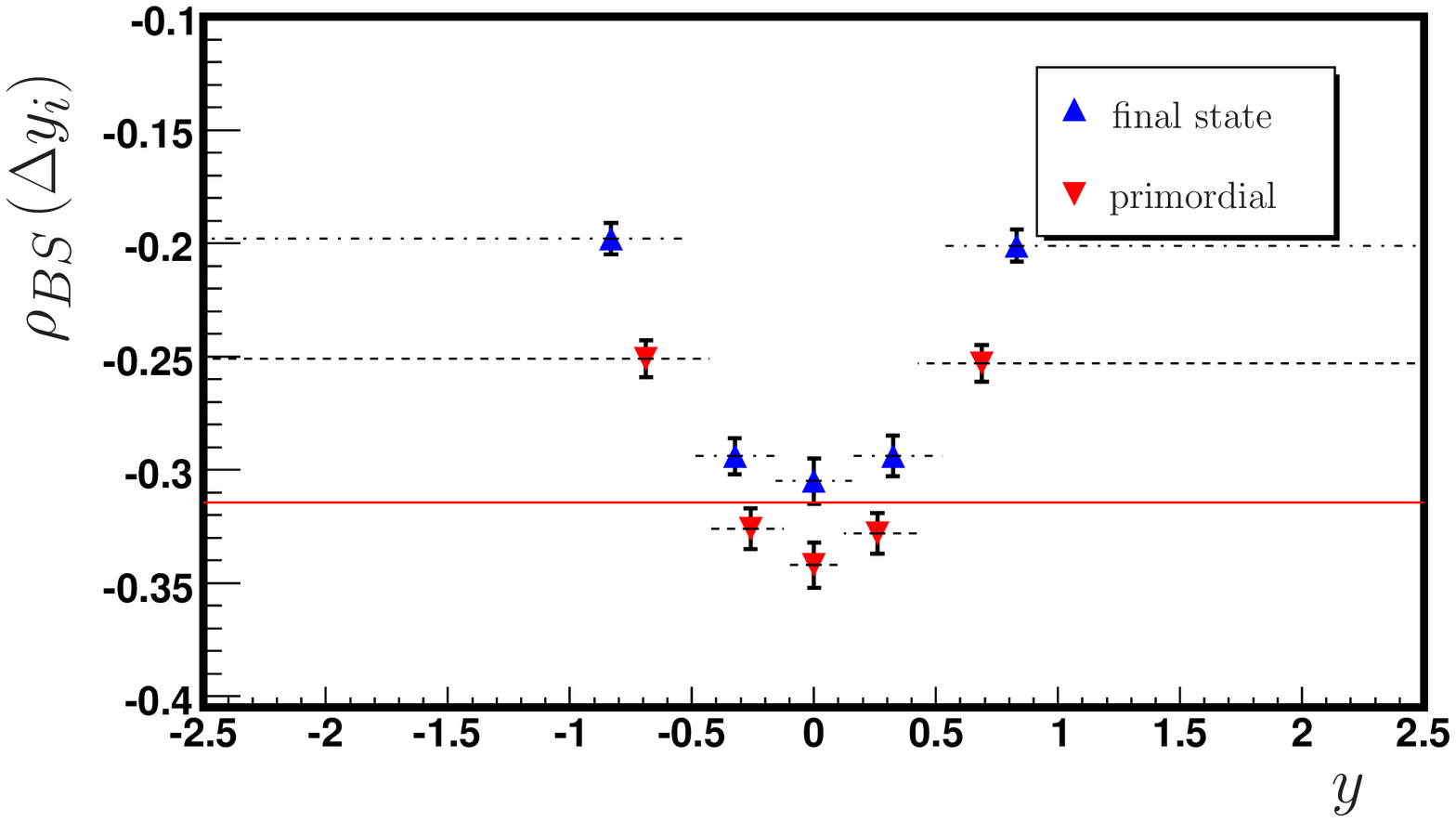,width=8.4cm,height=6.5cm}
  \caption{MCE baryon number - strangeness correlation coefficient $\rho_{BS}$ 
    in limited acceptance 
    windows, both primordial and final state.
    ({\it Left:}) transverse momentum bins  $\Delta p_{T,i}$. 
    ({\it Right:}) rapidity bins $\Delta y_i$.  
    Horizontal error bars indicate the width and position of the momentum bins 
    (And not an uncertainty!).
    Vertical error bars indicate the statistical uncertainty of the extrapolation 
    of $8 \cdot 20$ Monte Carlo runs of $10^5$ events each.
    The marker indicates the center of gravity of the corresponding bin.
    The solid lines show the fully phase space integrated GCE result.
  }  
  \label{lc_bs_mce}
\end{figure}

The $\Delta p_{T,i}$ dependence of the GCE covariance $\langle \Delta B \Delta S\rangle$
and the GCE correlation coefficient $\rho_{BS}$ in Fig.(\ref{CwL_corrCC}) is explained 
by the arguments of the previous subsection. Varying contributions of hadrons of different 
mass (and charge contents) to different parts of momentum space are responsible.

We now turn our attention to the extrapolation. MCE effects on the baryonic sector are felt most strongly in 
momentum space segments in which the baryonic contribution is strong (e.g. see the evolution of the last bin 
$\Delta p_{T,5}$ with $\lambda$ 
in Figs.(\ref{CwL_varCC},\ref{CwL_corrCC})). The correlation coefficient is 
not as strongly affected, in general, by MCE effects. 
  
In Fig.(\ref{lc_bs_mce}) we show the results of the extrapolation to the MCE limit of 
the baryon number-strangeness correlation coefficient $\rho_{BS}$ in acceptance bins 
$\Delta p_{T,i}$ and $\Delta y_i$, both primordial and final state. 
MCE values are closer to each other than corresponding GCE values, Fig.(\ref{lc_bs_0000}). 
The influence of globally applied conservation laws on charge correlations is 
less strong than for the multiplicity fluctuations 
and correlations discussed in the next section.

\section{Momentum Space Dependence of Multiplicity Fluctuations and Correlations}
\label{Sec_MultFluc}

Multiplicity fluctuations and correlations are qualitatively affected by the 
choice of ensemble and are directly sensitive to the fraction of the system observed. For vanishing
size of ones acceptance window, one would lose all information on how the
multiplicities of any two distinct groups $N_i$ and $N_j$ of particles are correlated, and 
 measure $\rho_{ij} =0$. This information, on the other hand, is to some extent 
preserved in $\rho_{BS}$, $\rho_{BQ}$, and $\rho_{SQ}$, i.e. the way in which quantum 
numbers are correlated, if at least 
occasionally a particle is detected  during an experiment.

We first sample the same GCE system, which we have discussed in the previous sections, 
and consider the effects of resonance decay. Next the joint distributions of 
positively and negatively charged particles in momentum bins $\Delta p_{T,i}$
and $\Delta y_i$ are constructed. Then we, in turn, extrapolate the GCE 
primordial and final state results on the scaled variance $\omega$, Eq.(\ref{omega}),
and the correlation coefficient $\rho$, Eq.(\ref{rho}), to the MCE limit.

\subsection{Grand Canonical Ensemble}
\label{none}

\begin{figure}[ht!]
  \epsfig{file=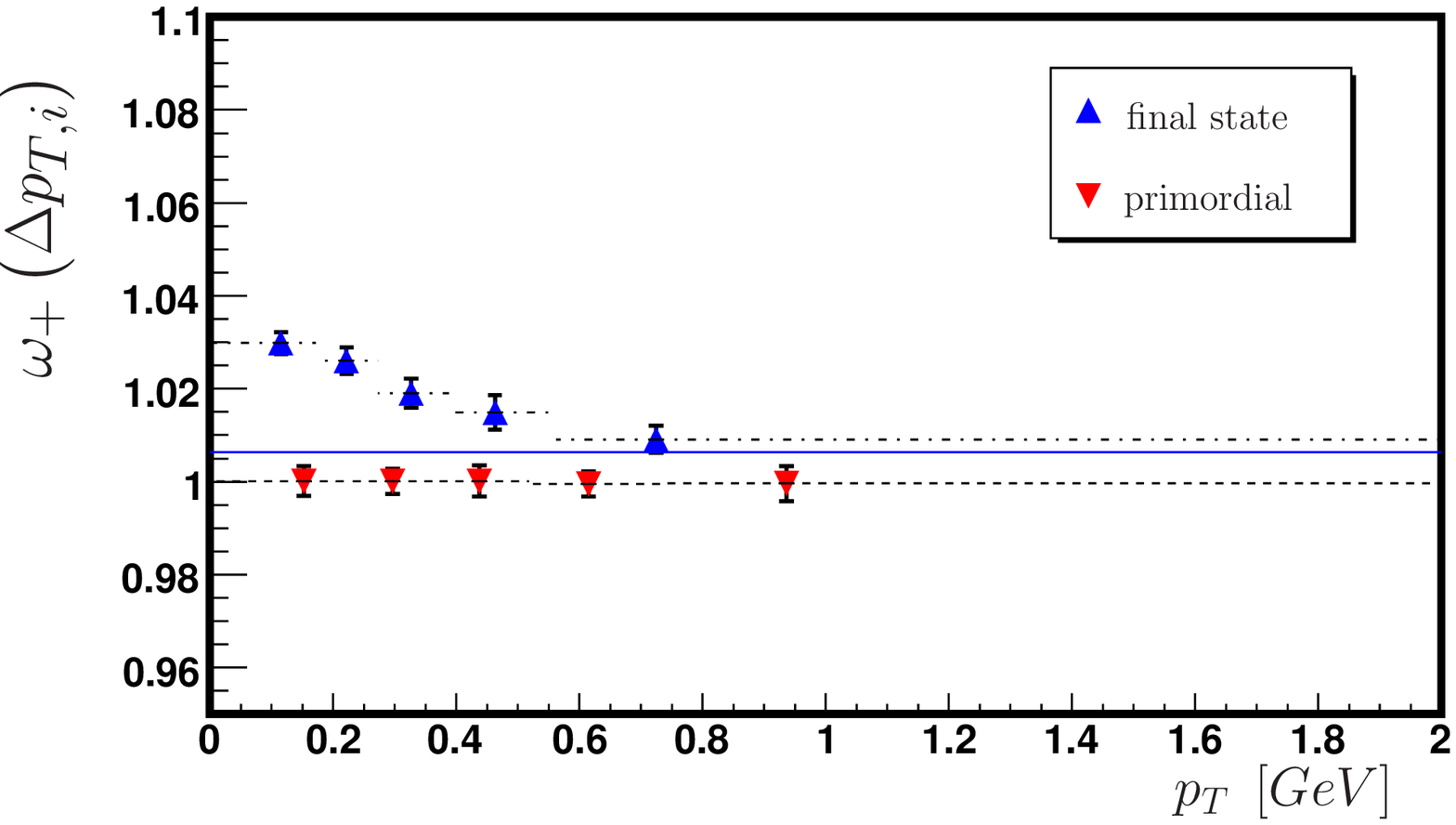,width=8.4cm,height=6.5cm}
  \epsfig{file=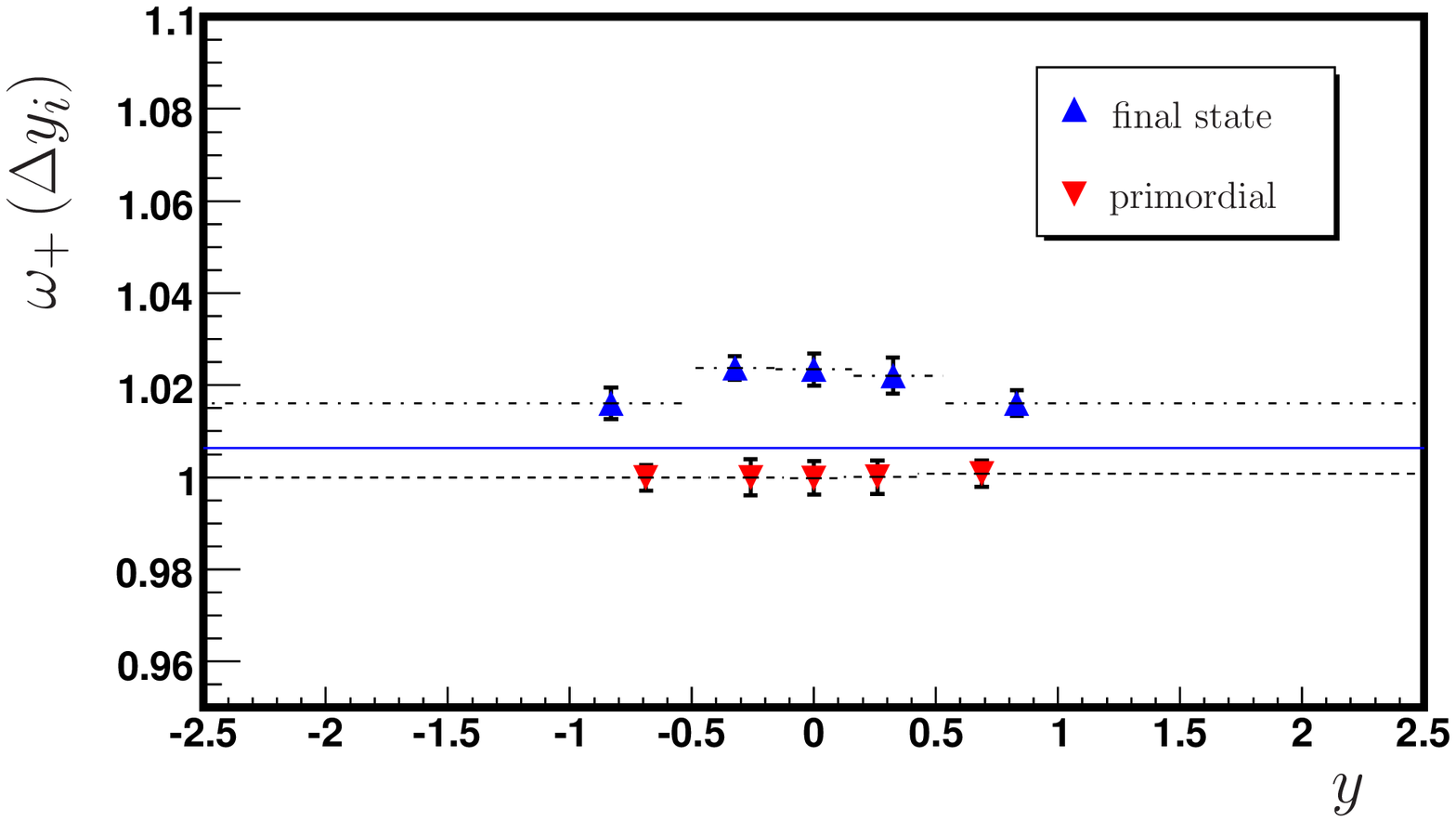,width=8.4cm,height=6.5cm}
  \caption{GCE scaled variance $\omega_+$ of multiplicity fluctuations of positively 
    charged hadrons, both primordial and final state, 
    in transverse momentum bins $\Delta p_{T,i}$ ({\it left}) and rapidity bins $\Delta y_i$ ({\it right}). 
    Horizontal error bars indicate the width and position of the momentum bins 
    (And not an uncertainty!).
    Vertical error bars indicate the statistical uncertainty of $20$ Monte Carlo runs 
    of $2\cdot 10^5$ events each.
    The markers indicate the center of gravity of the corresponding bin. 
    The solid line indicates the final state acceptance scaling estimate.
  }  
  \label{mult_pm_0000_omega}
\end{figure}

In Fig.(\ref{mult_pm_0000_omega}) we show the $\Delta p_{T,i}$ ({\it left}) and $\Delta y_i$
({\it right}) dependence of the GCE scaled variance $\omega_+$ of positively charged hadrons, both 
primordial and final state. In the primordial Boltzmann case one finds no dependence 
of multiplicity fluctuations on the position and size of the acceptance window.
The observed multiplicity distribution is, within error bars, a Poissonian 
with scaled variance $\omega_+=1$. In fact, in the primordial GCE 
Boltzmann case any selection of particles has $\omega = 1$.

\begin{figure}[ht!]
  \epsfig{file=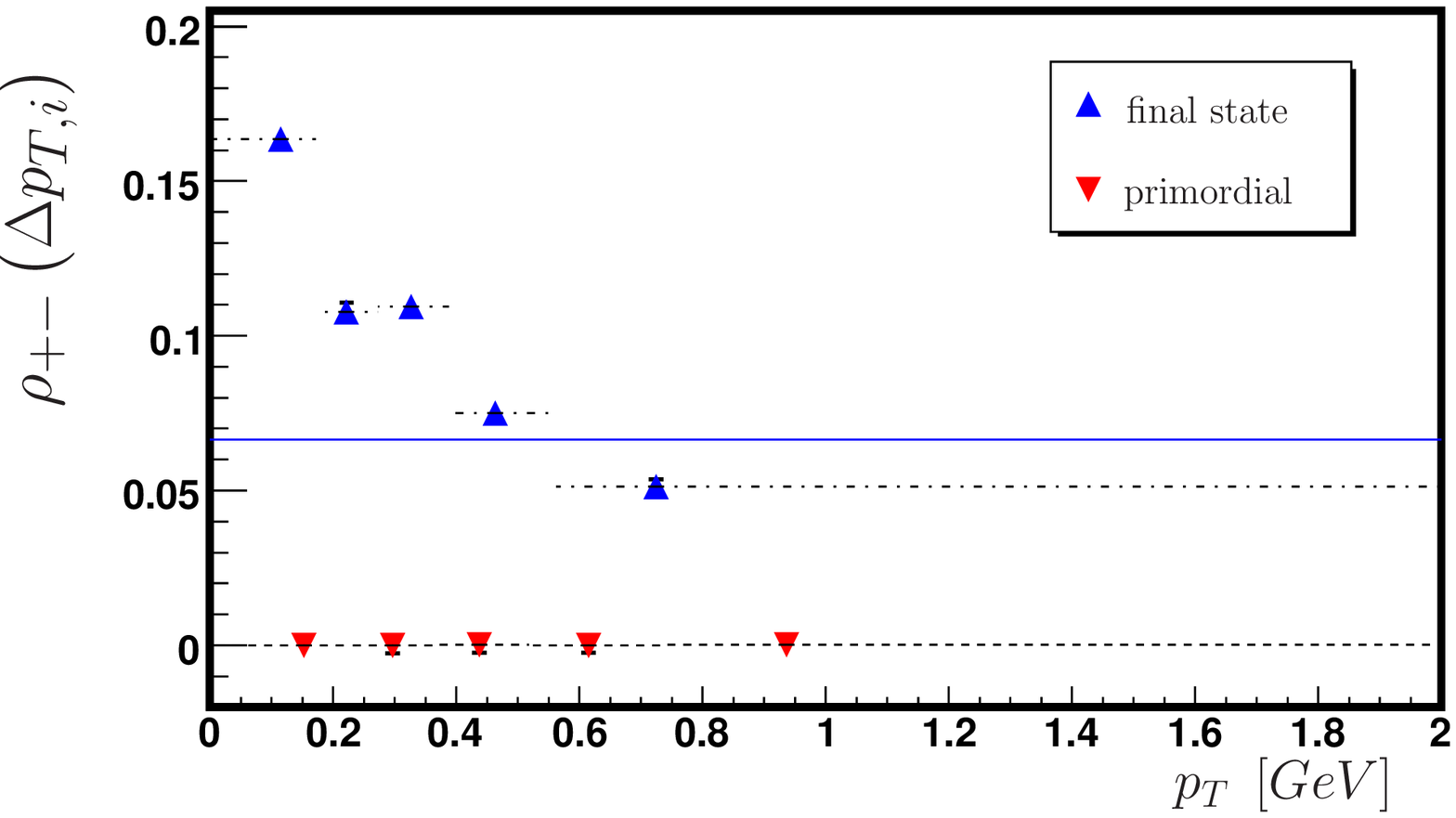,width=8.4cm,height=6.5cm}
  \epsfig{file=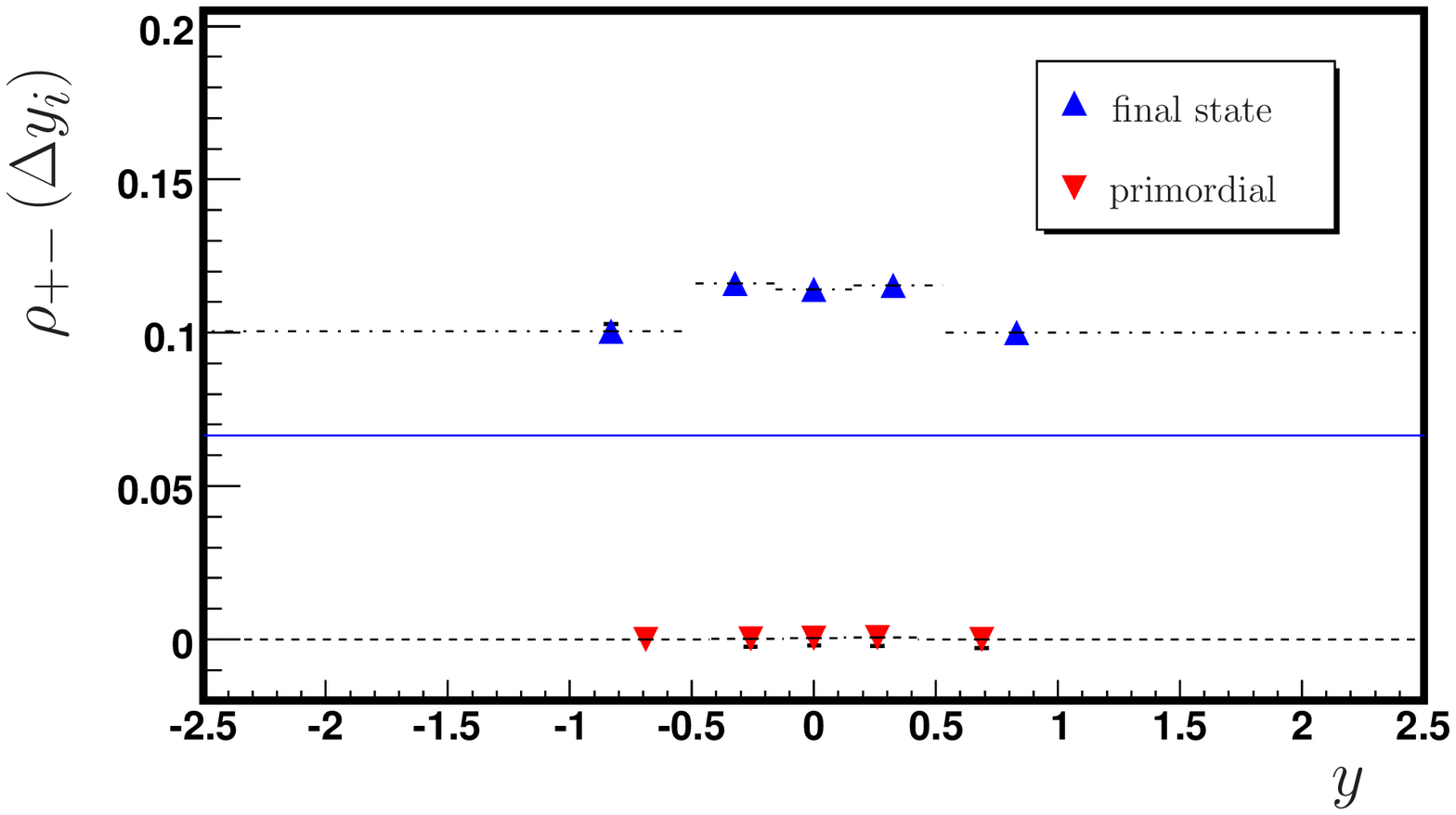,width=8.4cm,height=6.5cm}
  \caption{GCE multiplicity correlations $\rho_{+-}$ between positively and negatively 
    charged hadrons, both primordial and final state, in transverse momentum 
    bins $\Delta p_{T,i}$ ({\it left}) and rapidity bins  $\Delta y_i$ ({\it right}). 
    The rest as in Fig.(\ref{mult_pm_0000_omega}).}  
  \label{mult_pm_0000_rho}
\end{figure}

In Fig.(\ref{mult_pm_0000_rho}) we show the $\Delta p_{T,i}$ ({\it left}) and $\Delta y_i$
({\it right}) dependence of the GCE correlation coefficient $\rho_{+-}$ between positively and 
negatively charged hadrons, both primordial and final state. In the primordial Boltzmann 
case one finds also no dependence of multiplicity correlations on the position and size 
of the acceptance window. The observed joint multiplicity distribution is a product 
of two Poissonians with correlation coefficient $\rho_{+-}=0$.

Resonance decay is the only source of correlation in an ideal GCE Boltzmann gas.
Neutral hadrons decaying into two hadrons of opposite electric charge are the strongest 
contributors to the correlation coefficient $\rho_{+-}$. The chance that both 
(oppositely charged) decay products are dropped into the same momentum space bin is 
obviously highest at low transverse momentum (i.e. the correlation coefficient is 
strongest in $\Delta p_{T,1}$).
The rapidity dependence is somewhat milder again, because heavier particles (parents) 
are dominantly produced at mid-rapidity and spread their daughter 
particles over a range in rapidity. One notes that the scaled variances and 
correlation coefficients in the respective acceptance bins in 
Figs.(\ref{mult_pm_0000_omega},\ref{mult_pm_0000_rho}) are generally larger 
than the acceptance scaling procedure\footnote{For the acceptance scaling approximation 
it is assumed that particles are randomly detected with a certain probability $q=0.2$, 
independent of their momentum.} suggests, with the notable exception of 
$\rho_{+-}(\Delta p_{T,5})$. 

If one would construct now a larger and larger number of  momentum space bins of 
equal average particle multiplicities, one would successively lose more and 
more information about how multiplicities of distinct groups of 
particles are correlated. 

There is a simple relation connecting the scaled variance of the fluctuations of all charged 
hadrons $\omega_{\pm}$ to the fluctuations of only positively charged particles 
$\omega_+$ via the correlation coefficient $\rho_{+-}$ between positively and negatively 
charged hadrons in a neutral system:
\begin{equation}
\omega_{\pm} ~=~ \omega_+~\left(1~+~\rho_{+-}\right).
\end{equation}
We, therefore, find the effect of resonance decay on the $\Delta p_{T,i}$ 
dependence of $\omega_{\pm}$ to be considerably stronger than on that of $\omega_+$,
 and generally $\omega_{\pm} > \omega_+$, as the correlation coefficient 
$\rho_{+-}$ remains positive in the final state GCE. Compared to this, the final state 
values of  $\omega_{\pm}$, $\omega_+$ and $\rho_{+-}$ remain rather flat 
with $\Delta y_{i}$ in the GCE.

\subsection{Extrapolating to the MCE}
\label{none}

In the very same way that we extrapolated fully phase space integrated extensive quantities 
to the MCE limit in Section~\ref{Sec_ExtraMCE}, we now extrapolate multiplicity fluctuations 
$\omega_+$ and correlations $\rho_{+-}$ in transverse momentum bins $\Delta p_{T,i}$ 
and rapidity bins $\Delta y_i$ for a hadron resonance gas from the GCE $(\lambda =0)$ to the MCE 
($\lambda \rightarrow 1$).
Analytical primordial MCE results are done in the infinite volume 
approximation \cite{acc,baseline}.
We, hence, have some guidance as to further asses the accuracy of the extrapolation
scheme. For final state fluctuations and correlations in limited acceptance, 
on the other hand, no analytical results are available.

Mean values of particle numbers of positively charged hadrons $\langle N_+ \rangle$ 
and negatively charged hadrons $\langle N_- \rangle$ in the respective acceptance 
bins, defined in Table~\ref{accbins}, remain constant as $\lambda$ goes from~$0$~to~$1$, 
while the variances $\langle ( \Delta N_+)^2 \rangle$ and 
$\langle ( \Delta N_-)^2 \rangle$, and covariance $\langle \Delta N_+ \Delta N_- \rangle$ 
converge linearly to their respective MCE limits. The correlation coefficient $\rho_{+-}$
between positively and negatively charged hadrons, on the other hand, will not approach 
its MCE value linearly, as discussed in Section~\ref{Sec_ExtraMCE}.

\subsubsection{Primordial}
\label{none}

\begin{figure}[ht!]
  \epsfig{file=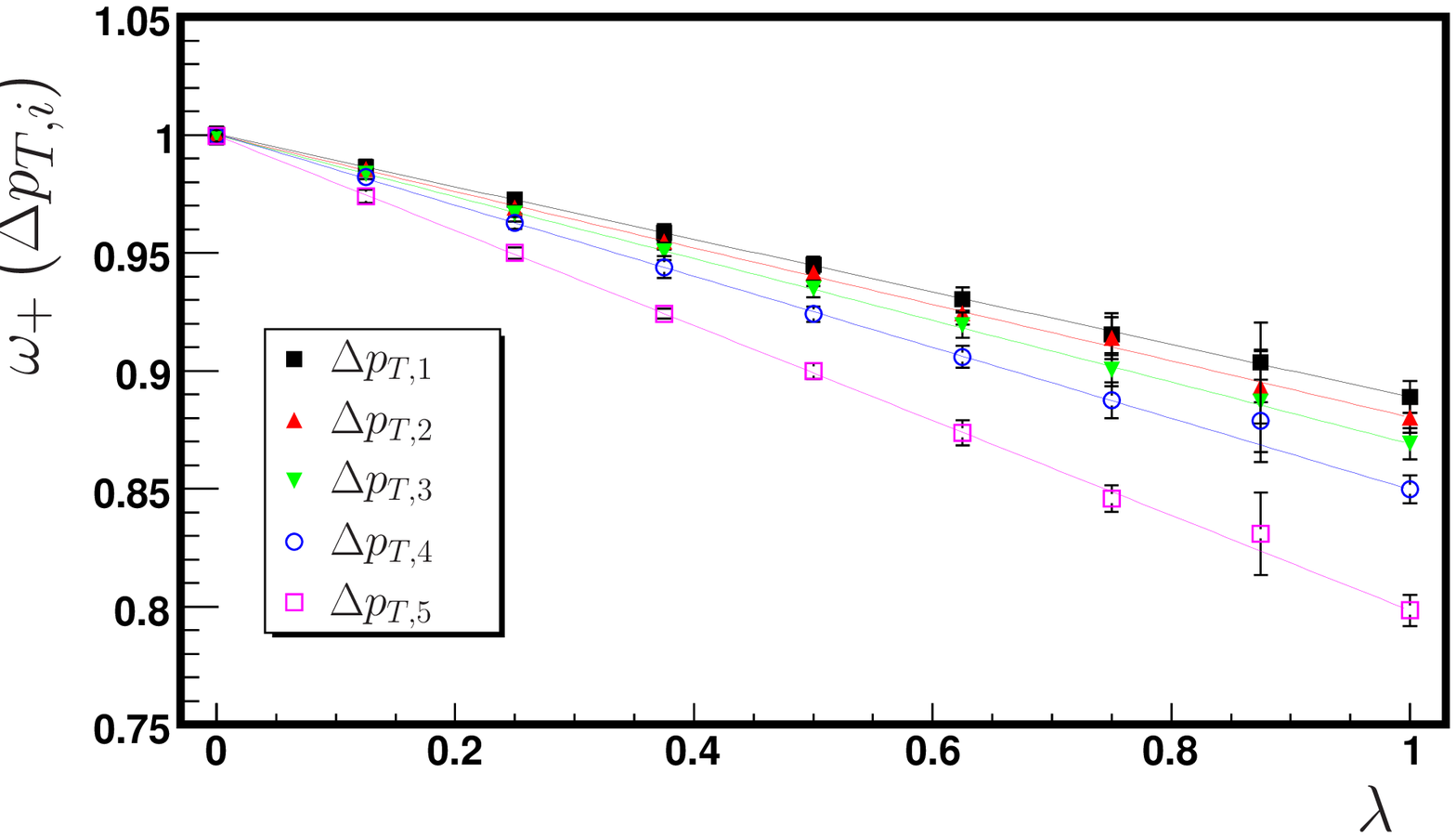,width=8.4cm,height=6.5cm}
  \epsfig{file=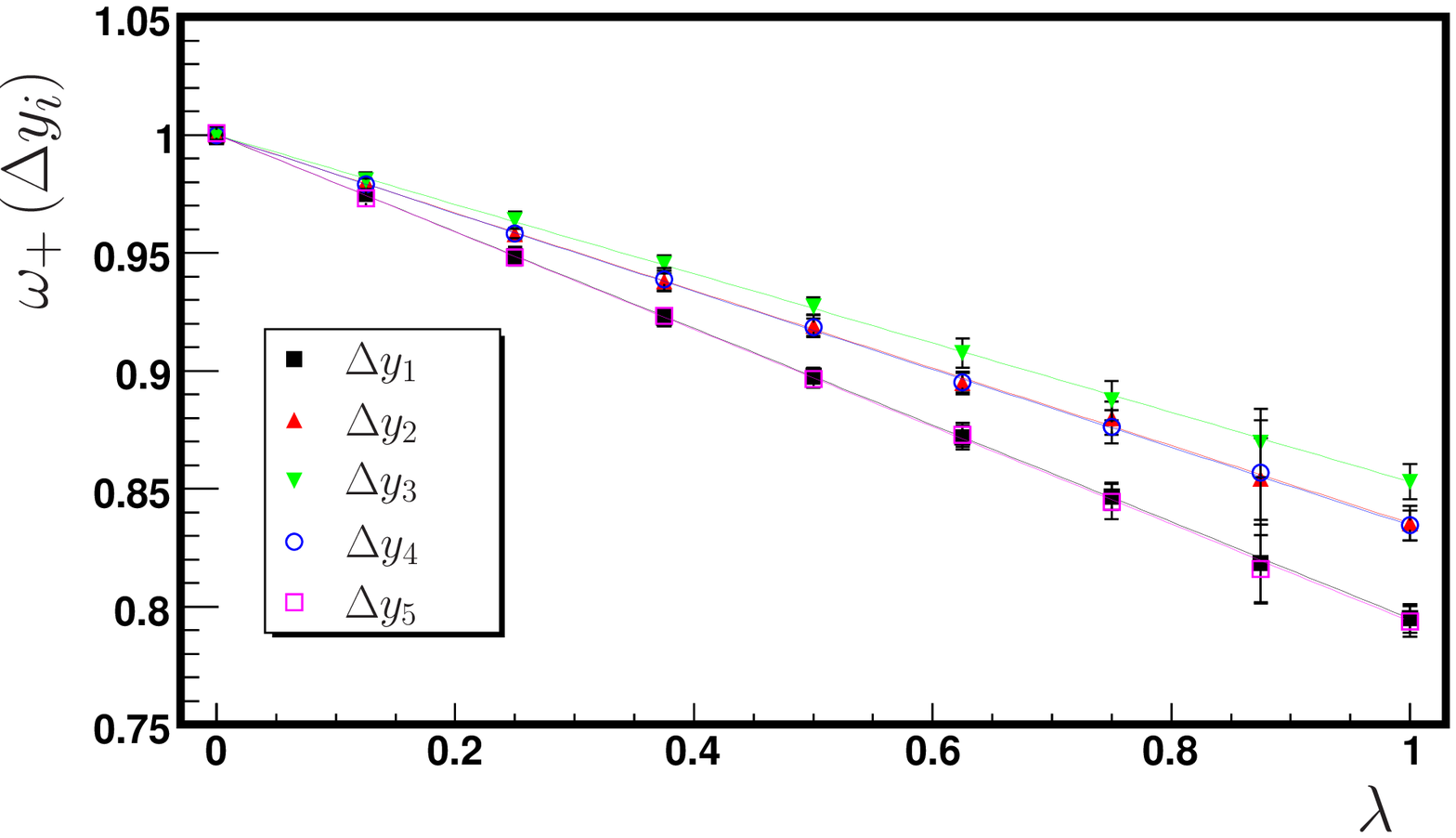,width=8.4cm,height=6.5cm}
  \caption{Evolution of the primordial scaled variance $\omega_+$ of positively charged hadrons 
    with the Monte Carlo parameter $\lambda = V_1/V_g$ for transverse momentum
    bins $\Delta p_{T,i}$ ({\it left}) and for rapidity bins $\Delta y_i$ ({\it right}). 
    The solid lines show an analytic extrapolation from GCE results ($\lambda =0$)
    to the MCE limit ($\lambda \rightarrow 1$).
    Each marker and its error bar except the last represents the result of $20$ Monte Carlo runs 
    of $2 \cdot 10^5$ events. 
    $8$ different equally spaced values of $\lambda$ have been investigated.
    The last marker denotes the result of the extrapolation.
  }  
  \label{conv_lambda_omega_prim}
\end{figure}

\begin{figure}[ht!]
  \epsfig{file=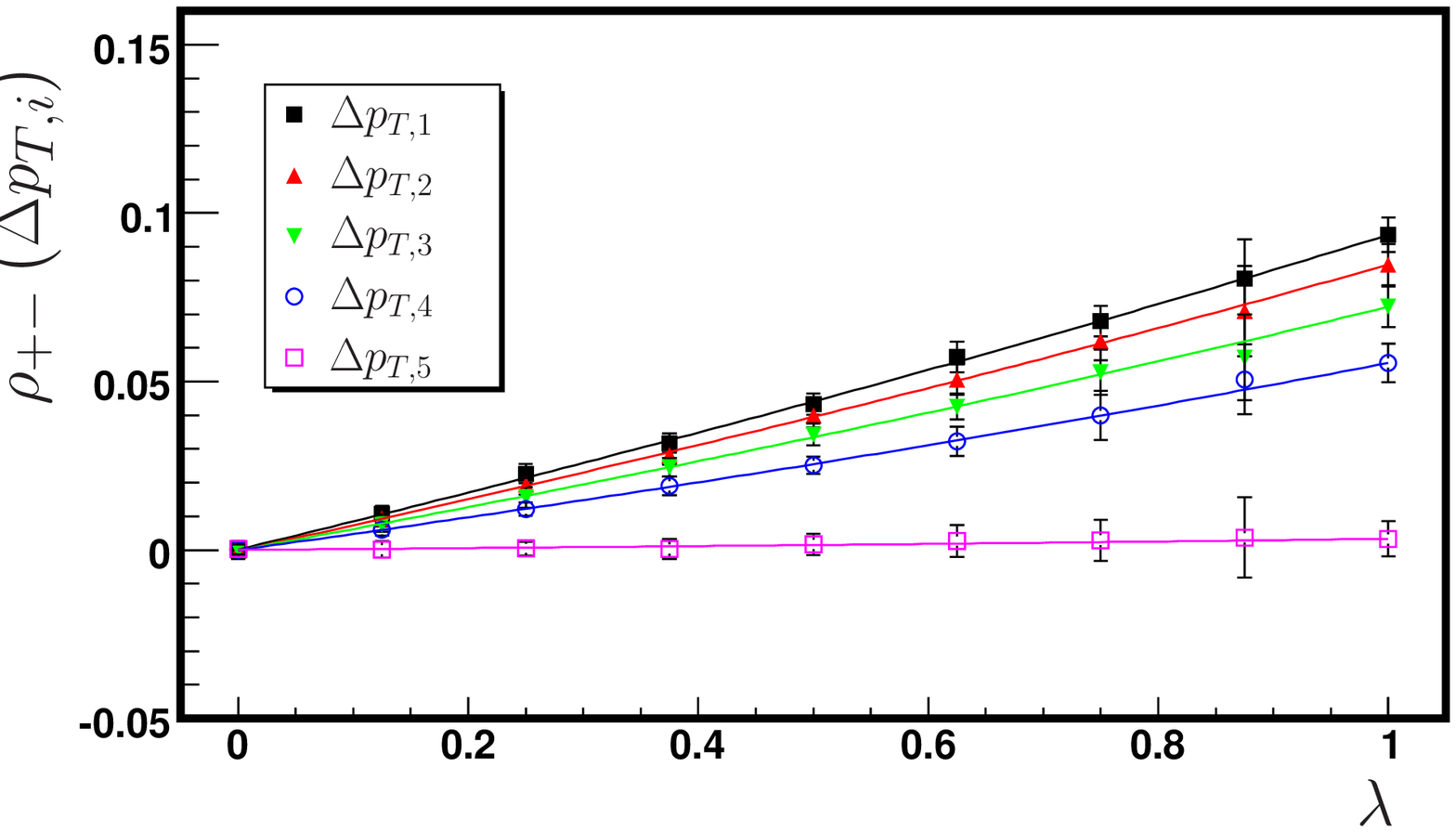,width=8.4cm,height=6.5cm}
  \epsfig{file=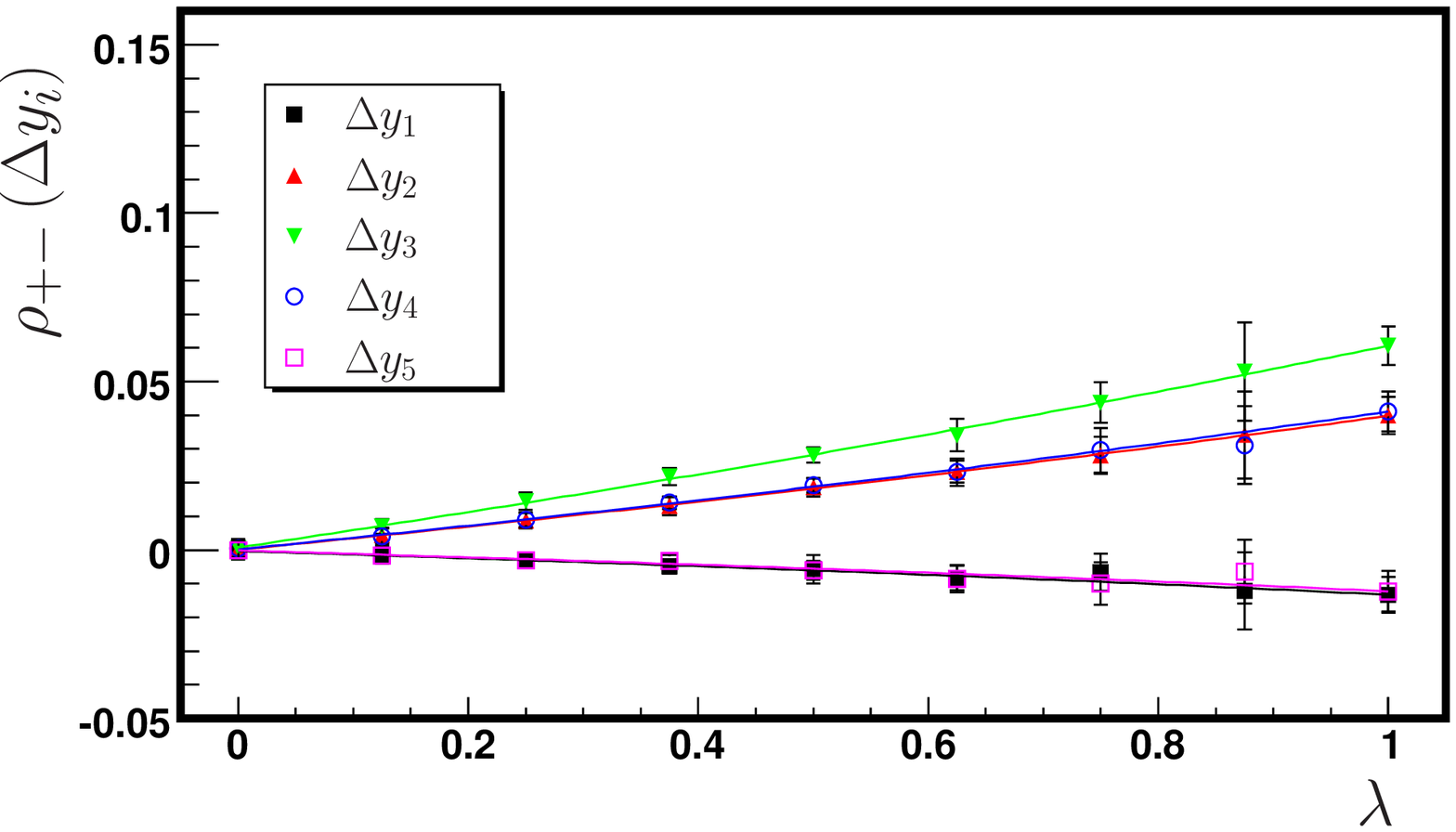,width=8.4cm,height=6.5cm}
  \caption{Evolution of the primordial correlation coefficient $\rho_{+-}$ between 
    positively and negatively charged hadrons with the Monte Carlo parameter 
    $\lambda = V_1/V_g$ for transverse momentum bins $\Delta p_{T,i}$ ({\it left}) and 
    for rapidity bins $\Delta y_i$ ({\it right}).
    The rest as in Fig.(\ref{conv_lambda_omega_prim}).}  
  \label{conv_lambda_rho_prim}
\end{figure}

In Fig.(\ref{conv_lambda_omega_prim}) we show the primordial scaled variance $\omega_+$
of positively charged hadrons in transverse momentum bins $ \Delta p_{T,i}$ ({\it left}) 
and rapidity bins $\Delta y_i$ ({\it right}) as a function of the size of the bath 
$\lambda= V_1/V_g$, while in Fig.(\ref{conv_lambda_rho_prim}) we show the dependence of 
the primordial correlation coefficient $\rho_{+-}$ between positively and negatively 
charged hadrons in transverse momentum bins $\Delta p_{T,i}$ ({\it left}) and rapidity 
bins $\Delta y_i$ ({\it right}) on $\lambda$.

The results of $8 \cdot 20$ Monte Carlo runs of $2 \cdot 10^5$ events each are summarized in
Table \ref{accbins_Mult_prim}. The system sampled was assumed to be neutral 
$\mu_j = (0,0,0)$ and static $u_{\mu} = (1,0,0,0)$ with local 
temperature $\beta^{-1} = 0.160GeV$ and a system volume of $V_1 = 2000 fm^3$. 
8 different values of $\lambda$ have been studied. 
The last marker $(\lambda = 1)$ denotes the result of the extrapolation.
Only primordial hadrons are analyzed. 
Values for both $\Delta p_{T,i}$ and $\Delta y_i$ bins are listed. 
Analytical numbers  are calculated according to the method 
developed in \cite{acc,baseline}, using the acceptance bins defined in Table~\ref{accbins}, 
and are shown for comparison.

The effects of energy-momentum and charge conservation on primordial 
multiplicity fluctuations and correlations in finite acceptance have been 
discussed in \cite{acc,baseline}. A few words attempt to summarize.

Let us first attend to fully phase space integrated results. The scaled variance 
of multiplicity fluctuations is lowest in the MCE due to the requirement of exact energy  
and charge conservation, somewhat larger in the CE, and largest 
in the GCE, as now all constraints on the microstates of the system have been 
dropped~\cite{MCEvsData,Res,clt}. 
The fully phase space integrated MCE and CE correlation coefficients between oppositely 
charged particles are rather close to 1. Doubly charged particles
allow for mild deviation, as also the $\Delta^{++}$ resonance is counted 
as only one particle. 

The transverse momentum dependence can be understood as follows: 
a change in particle number at high transverse momentum involves 
a large amount of energy. I.e., in order to balance the energy record, one 
needs to create (or annihilate) either a lighter particle with more kinetic energy, or two 
particles at lower $p_T$. This leads to suppressed multiplicity fluctuations in
high $\Delta p_{T,i}$ bins compared to low $\Delta p_{T,i}$ bins. By the same argument, it seems
favorable, due to the constraint of energy and charge conservation, to balance 
electric charge, by creating (or annihilating) pairs of oppositely charged particles,
predominantly in lower $\Delta p_{T,i}$ bins, while allowing for a more un-correlated 
multiplicity distribution, i.e. also larger net-charge ($\delta Q = N_+-N_-$) fluctuations, 
in higher $\Delta p_{T,i}$ bins. 

For the rapidity  dependence similar arguments hold. Here, however, the strongest 
role is played by longitudinal momentum conservation. A change in particle number 
at high $y$ involves now, in addition to a large amount of energy, a large momentum
$p_z$ to be balanced. The constraints of global $P_z$ conservation are, hence, felt 
least severely around $|y| \sim 0$, and it becomes favorable to balance charge 
predominantly at mid-rapidity ($\rho_{+-}$ larger) and allow for stronger 
multiplicity fluctuations ($\omega_+$ larger) compared to forward and backward
rapidity bins.

\begin{table}[h!]
  \begin{center}
    \begin{tabular}{||c||c|c|c|c|c||}\hline
      ~primordial~ & $\Delta p_{T,1}$ & $\Delta p_{T,2}$ & $\Delta p_{T,3}$ & $\Delta p_{T,4}$ 
      & $\Delta p_{T,5}$   \\
      \hline
      ~$\omega^{gce}_+$ ~&~ $1.000 \pm 0.002$ ~&~ 
      $1.000 \pm 0.002$ ~&~ $1.000 \pm 0.002$ ~&~ 
      $1.000 \pm 0.002$ ~&~ $1.000 \pm 0.002$~ \\
      ~$\omega^{mce}_+$ ~&~ $0.889 \pm 0.007$ ~&~ 
      $0.880 \pm 0.007$ ~&~ $0.869 \pm 0.007$ ~&~ 
      $0.850 \pm 0.006$ ~&~ $0.798 \pm 0.007$~ \\
      ~$\omega^{mce,c}_+$ ~&~ $0.8886$ ~&~ $0.8802$ ~&~ $0.8682$ ~&~ $0.8489$ ~&~ $0.7980$~ \\
      \hline
      ~$\rho^{gce}_{+-}$ ~&~ $0.000 \pm 0.002$ ~&~ 
      $-0.000 \pm 0.002$ ~&~ $-0.000 \pm 0.002$ ~&~ 
      $0.000 \pm 0.002$ ~&~ $0.000 \pm 0.001$~ \\
      ~$\rho^{mce}_{+-}$ ~&~ $0.094 \pm 0.005$ ~&~ 
      $0.085 \pm 0.006$ ~&~ $0.072 \pm 0.006$ ~&~ 
      $0.056 \pm 0.006$ ~&~ $0.003 \pm 0.005$~ \\
      ~$\rho^{mce,c}_{+-}$ ~&~ $0.0935$ ~&~ $0.0844$ ~&~ $0.0730$ ~&~ $0.0554$ ~&~ $0.0040$~ \\
      \hline \hline
      ~primordial~ & $\Delta y_1$  & $\Delta y_2$  & $\Delta y_3$  & $\Delta y_4$  
      & $\Delta y_5$ \\
      \hline 
      ~$\omega^{gce}_+$ ~&~ $1.000 \pm 0.002$ ~&~ 
      $1.000 \pm 0.002$ ~&~ $1.000 \pm 0.003$ ~&~ 
      $1.000 \pm 0.002$ ~&~ $1.000 \pm 0.002$~ \\
      ~$\omega^{mce}_+$ ~&~ $0.795 \pm 0.006$ ~&~ 
      $0.835 \pm 0.007$ ~&~ $0.853 \pm 0.008$ ~&~ 
      $0.834 \pm 0.006$ ~&~ $0.794 \pm 0.007$~ \\
      ~$\omega^{mce,c}_+$ ~&~ $0.7950$ ~&~ $0.8350$ ~&~ $0.8521$ ~&~ $0.8351$ ~&~ $0.7949$~ \\
      \hline 
      ~$\rho^{gce}_{+-}$ ~&~ $-0.000 \pm 0.001$ ~&~ 
      $0.000 \pm 0.002$ ~&~ $0.001 \pm 0.002$ ~&~ 
      $0.000 \pm 0.002$ ~&~ $-0.000 \pm 0.002$~ \\
      ~$\rho^{mce}_{+-}$ ~&~ $-0.013 \pm 0.005$ ~&~ 
      $0.040 \pm 0.006$ ~&~ $0.061 \pm 0.006$ ~&~ 
      $0.041 \pm 0.006$ ~&~ $-0.012 \pm 0.006$~ \\
      ~$\rho^{mce,c}_{+-}$ ~&~ $-0.0135$ ~&~ $0.0406$ ~&~ $0.0616$ ~&~ $0.0406$ ~&~ $-0.0135$~ \\
      \hline
    \end{tabular}
    \caption{Summary of the primordial scaled variance $\omega_+$ of positively charged hadrons 
      and the correlation coefficient $\rho_{+-}$ between positively and negatively charged 
      hadrons in transverse momentum bins $\Delta p_{T,i}$ and rapidity bins $\Delta y_i$. 
      Both the GCE result ($\lambda = 0$) and the extrapolation to MCE ($\lambda = 1$) 
      are shown.
      The uncertainty quoted corresponds to $20$ Monte Carlo runs of $2 \cdot 10^5$ events
      (GCE) or is the result of the extrapolation (MCE). 
      Analytic MCE results $\omega^{mce,c}_+$ and $\rho^{mce,c}_{+-}$ are listed too.
    } 
    \label{accbins_Mult_prim}
  \end{center}
\end{table}

In a somewhat casual way one could say: events of a neutral hadron resonance gas 
with values of extensive quantities $B$, $S$, $Q$, $E$ and $P_z$ in the vicinity 
of $\langle \mathcal{Q}_1^l \rangle$ have a tendency to have similar numbers of positively 
and negatively charged particles at low transverse momentum $p_T$ and rapidity  $y$ 
and less strongly so at high $p_T$ and $|y|$.
 
The statistical error on the `data` points grows as $\lambda \rightarrow 1$, as can be seen 
from Figs.(\ref{conv_lambda_omega_prim},\ref{conv_lambda_rho_prim}).
The extrapolation helps greatly to keep the statistical uncertainty on the MCE limit low, 
as summarized in Table~\ref{accbins_Mult_prim}, and can be seen from a comparison of the last 
two data points in  Figs.(\ref{conv_lambda_omega_prim},\ref{conv_lambda_rho_prim}). The
last point and its error bar denote the result of a linear extrapolation of variances 
and covariances, while the second to last data point and its error bar are the result
of $20$ Monte Carlo runs with $\lambda = 0.875$. 
The analytical MCE values are well within error bars
of extrapolated Monte Carlo results, and 
agree surprisingly well, given the large number of ``conserved'' quantities (5) and
a relatively small sample size of $8 \cdot 20 \cdot 2 \cdot 10^5 = 3.2 \cdot 10^7$ events.
In a sample-reject type of approach this sample size would yield a substantially larger
statistical error, as only events with exact values of extensive quantities are 
kept for the analysis. As the system size is increased, a sample-reject formalism, hence, 
becomes increasingly inefficient, while the extrapolation method still yields
good results. For a further discussion see Appendix~\ref{App_ConvStudy}.

\subsubsection{Final State}
\label{none}

We now attend to the extrapolation of final state multiplicity
fluctuations and correlations to the MCE limit. An independent Monte Carlo run for the same 
physical system was done, but now with only stable final state particles `detected'.

\begin{figure}[ht!]
  \epsfig{file=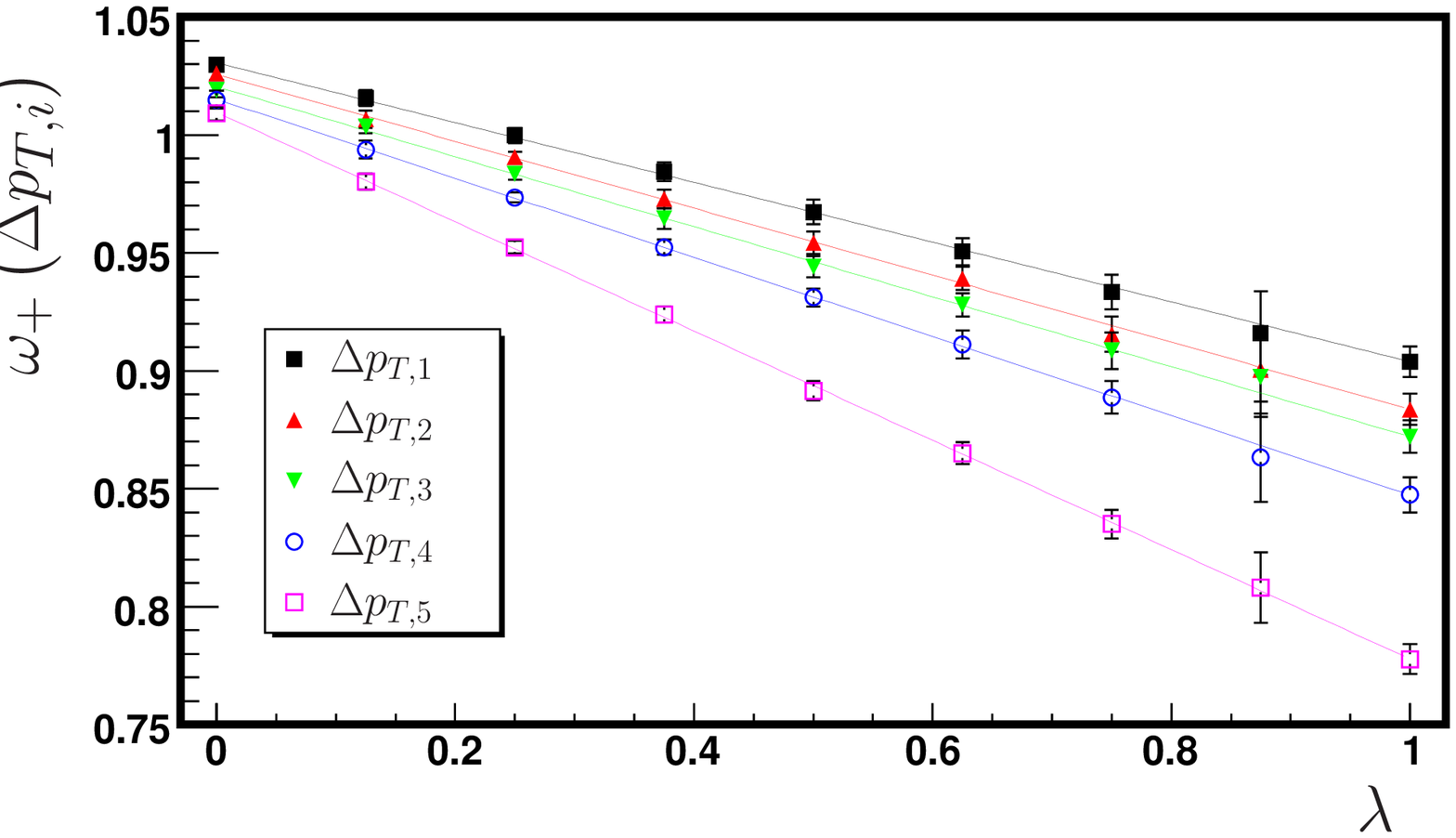,width=8.4cm,height=6.5cm}
  \epsfig{file=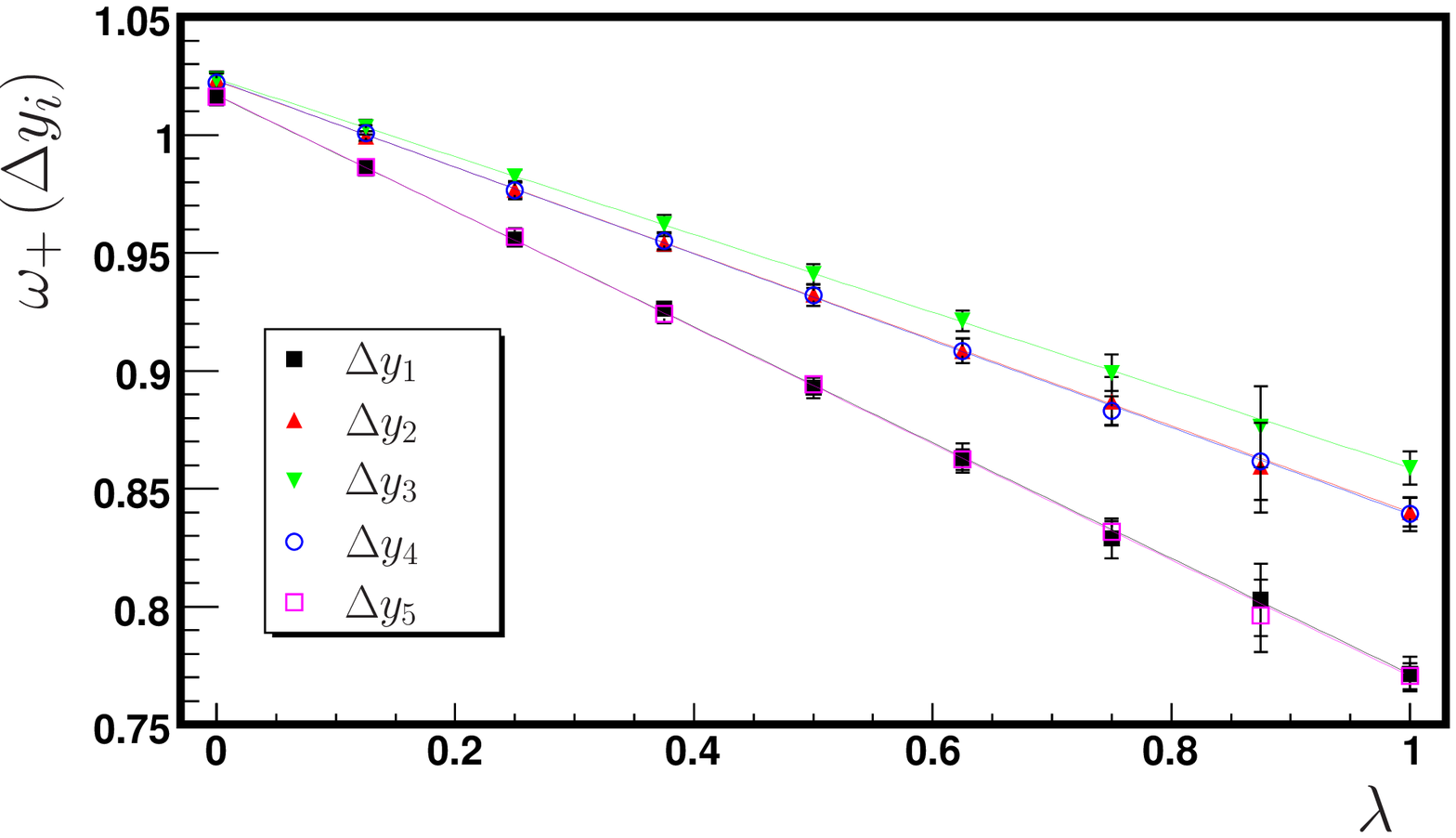,width=8.4cm,height=6.5cm}
  \caption{Evolution of the final state scaled variance $\omega_+$ of positively charged 
    hadrons with the Monte Carlo parameter $\lambda = V_1/V_g$ for transverse momentum
    bins $\Delta p_{T,i}$ ({\it left}) and for rapidity bins $\Delta y_i$ ({\it right}). 
    The solid lines show an analytic extrapolation from GCE results ($\lambda =0$)
    to the MCE limit ($\lambda \rightarrow 1$).
    Each marker except the last represents the result of $20$ Monte Carlo 
    runs of $2 \cdot 10^5$ events. 
    $8$~different equally spaced values of $\lambda$ have been investigated.
    The last marker denotes the result of the extrapolation.
  }  
  \label{conv_lambda_omega_final}
\end{figure}

\begin{figure}[ht!]
  \epsfig{file=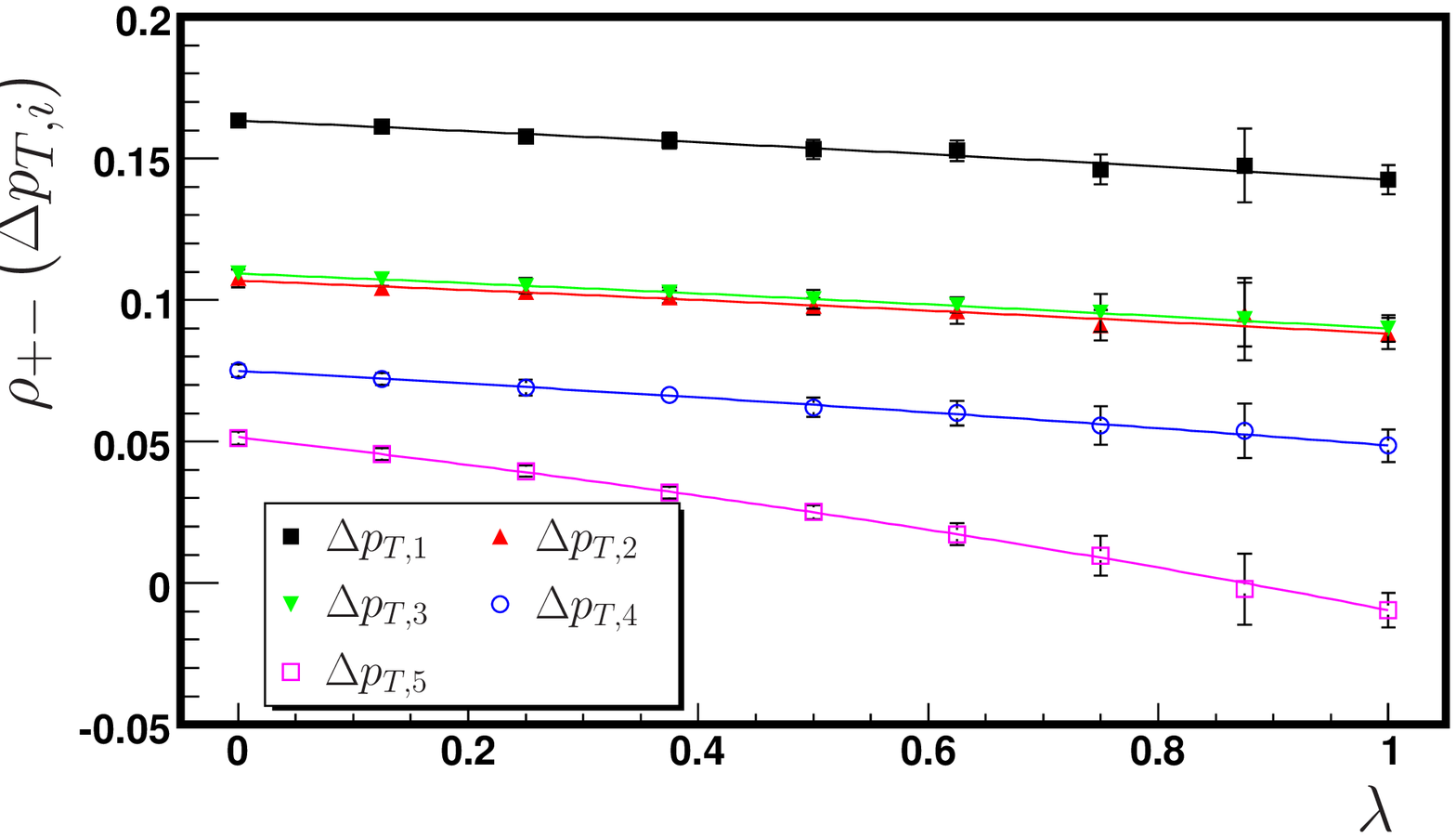,width=8.4cm,height=6.5cm}
  \epsfig{file=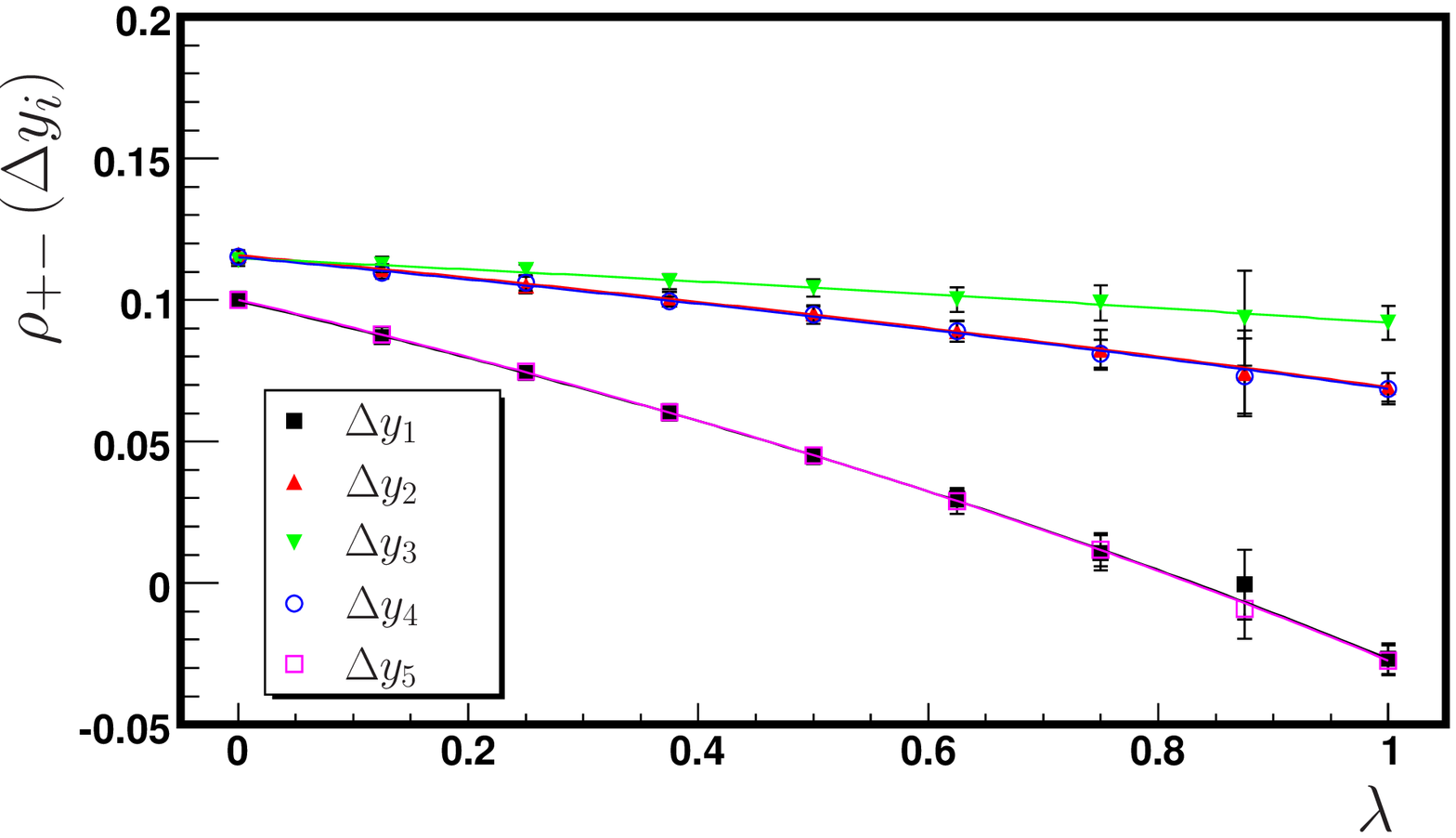,width=8.4cm,height=6.5cm}
  \caption{Evolution of the final state correlation coefficient $\rho_{+-}$ between 
    positively and negatively charged hadrons with the Monte Carlo parameter 
    $\lambda = V_1/V_g$ for transverse momentum bins $\Delta p_{T,i}$ ({\it left}) and 
    for rapidity bins $\Delta y_i$ ({\it right}). 
    The rest as in Fig.(\ref{conv_lambda_omega_final}).}  
  \label{conv_lambda_rho_final}
\end{figure}

In Fig.(\ref{conv_lambda_omega_final}) we show the final state scaled variance $\omega_+$
of positively charged hadrons in transverse momentum bins $\Delta p_{T,i}$ ({\it left}) 
and rapidity bins $\Delta y_i$ ({\it right}) as a function of $\lambda$, while in 
Fig.(\ref{conv_lambda_rho_final}) we show the dependence of the final state correlation
coefficient $\rho_{+-}$ between positively and negatively charged hadrons in transverse
momentum bins $\Delta p_{T,i}$ ({\it left}) and rapidity bins $\Delta y_i$ ({\it right}) on
the size of the bath $\lambda = V_1 / V_g$.

The $\Delta p_{T,i}$ and $\Delta y_i$ dependence on $\lambda$ of the final state MCE 
scaled variance $\omega_+$ is qualitatively similar to that of the primordial 
versions, Fig.(\ref{conv_lambda_omega_prim}), and is essentially 
also explained by the arguments 
of the previous section. The effects of charge and energy-momentum conservation work in pretty 
much the same way as before, and it still seems favorable to have events with 
wider multiplicity distributions at low $p_T$ and low $y$, and narrower distributions at 
larger $p_T$ and larger $|y|$.  The dependence of the final state 
correlation coefficients $\rho_{+-}$ on $\lambda$, Fig.(\ref{conv_lambda_rho_final}), 
is a bit different to the primordial case, Fig.(\ref{conv_lambda_rho_prim}). 
However, in the MCE limit, events still tend to have more similar numbers of 
oppositely charged particles at low $p_T$ and low $y$, than at large $p_T$ and large $|y|$.

The effects of resonance decay are qualitatively different in the MCE, CE, and GCE. 
Let us again first attend to fully phase space integrated multiplicity fluctuations 
discussed  in~\cite{MCEvsData,Res}. 
The final state scaled variance increases in the GCE and CE compared to the primordial scaled 
variance. Multiplicity fluctuations of neutral mesons remain unconstrained by 
conservation laws. However, they often decay into oppositely charged particles, which increases 
multiplicity fluctuations of pions, for instance. In the MCE, due to the constraint of 
energy conservation, the event-by-event fluctuations of primordial pions are correlated to the 
event-by-event fluctuations of, in general, primordial parent particles, and 
$\omega^{final}< \omega^{prim}$ is possible in the MCE.

\begin{figure}[ht!]
  \epsfig{file=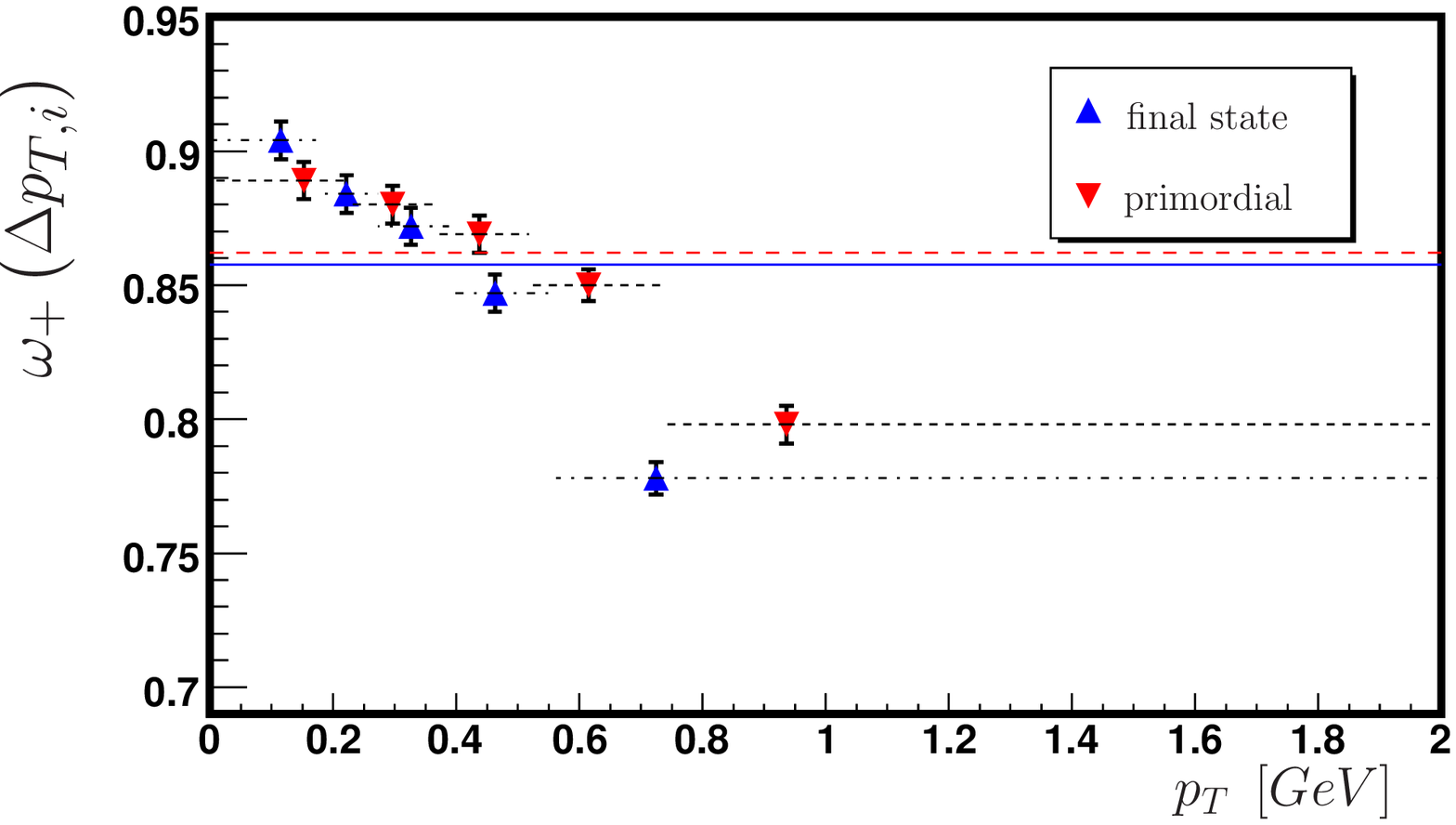,width=8.4cm,height=6.5cm}
  \epsfig{file=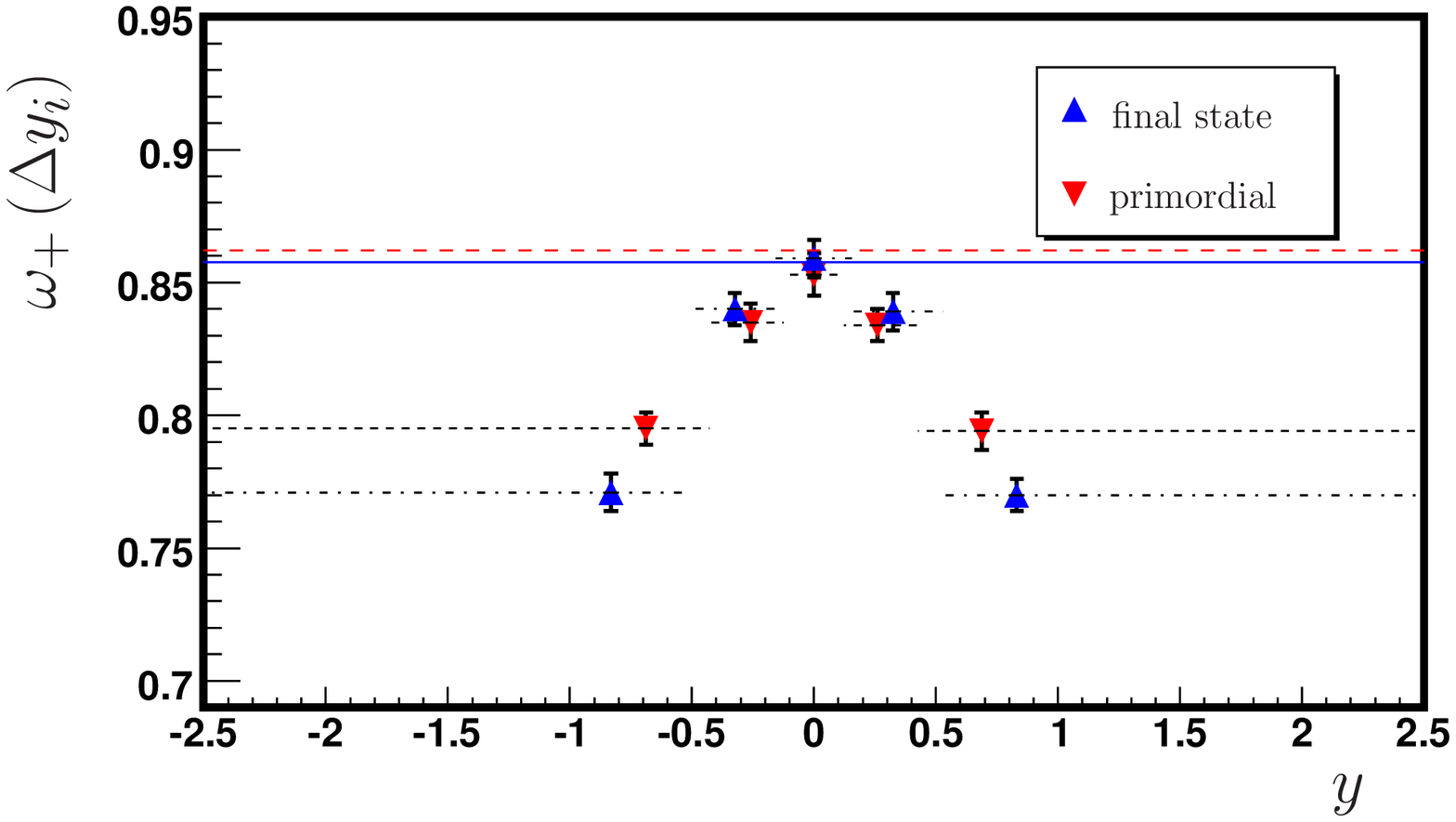,width=8.4cm,height=6.5cm}
  \caption{MCE scaled variance $\omega_+$ of multiplicity fluctuations of positively 
    charged hadrons, both primordial and final state, in transverse momentum bins 
    $\Delta p_{T,i}$ ({\it left}) and rapidity bins $\Delta y_i$ ({\it right}). 
    Horizontal error bars indicate the width and position of the momentum bins 
    (And not an uncertainty!).
    Vertical error bars indicate the statistical uncertainty quoted in 
    Table~\ref{accbins_Mult_final}.
    The markers indicate the center of gravity of the corresponding bin.
    The solid and the dashed lines show final state and primordial acceptance 
    scaling estimates respectively.
  }
  \label{mult_pm_MCE_omega}
\end{figure}

\begin{figure}[ht!]
  \epsfig{file=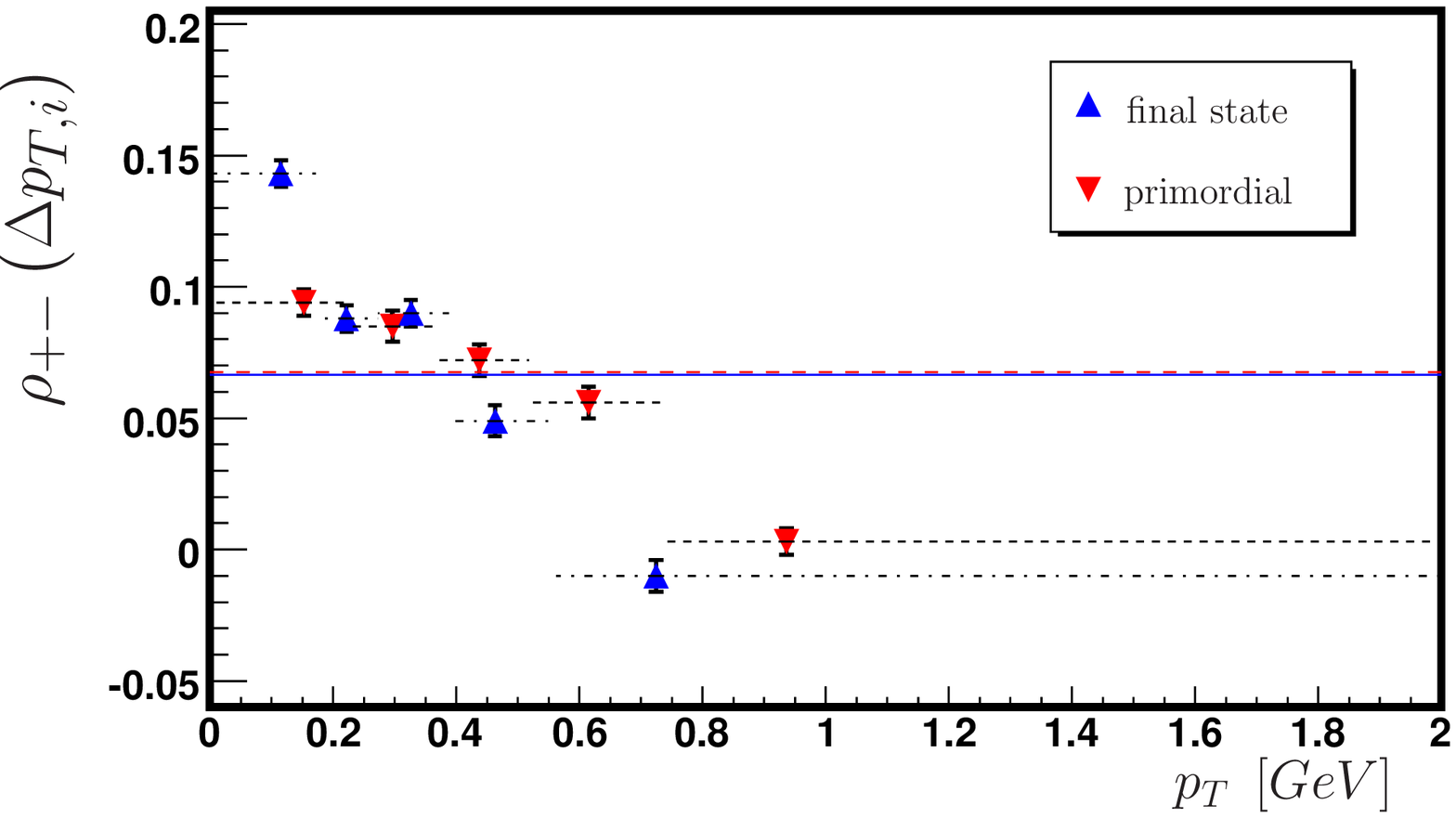,width=8.4cm,height=6.5cm}
  \epsfig{file=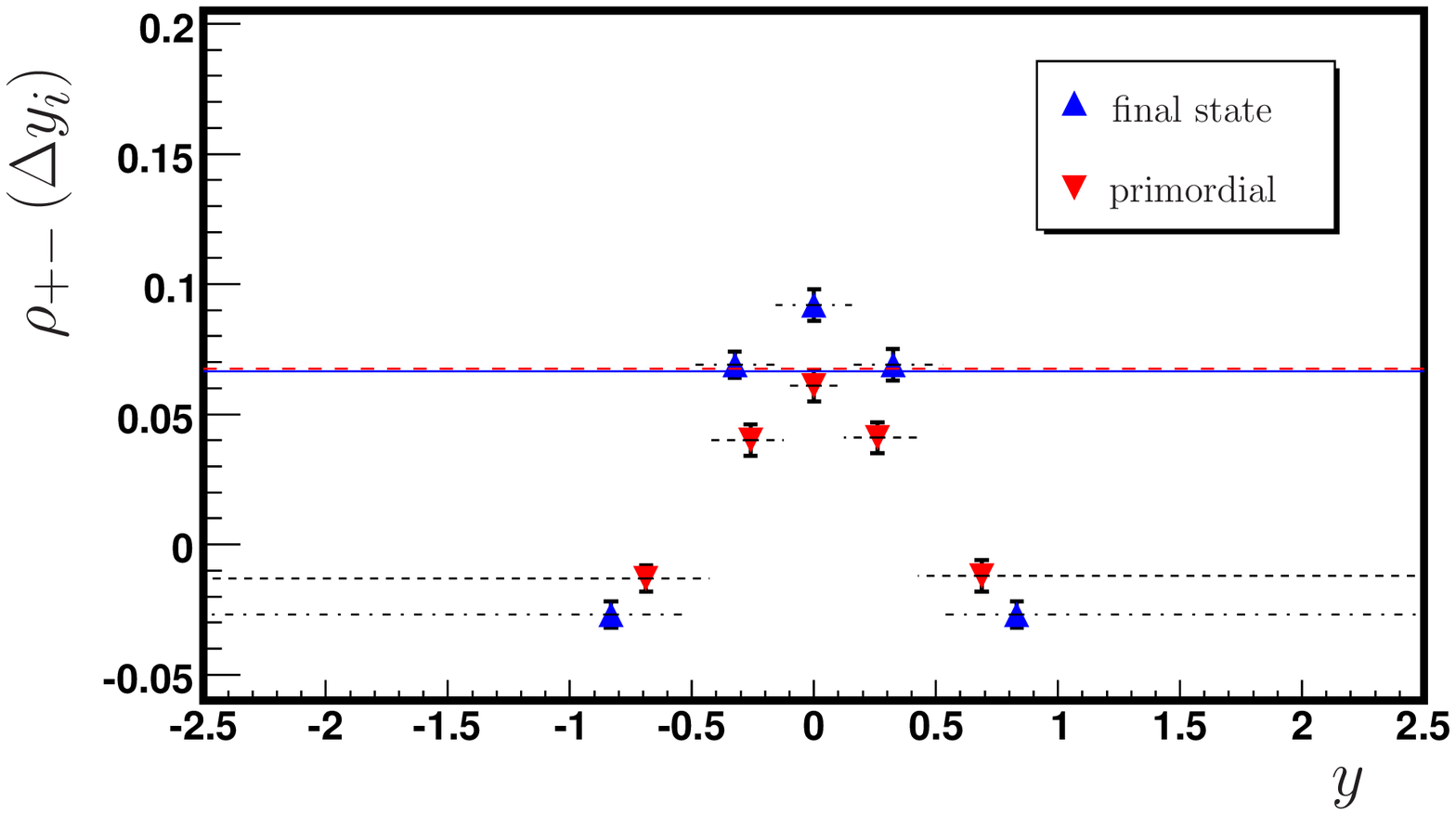,width=8.4cm,height=6.5cm}
  \caption{MCE multiplicity correlation coefficient $\rho_{+-}$ between positively 
    and negatively charged hadrons, both primordial and final state, in transverse 
    momentum bins $\Delta p_{T,i}$ ({\it left}) and rapidity bins $\Delta y_i$ ({\it right}). 
    The rest as in Fig.(\ref{mult_pm_MCE_omega}).
  }  
  \label{mult_pm_MCE_rho}
\end{figure}

In Fig.(\ref{mult_pm_MCE_omega}) and Fig.(\ref{mult_pm_MCE_rho}) we compare 
the final state $\Delta p_{T,i}$ ({\it left}) 
and $\Delta y_i$ ({\it right}) dependence of the MCE scaled variance $\omega_+$ and
the MCE correlation coefficient $\rho_{+-}$ respectively to 
their primordial counterparts. The results of $8 \cdot 20$ Monte 
Carlo runs of $2\cdot 10^5$ events each for a static and neutral hadron resonance gas 
 with $T=0.160 GeV$ are summarized in Table (\ref{accbins_Mult_final}). 

\begin{table}[h!]
  \begin{center}
    \begin{tabular}{||c||c|c|c|c|c||}\hline
      ~final state~& $\Delta p_{T,1}$ & $\Delta p_{T,2}$ & $\Delta p_{T,3}$ & $\Delta p_{T,4}$ 
      & $\Delta p_{T,5}$   \\
      \hline
      ~$\omega^{gce}_+$ ~&~ $1.031 \pm 0.002$ ~&~ 
      $1.026 \pm 0.002$ ~&~ $1.020 \pm 0.002$ ~&~ 
      $1.015 \pm 0.002$ ~&~ $1.010 \pm 0.002$~ \\
      ~$\omega^{mce}_+$ ~&~ $0.904 \pm 0.007$ ~&~ 
      $0.884 \pm 0.007$ ~&~ $0.872 \pm 0.007$ ~&~ 
      $0.847 \pm 0.007$ ~&~ $0.778 \pm 0.006$~ \\
      \hline
      ~$\rho^{gce}_{+-}$ ~&~ $0.163 \pm 0.001$ ~&~ 
      $0.107 \pm 0.001$ ~&~ $0.109 \pm 0.001$ ~&~ 
      $0.075 \pm 0.002$ ~&~ $0.052 \pm 0.002$~ \\
      ~$\rho^{mce}_{+-}$ ~&~ $0.143 \pm 0.005$ ~&~ 
      $0.088 \pm 0.005$ ~&~ $0.090 \pm 0.005$ ~&~ 
      $0.049 \pm 0.006$ ~&~ $-0.010 \pm 0.006$~ \\
      \hline \hline
      ~final state~& $\Delta y_1$  & $\Delta y_2$  & $\Delta y_3$  & $\Delta y_4$  
      & $\Delta y_5$ \\
      \hline 
      ~$\omega^{gce}_+$ ~&~ $1.017 \pm 0.002$ ~&~ 
      $1.023 \pm 0.002$ ~&~ $1.024 \pm 0.002$ ~&~ 
      $1.023 \pm 0.003$ ~&~ $1.017 \pm 0.002$~ \\
      ~$\omega^{mce}_+$ ~&~ $0.771 \pm 0.007$ ~&~ 
      $0.840 \pm 0.006$ ~&~ $0.859 \pm 0.007$ ~&~ 
      $0.839 \pm 0.007$ ~&~ $0.770 \pm 0.006$~ \\
      \hline 
      ~$\rho^{gce}_{+-}$ ~&~ $0.100 \pm 0.001$ ~&~ 
      $0.116 \pm 0.001$ ~&~ $0.115 \pm 0.002$ ~&~ 
      $0.115 \pm 0.002$ ~&~ $0.100 \pm 0.001$~ \\
      ~$\rho^{mce}_{+-}$ ~&~ $-0.027 \pm 0.005$ ~&~ 
      $0.069 \pm 0.005$ ~&~ $0.092 \pm 0.006$ ~&~ 
      $0.069 \pm 0.006$ ~&~ $-0.027 \pm 0.005$~ \\
      \hline
    \end{tabular}
    \caption{Summary of the final state scaled variance $\omega_+$ of positively charged hadrons
      and the correlation coefficient $\rho_{+-}$ between positively and negatively charged 
      hadrons in transverse momentum bins $\Delta p_{T,i}$ and rapidity bins $\Delta y_i$. 
      Both the GCE result ($\lambda = 0$) and the extrapolation to the MCE ($\lambda = 1$) are shown.
      The uncertainty quoted corresponds to $20$ Monte Carlo runs of $2 \cdot 10^5$ events
      (GCE) or is the result of the extrapolation (MCE). 
    } 
    \label{accbins_Mult_final}
  \end{center}
\end{table}

A few words to summarize Figs.(\ref{mult_pm_MCE_omega},\ref{mult_pm_MCE_rho}): resonance decay and 
(energy) conservation laws work in the same direction, as far as the transverse momentum
dependence of the  scaled variance $\omega_+$  and the correlation coefficient 
$\rho_{+-}$ is concerned. Both effects lead to increased multiplicity fluctuations and 
an increased correlation between the
multiplicities of  oppositely charged particles in the low $p_T$ region, compared 
to the high $p_T$ domain.

Compared to this, the MCE $\Delta y_i$ dependence of $\omega_+$ and $\rho_{+-}$
is mainly dominated by global conservation of $P_z$. Resonance decay effects, see 
Figs.(\ref{mult_pm_0000_omega},\ref{mult_pm_0000_rho}), are more equal across rapidity, 
than in transverse momentum.
 
Again, we find the scaled variance of all charged particles larger than the scaled 
variance of only positively charged hadrons $\omega_{\pm} > \omega_+$, except 
for when $\rho_{+-} <0$, i.e when the multiplicities of oppositely charged particles
are anti-correlated, as for instance in $\Delta p_{T,5}$, $\Delta y_{1}$, and $\Delta y_{5}$.  
In contrast to that, we narrowly find $\omega_{\pm} > 1$ in the lowest transverse 
momentum bin $\Delta p_{T,1}$.

The qualitative picture presented in Fig.(\ref{mult_pm_MCE_omega}) could be 
compared to similar analysis of UrQMD transport simulation data \cite{beni_urqmd},
or recently published NA49 data on multiplicity fluctuations in limited momentum
bins \cite{beni_data}. We, however, do not claim that the effects discussed above are the sole effects 
leading to the qualitative agreement with either of the two.

\section{Summary}
\label{Sec_Summary}

We have presented a recipe for a thermal model Monte Carlo event generator
capable of extrapolating fluctuation and correlation observables for 
Boltzmann systems of large volume from their  GCE values to the MCE limit.
Our approach has a strong advantage compared to analytical approaches or
standard microcanonical sample-and-reject Monte Carlo techniques, 
in that it can handle resonance decays as well 
as (very) large system sizes at the same time.

To introduce our scheme, 
we have conceptually divided a microcanonical system into two subsystems.
These subsystems are assumed to be in equilibrium with each other,
and subject to the constraints of joint energy-momentum and charge conservation.
Particles are only measured in one subsystem, while the second subsystem
provides a thermodynamic bath.
By keeping the size of the first subsystem fixed, while varying the size of the second, 
one can thus study the dependence of statistical properties of an ensemble 
on the fraction of the system observed (i.e. assess their sensitivity 
to globally applied conservation laws). 
The ensembles generated are thermodynamically equivalent
in the sense that mean values in the observed subsystem remain unchanged
when the size of the bath is varied, provided the combined system is sufficiently large. 

The Monte Carlo process can be divided into four steps. 
In the first two steps primordial particle multiplicities for each species,
and momenta for each particle, are generated for each event 
by sampling the grand canonical partition function. 
In the third step resonance decay of unstable particles is performed. 
Lastly the values of extensive quantities are
calculated for each event and a corresponding weight factor is assigned. 
All events with the same set of extensive quantities hence still 
have `a priori equal probabilities'. In the limit of an infinite bath, all events have 
a weight equal to unity. In the opposite limit of a vanishing bath, only events with an exactly 
specified set of extensive quantities have non-vanishing weight. In between, 
we extrapolate in a controlled manner. 
The method is even rather efficient for large volume, inaccessible 
to sample-and-reject procedures, 
and agrees well, where available, 
with analytic asymptotic microcanonical solutions. 

Given the success of  the hadron resonance gas model in describing experimentally 
measured average hadron yields, and its ability to reproduce 
low temperature lattice susceptibilities, 
the question arises as to whether fluctuation and correlation observables 
also follow its main line. 
In particular, three effects are nicely discussable: 
Resonance decay, conservation laws, and limited acceptance effects.
Due to the Monte Carlo nature, data can be analyzed in close relation to 
experimental analysis techniques. The hadron resonance gas is an ideal testbed 
for this type of study, in that it is simple and intuitive.

The statistical properties of a sample of hadron resonance gas events 
show a systematic dependence on what part of the momentum distribution and
what fraction of the system is observed. 
Two examples served to illustrate: grand canonical charge-charge correlations, 
and microcanonical multiplicity fluctuations and correlations.
In the case of charge-charge correlations, momentum space effects are caused by different
masses of hadrons and, hence, their varying contribution to different 
parts of the momentum spectra.
Although microcanonical effects on the (co)variances of the joint 
baryon number - strangeness - electric charge distribution are considerable, they remain weak
for the correlation coefficients between these quantum numbers. 
In contrast to this, momentum space effects on multiplicity fluctuations and correlations 
arise due to conservations laws. For an ideal primordial grand canonical ensemble 
in the Boltzmann approximation (our starting point), 
multiplicity distributions are just uncorrelated Poissonians, regardless of the 
acceptance cuts applied, as particles are assumed to be produced independently. 
The requirement of energy-momentum and charge conservation leads to suppressed 
fluctuations and enhanced correlations between the multiplicities of two distinct 
groups of particles at the `high momentum' end of the momentum spectrum, provided 
some fraction of an isolated system is observed.
Resonance decay does not change these trends.
The arguments on which the explanation of this particular dependence are based seem
general enough to hope that they might hold too in non-equilibrium systems, such as 
real heavy ion data or theoretical transport simulations. 

A direct comparison with experimental data seems problematic at the moment. 
The static global thermal and chemical equilibrium assumption made here is 
certainly insufficient. The model presented here is far from complete. 
Several interesting aspects deserve attention. 
They include the sampling of Fermi-Dirac or Bose-Einstein particles, 
for which low transverse momentum is particularly sensitive; 
finite volume corrections could be done 
(possible if one has a good approximation to $\mathcal{W}$);
the convergence properties (at fixed $\lambda$, and as a function of $\lambda$) fall 
basically into the same direction;
so far we also have not derived a thermodynamic potential for our ensembles;
one could also consider more general forms of $\mathcal{W}$;
one could ask how to couple two systems of different densities, 
or altogether depart from the local equilibrium assumption.
There are also several interesting things that the model could do in its present form.
Examples include mean transverse momentum fluctuations, correlation between 
transverse momentum and particle number, or even 2 and 3 particle correlation functions. 
This should be the subject of future work.

\begin{acknowledgments}
We would like to thank F.~Becattini, E.~Bratkovskaya, W.~Cassing, J.~Cleymans, 
M.~Gazdzicki, M.~Gorenstein, J.~Manninen, J.~Randrup, and K.~Redlich for fruitful discussions.
Special thanks goes to W.~Broniowski for his contribution to the very idea 
which started this project. The computational work was done on the CARMEN cluster 
of the UCT physics department. We would also like to thank G.~de Vaux for
valuable help with many aspects of running the code. 
\end{acknowledgments}


\appendix

\section{Convergence Study}
\label{App_ConvStudy}
Not only for the sake of completeness we discuss in this section 
the convergence of various quantities
with the sample size, i.e. the number of events, $N_{events}$, in our Monte Carlo
scheme. Here we analyze final state (stable against electromagnetic and weak decays) particles 
only. We mainly take a closer look at the data sub-set of $20 \cdot 2 \cdot 10^5$ events,  
with ${\lambda=V_1/V_g=0.875}$ for the size of the bath, which already has been discussed in 
Section \ref{Sec_MultFluc}. 

There is a degree of freedom at so how to estimate the statistical uncertainty on the 
moments of a distribution of observables of a finite sample. The approach taken
here is straight forward, but could, however, certainly be improved.

\begin{figure}[ht!]
  \epsfig{file=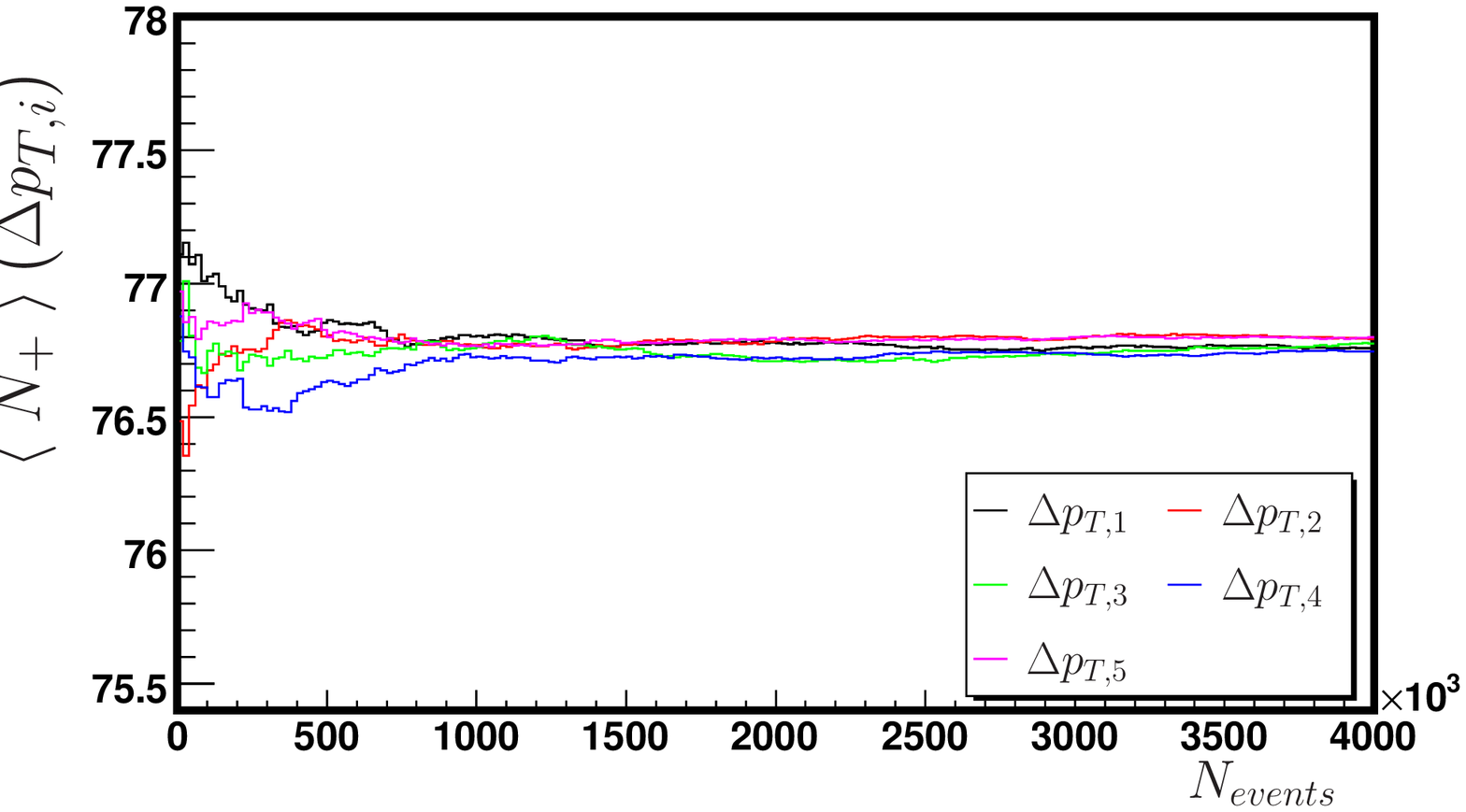,width=8.4cm,height=6.5cm}
  \epsfig{file=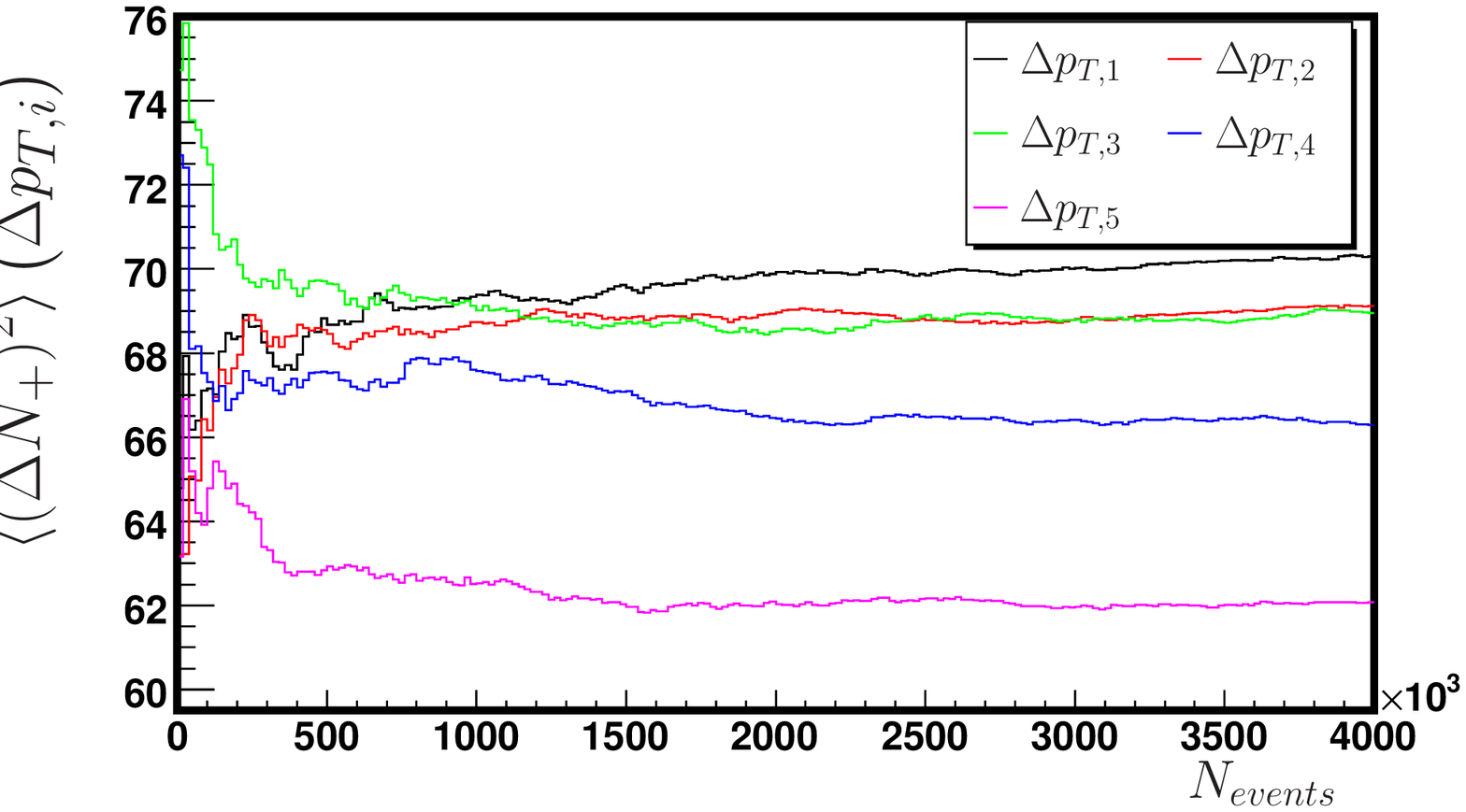,width=8.4cm,height=6.5cm}
  \caption{Step histogram showing the convergence of the mean values $\langle N_+ \rangle$ ({\it left})
    and variances $\langle \left( \Delta N_+ \right)^2 \rangle$ ({\it right}) for positively 
    charged final state hadrons in transverse momentum bins $\Delta p_{T,i}$ for 
    a hadron resonance gas with $\lambda = V_1/ V_g =0.875 $.}  
  \label{stephist_N_DeltaNSq}
\end{figure}

\begin{figure}[ht!]
  \epsfig{file=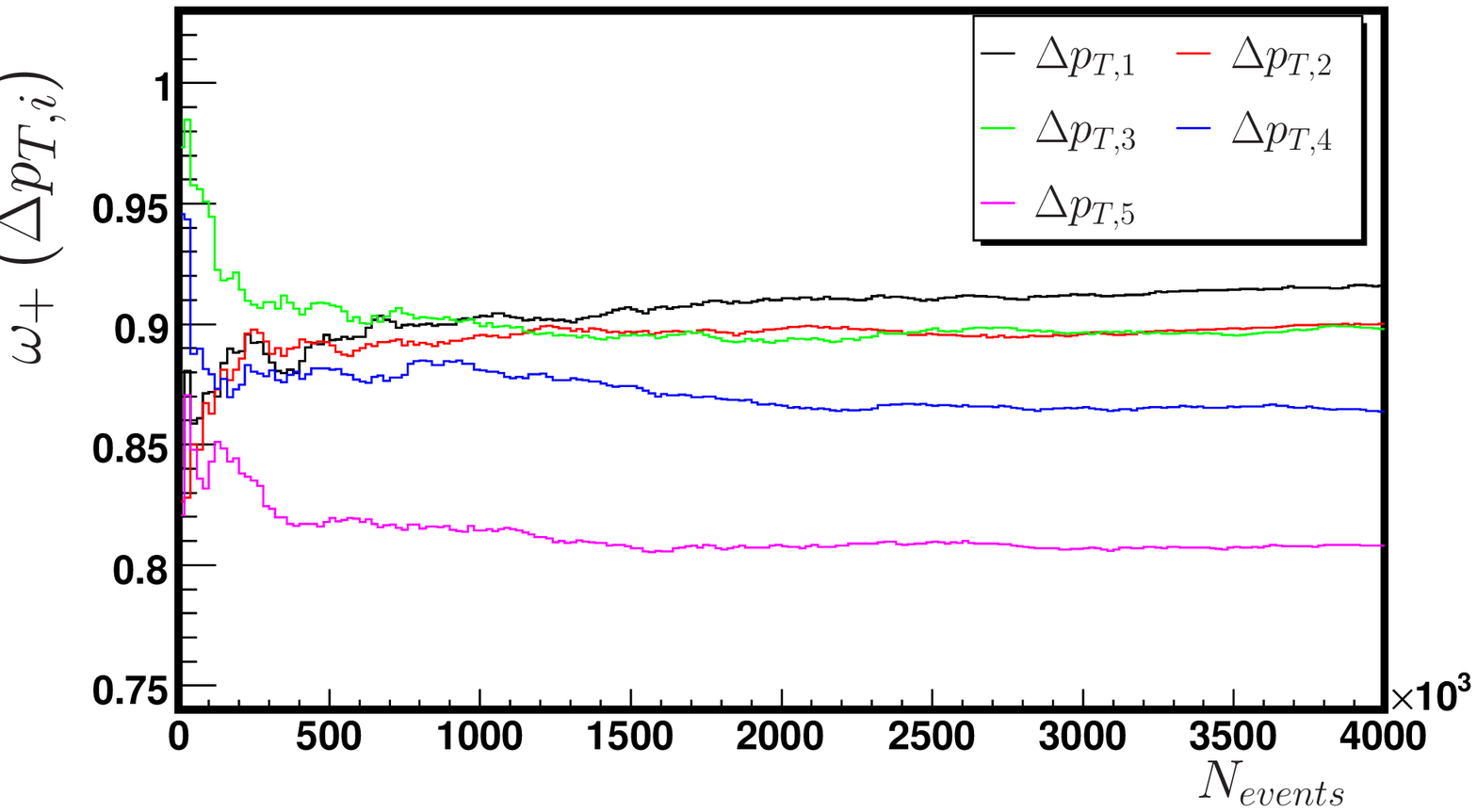,width=8.4cm,height=6.5cm}
  \epsfig{file=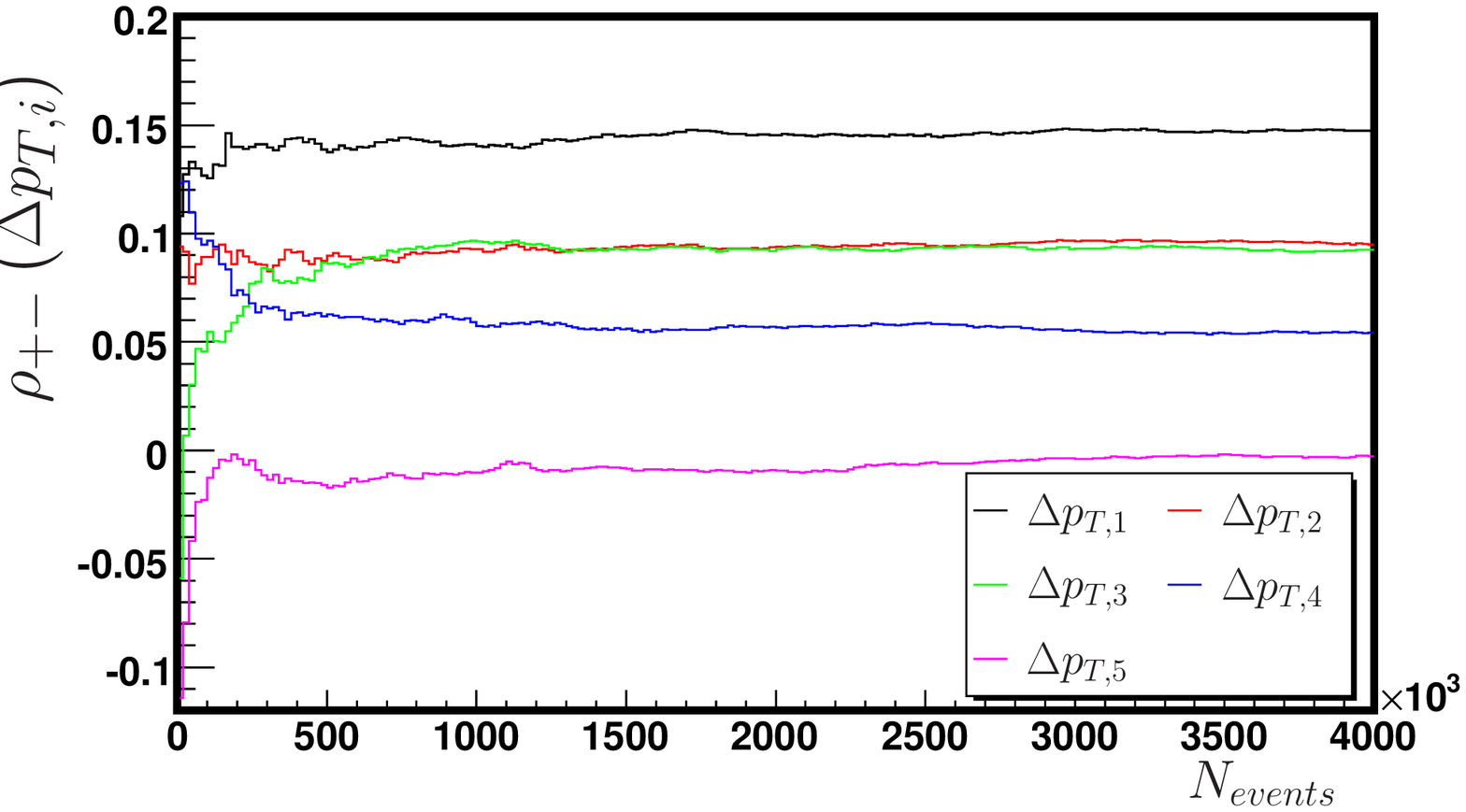,width=8.4cm,height=6.5cm}
  \caption{Step histogram showing the convergence of the scaled variance $\omega_+$ ({\it left})
    of positively charged hadrons and the correlation coefficient $\rho_{+-}$ between positively 
    and negatively charged hadrons ({\it right}) in transverse momentum bins $\Delta p_{T,i}$ for 
    a final state hadron resonance gas with $\lambda = V_1/ V_g =0.875 $.}  
  \label{stephist_omega_rho}
\end{figure}

\begin{figure}[ht!]
  \epsfig{file=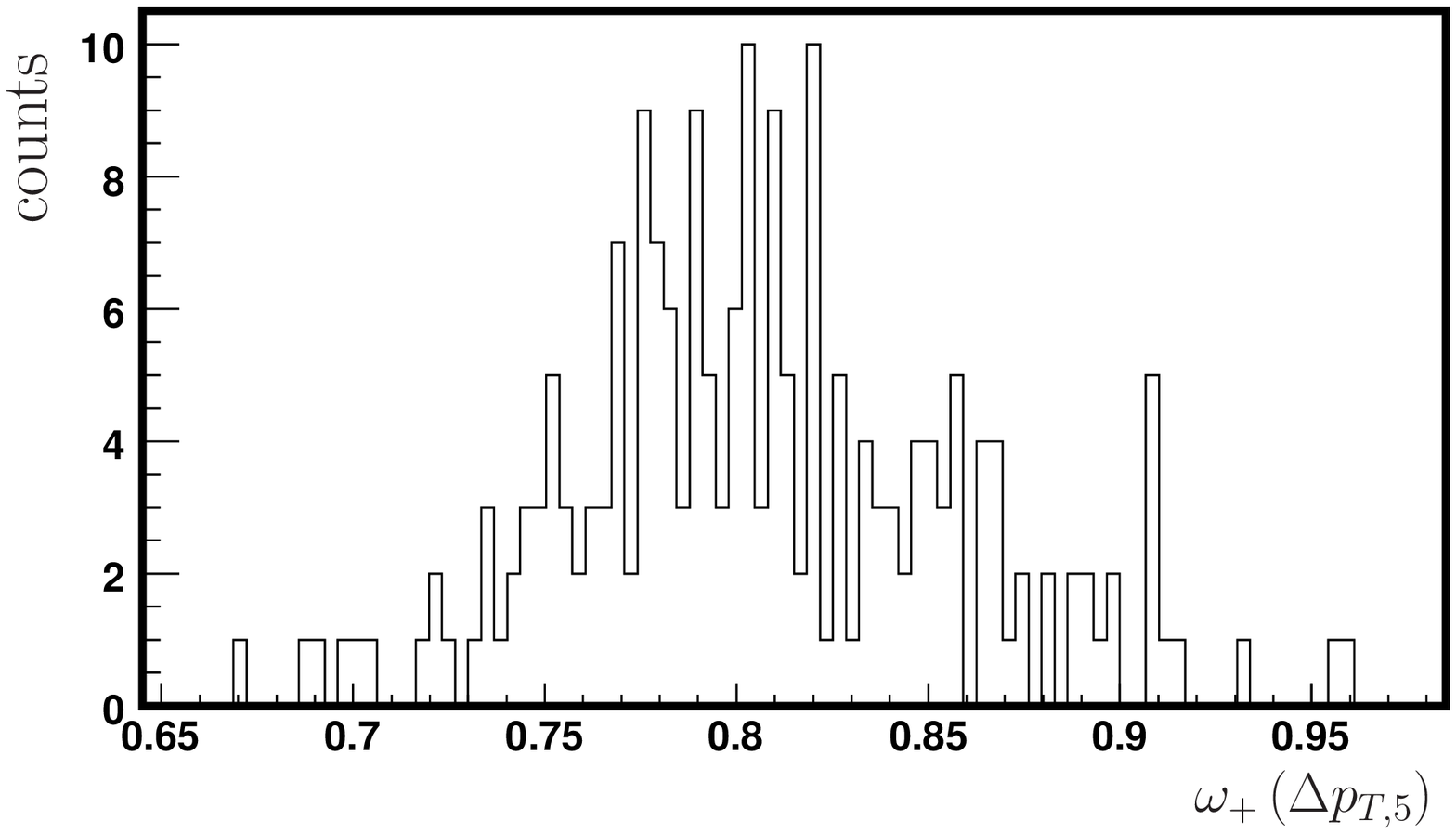,width=8.4cm,height=6.5cm}
  \epsfig{file=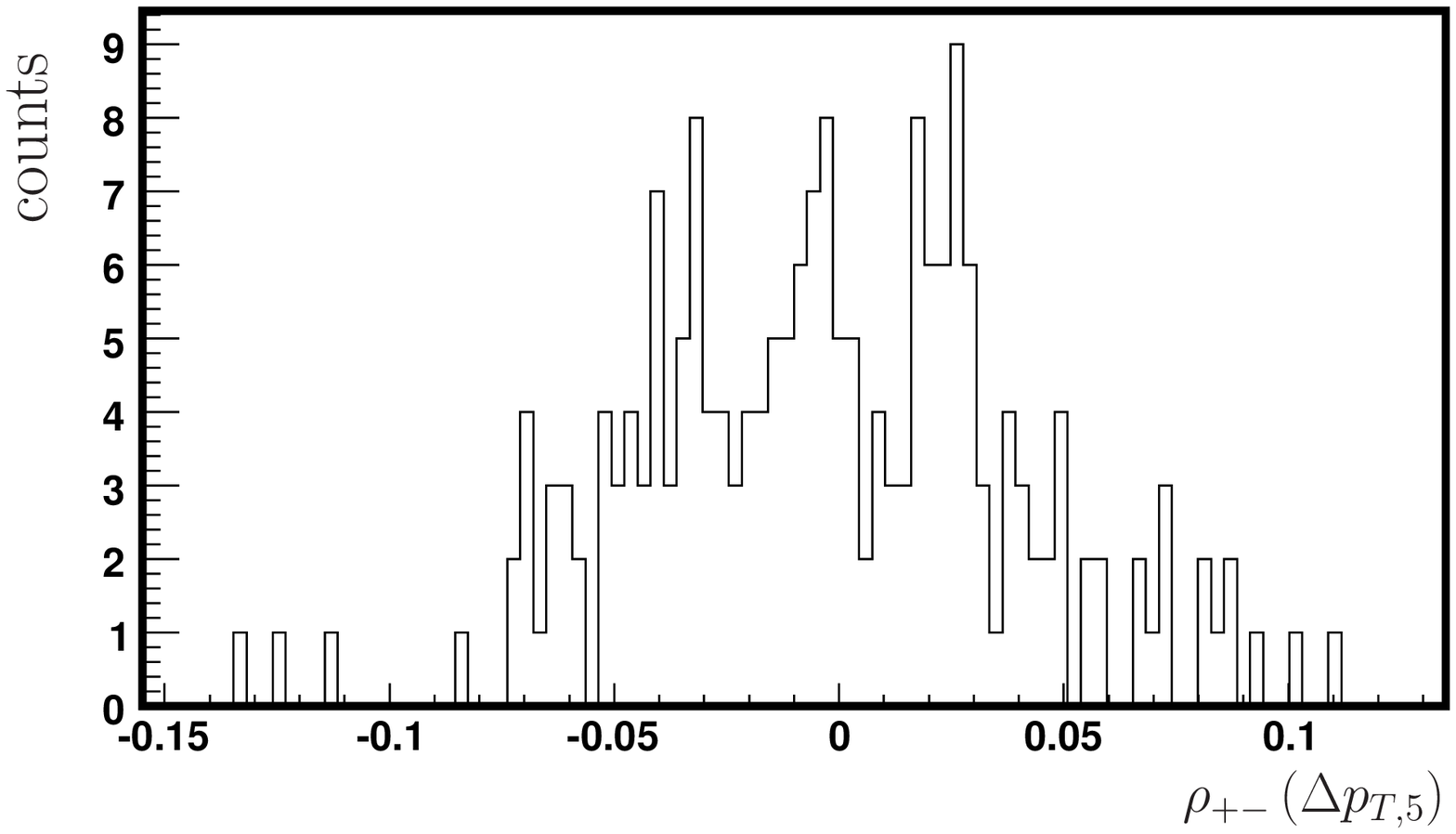,width=8.4cm,height=6.5cm}
  \caption{Histogram showing the results for the scaled variance $\omega_+$ ({\it left})
    of positively charged hadrons and the correlation coefficient $\rho_{+-}$ between positively 
    and negatively charged hadrons ({\it right}) in the transverse momentum bin $\Delta p_{T,5}$ for 
    a final state hadron resonance gas with $\lambda = V_1/ V_g =0.875 $. 
    200 Monte Carlo runs of $2\cdot 10^4$ events each are analyzed.}  
  \label{stat_error}
\end{figure}

In Fig.(\ref{stephist_N_DeltaNSq}) we show the evolution of the mean values 
$\langle N_+ \rangle$ ({\it left}) and the variances $\langle (\Delta N_+)^2\rangle$ 
({\it right}) of the distributions of positively charged hadrons for the 5 transverse 
momentum bins $\Delta p_{T,i}$, defined in Table~\ref{accbins}, with the sample size. 
Mean values of particle multiplicities in respective bins are in rather good approximation
equal to each other, but are, however, not identical due to finite resolution on the underlying 
momentum spectrum, even for $\lambda=0.875$ (bins were constructed 
using GCE events from an independent run). Variances converge 
steadily and are different in different bins, see Section \ref{Sec_MultFluc}.
The event output was iteratively stored in histograms, which were then evaluated 
after steps of $2\cdot10^4$~events. 

In Fig.(\ref{stephist_omega_rho}) we show the evolution of the scaled variance 
$\omega_+$ of positively charged final state particles ({\it left}) and the correlation
coefficient $\rho_{+-}$ between positively and negatively charged particles ({\it right}).
The results for the respective transverse momentum bins can be compared to the second to 
last markers Figs.(\ref{conv_lambda_omega_final},\ref{conv_lambda_rho_final}), 
left panels, which denote the corresponding results of grouping the same data into 
$20$ Monte Carlo sets of $2 \cdot 10^5$ events each.

\begin{figure}[ht!]
  \epsfig{file=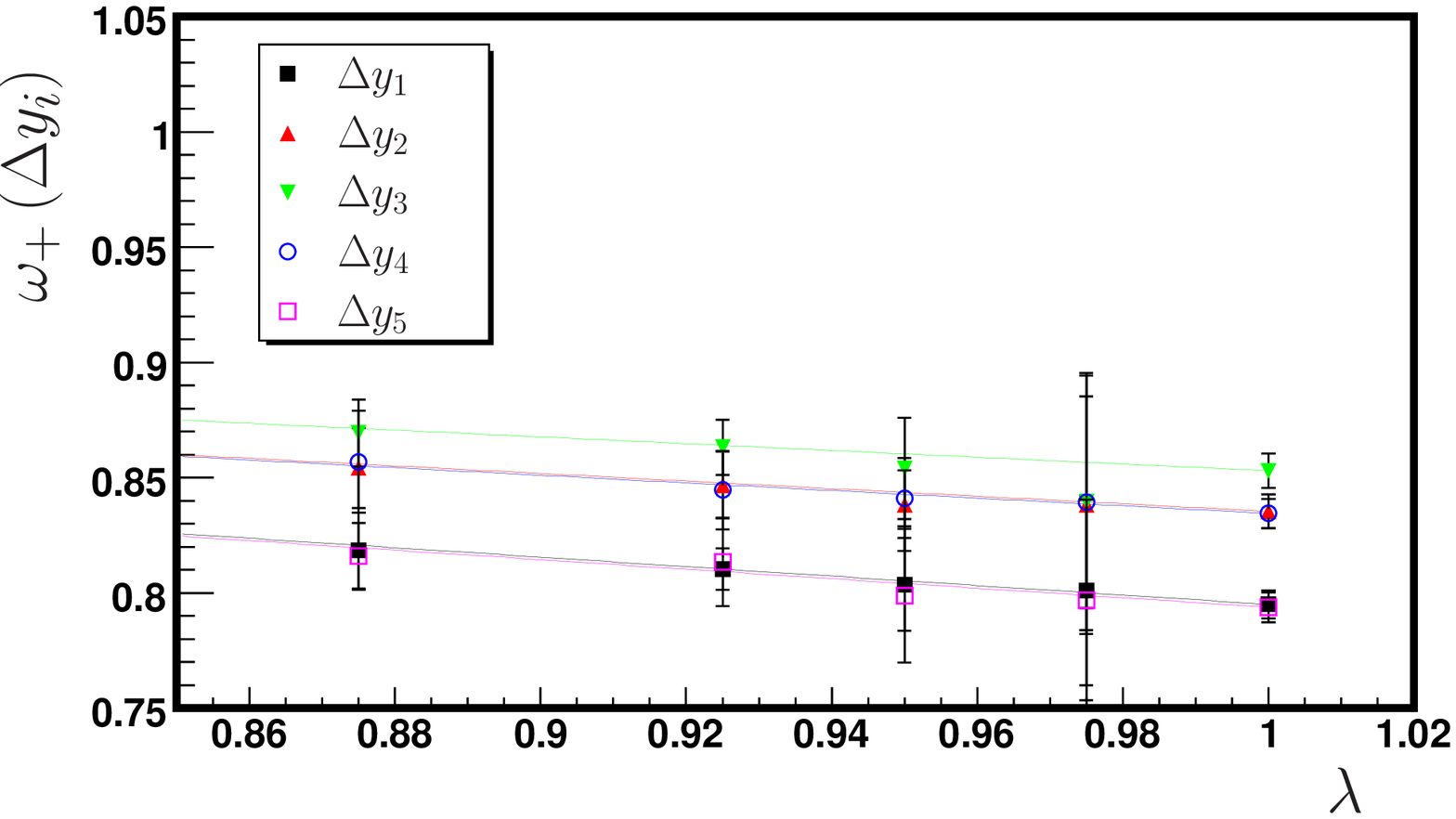,width=8.4cm,height=6.4cm}
  \epsfig{file=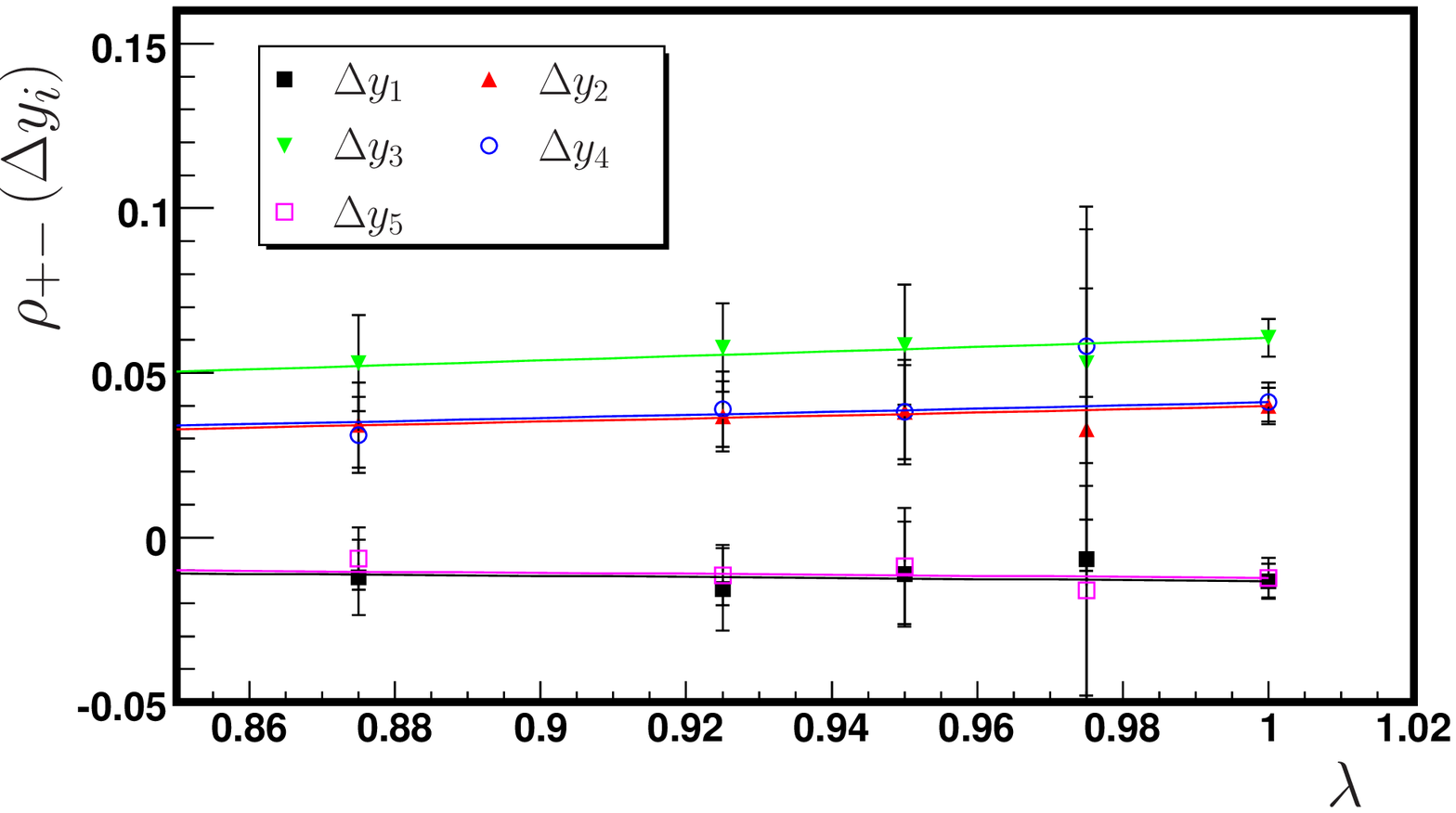,width=8.4cm,height=6.4cm}
  \caption{Evolution of the primordial scaled variance $\omega_+$ of positively 
    charged hadrons ({\it left}) and the primordial correlation coefficient $\rho_{+-}$
    between positively and negatively charged hadrons ({\it right}) with the Monte Carlo 
    parameter $\lambda = V_1/V_g$ in different rapidity bins $\Delta y_i$. 
    The solid lines show an analytic extrapolation from GCE results ($\lambda =0$)
    to the MCE limit ($\lambda \rightarrow 1$).
    The $4$ leftmost markers and their error bars represent the results of $20$ Monte Carlo runs 
    of $2 \cdot 10^5$ events. 
    $3$ additional values of $\lambda$ have been investigated with $20$ Monte Carlo runs 
    of $1 \cdot 10^7$ events. 
    The rightmost markers denote the results of the extrapolation.
   }  
  \label{largelambda}
\end{figure}

In Fig.(\ref{stat_error}) we show the distribution of scaled variances of positively 
charged particles 
$\omega_+$ ({\it left}) and correlation coefficients between positively and 
negatively charged particles $\rho_{+-}$ ({\it right}), resulting from grouping again
the same data set into $200$ samples of $2\cdot 10^4$ events each. 
We chose the transverse momentum bin 
$\Delta p_{T,5}$ for a final state hadron resonance gas with $\lambda = V_1/ V_g =0.875 $. 

Monte Carlo results for $\lambda = 0.875$ of the analysis shown in Fig.(\ref{stat_error}), 
are for the scaled variance $\omega_+ (\Delta p_{T,5}) = 0.8069  \pm 0.0514$, and
the correlation coefficient $\rho_{+-} (\Delta p_{T,5}) = -0.0026 \pm 0.0421$. 
They are nicely scattered around the mean values, denoted by the bottom lines in 
Fig.(\ref{stephist_omega_rho}), $\omega_+ (\Delta p_{T,5}) = 0.8082$, and 
$\rho_{+-} (\Delta p_{T,5}) = -0.0028$ respectively.

They are also compatible with the analysis shown in 
Figs.(\ref{conv_lambda_omega_final},\ref{conv_lambda_rho_final}),
of Section \ref{Sec_MultFluc}, 
$\omega_+ (\Delta p_{T,5}) = 0.8081 \pm 0.0149$, and
$\rho_{+-} (\Delta p_{T,5}) = -0.0022 \pm 0.0125$,
at the same value of $\lambda$. The comparatively large statistical error 
on the analysis in Fig.(\ref{stat_error}) is due to the splitting up into many
small sub-samples. The mean values of different analyses agree rather well.

Lastly, we show in Fig.(\ref{largelambda}) the results of additional Monte Carlo runs
for values of $\lambda$ closer to unity. This time we have performed $20$ runs 
of $1 \cdot 10^7$ primordial events for $\lambda =0.925$, $0.950$, and $0.975$. 
As discussed above, error bars diverge, but convergence 
seems to be rather good. The additional data has not been used for the extrapolation, so it 
can serve as an un-biased cross-check.

\section{The Canonical Boltzmann Gas}
\label{App_CBG}
An analytical and instructive example is the canonical classical relativistic 
particle anti-particle gas discussed in~\cite{CEfirst,Turko,CEsecond}. 
We use this example
to show that, although the procedure is formally independent of one's choice 
of Lagrange multipliers, it is most efficient for those defined by
Maxwell's relations. We start off with Eqs.(\ref{eq_one}), and then 
discuss, in turn, the first and second moments of the multiplicity 
distribution of particles, and the first four moments of the 
Monte Carlo weight factor. 

The canonical partition function $Z_{N_1}(V_1,\beta,Q_1)$ 
of a system with volume $V_1$, temperature ${T=\beta^{-1}}$, 
charge $Q_1$, particle number $N_1$, 
and anti-particle number $M_1=N_1-Q_1$,  is given by:
\begin{equation}
Z_{N_1}(V_1,\beta,Q_1) ~=~ \frac{\left( V_1 \psi \right)^{N_1}}{N_1!} 
~ \frac{\left( V_1 \psi \right)^{N_1-Q_1}}{\left( N_1 -Q_1 \right)!}~.
\end{equation}
The single particle partition function is given by Eq.(\ref{psi_meanN}), 
${\psi=\frac{g}{2\pi^2}~m^2~\beta^{-1}~K_2\left(m \beta\right)}$. 
The canonical partition function with arbitrary particle number, 
but still fixed charge $Q_1$, is obtained by:
\begin{equation}
Z(V_1,\beta,Q_1) ~=~ \sum_{N_1=Q_1}^{\infty} ~ Z_{N_1}(V_1,\beta,Q_1)  
~=~ I_{Q_1} \left(2~V_1~ \psi \right)~.
\end{equation}
Here $I_{Q_1}$ is a modified Bessel function. 
Temperature is the same in both subsystems; the bath and the observable part.
The partition function of the bath is therefore:
\begin{equation}
Z(V_2,\beta,Q_2) ~=~  I_{Q_2} \left( 2~ V_2~\psi \right)~.
\end{equation}
Imposing the constraints $V_2=V_g-V_1$, and $Q_2=Q_g-Q_1$, similar to 
Eq.(\ref{constraint_Q}), we find \cite{Abramowitz} for the canonical 
partition function, Eq.(\ref{PF_combined}), of the combined system:
\begin{equation}
Z(V_g,\beta,Q_g) ~=~ 
\sum_{Q_1= - \infty}^{\infty}  I_{Q_1} \left(2~V_1~ \psi \right) 
~I_{Q_g-Q_1} \Big(2 \left( V_g -V_1\right)\psi \Big) 
~=~ I_{Q_g} \left(2~V_g~ \psi \right)~,
\end{equation}
as required. The weight factor is then:
\begin{equation}\label{appendix_W}
  W(V_1,Q_1;V_g,Q_g|\beta) ~=~ 
  \frac{ I_{Q_g-Q_1} \big( 2 \left(V_g-V_1\right) \psi \big) }{ 
    I_{Q_g}\left(2V_g \psi \right)}~.
\end{equation}
Analogous to Eq.(\ref{basic}) we find for the joint particle multiplicity 
and charge distribution:
\begin{eqnarray}\label{BP_Q1N1_Qg}
P(Q_1,N_1) &=~& W(V_1,Q_1;V_g,Q_g|\beta) ~ Z_{N_1}(V_1,\beta,Q_1)~.
\end{eqnarray}

\subsection{Monte Carlo Weight}
We next introduce Eq.(\ref{Pgce}), 
the joint GCE distribution of charges and particle multiplicity:
\begin{equation}
P_{gce}(Q_1,N_1)~=~ \frac{e^{Q_1 \mu \beta }}{Z(V_1,\beta,\mu)}
~Z_{N_1}(V_1,\beta,Q_1)~.
\end{equation}
The Monte Carlo weight, Eq.(\ref{simpleW}), is then given by:
\begin{equation}\label{appendix_W_G}
\mathcal{W}^{Q_1;Q_g}(V_1;V_g|\beta,\mu)~\equiv~ 
W(V_1,Q_1;V_g,Q_g|\beta)~Z(V_1,\beta,\mu) ~ e^{-Q_1 \mu \beta}~.
\end{equation}
In accordance with Eq.(\ref{thetrick}), 
the distribution Eq.(\ref{BP_Q1N1_Qg}) is then equivalently written as:
\begin{eqnarray}\label{BP_Q1N1_Qg_G}
P(Q_1,N_1) &=~& \mathcal{W}^{Q_1;Q_g}(V_1;V_g|\mu,\beta)
~P_{gce}(Q_1,N_1)~.
\end{eqnarray}
The GCE partition function is: 
\begin{equation}
Z(V_1,\beta,\mu) ~=~ \sum \limits_{Q_1=-\infty}^{\infty} e^{Q_1 \mu \beta  }
~ Z(V_1,\beta,Q_1) ~=~ \exp \big[ V_1 2 \cosh(\beta \mu) \big]~.
\end{equation}

\subsection{Moments of Distributions}
To define the multiplicity moments of the distributions Eq.(\ref{BP_Q1N1_Qg}) 
or Eq.(\ref{BP_Q1N1_Qg_G}) we write:
\begin{equation}\label{mom_N}
\langle N_1^n \rangle ~\equiv~ \sum \limits_{N_1=0}^{\infty} 
\sum \limits_{Q_1=-\infty}^{\infty} ~N_1^n ~P(N_1,Q_1)~.
\end{equation}
Additionally we define the moments of the weight Eq.(\ref{appendix_W}):
\begin{equation}\label{mom_W}
\langle W^n \rangle ~\equiv~ \sum \limits_{N_1=0}^{\infty} 
\sum \limits_{Q_1=-\infty}^{\infty} 
~\Big[  W(V_1,Q_1;V_g,Q_g|\beta) \Big]^n ~  Z_{N_1}(V_1,\beta,Q_1)~,
\end{equation}
and of the Monte Carlo weight Eq.(\ref{appendix_W_G}): 
\begin{equation}\label{mom_gen_W}
\langle\mathcal{W}^n \rangle ~\equiv~ \sum \limits_{N_1=0}^{\infty} 
\sum \limits_{Q_1=-\infty}^{\infty} 
~\Big[ \mathcal{W}^{Q_1;Q_g}(V_1;V_g|\beta,\mu) \Big]^n ~   P_{gce}(Q_1,N_1)~.
\end{equation}
We first attend to the first two moments of the multiplicity distribution.
Substituting Eq.(\ref{BP_Q1N1_Qg}) or Eq.(\ref{BP_Q1N1_Qg_G}) 
into Eq.(\ref{mom_N}) yields:
\begin{equation}\label{mean_N}
\langle N_1 \rangle ~=~ \left( V_1 \psi \right) ~ \frac{
  I_{Q_g-1}\left(2V_g \psi \right)}{ I_{Q_g}\left(2V_g \psi \right)}~,
\end{equation}
and
\begin{equation}\label{N_Sq}
\langle N^2_1 \rangle ~=~ \left( V_1 \psi \right) ~ \frac{
  I_{Q_g-1}\left(2V_g \psi \right)}{ I_{Q_g}\left(2V_g \psi \right)} ~+~
\left( V_1 \psi \right)^2 ~ \frac{ I_{Q_g-2}\left(2V_g \psi \right)}{
  I_{Q_g}\left(2V_g \psi \right)}~.
\end{equation}
Canonical suppression of yields and fluctuations acts on the global volume $V_g$.
In the GCE the first two moments are 
$\langle N_1 \rangle =  V_1 \psi e^{\mu \beta }$, and 
$\langle N^2_1 \rangle = \langle N_1 \rangle^2 + \langle N_1 \rangle$,
respectively. The CE limit is obtained by $V_g \rightarrow V_1$, and 
$Q_g = \langle Q_1 \rangle$.
Substituting Eq.(\ref{mean_N}) and Eq.(\ref{N_Sq}) into Eq.(\ref{omega}),
and using Eq.(\ref{lambda_def}), $\lambda = V_1/V_g$, yields:
\begin{eqnarray}\label{svar}
\omega &=& \lambda ~ \omega_{ce} +
\left(1-\lambda \right) ~ \omega_{gce}~,
\end{eqnarray} 
where the CE scaled variance $\omega_{ce}$ of the combined system 
is given by \cite{CEfirst,CEsecond}: 
\begin{equation}\label{omega_ce}
\omega_{ce} ~=~ 1 ~-~ \left(V_g \psi \right) ~ \left[ ~
\frac{ I_{Q_g-1}\left(2V_g \psi \right)}{ I_{Q_g}\left(2V_g \psi \right)}
~-~ \frac{ I_{Q_g-2}\left(2V_g \psi \right)}{ I_{Q_g-1}\left(2V_g \psi \right)}
~\right] ~,
\end{equation}
and $\omega_{gce}=1$ is the GCE scaled variance, as the particle number 
distribution is a Poissonian. 

We next apply our Monte Carlo scheme to an observable 
subsystem of volume $V_1=50fm^3$ embedded into a system of
volume $V_g=75fm^3$, charge $Q_g=10$, and temperature 
$T=\beta^{-1}=0.160GeV$. Particles and anti-particles have 
mass $m=0.140GeV$ and degeneracy factor $g=1$. The average
charge content in the observable subsystem is then 
$\langle Q_1 \rangle \simeq 6.667$. The mean particle multiplicity, 
Eq.(\ref{mean_N}), is $\langle N_1 \rangle \simeq 7.335$, 
and the scaled variance of particle number fluctuations, 
Eq.(\ref{svar}), is $\omega \simeq 0.3896$.
We will sample the GCE in $V_1$ for various values of $\mu_Q$
and use the Monte Carlo weight, Eq.(\ref{appendix_W_G}), to transform these
samples to have the statistical properties required by 
Eq.(\ref{BP_Q1N1_Qg}) or Eq.(\ref{BP_Q1N1_Qg_G}).
For each value of $\mu_Q$ we have generated $50$ samples of
$2000$ events each to allow for calculation of a statistical uncertainty 
estimate. 

\begin{figure}[ht!]
  \epsfig{file=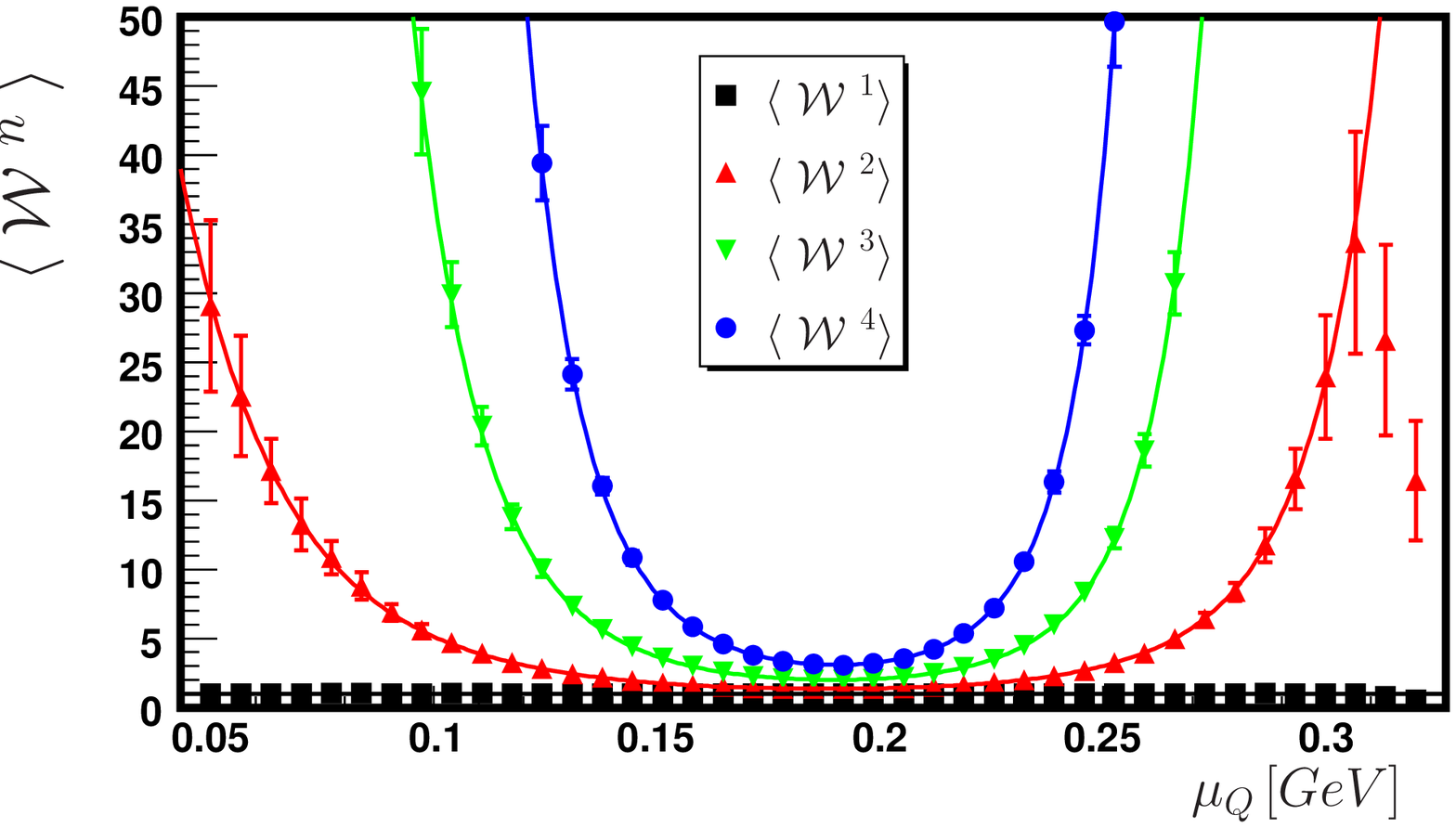,width=8.4cm,height=6.5cm}
  \epsfig{file=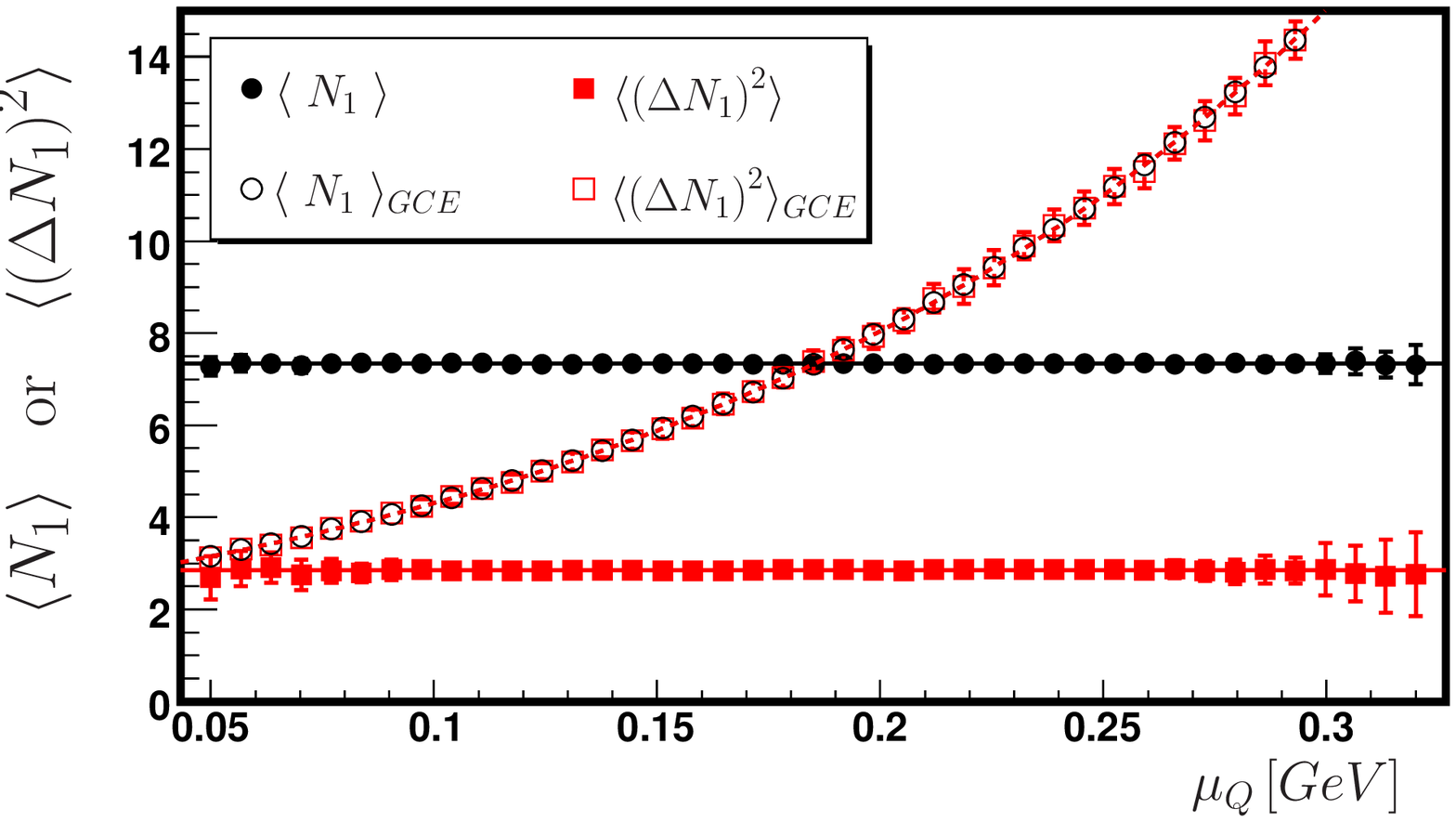,width=8.4cm,height=6.5cm}
  \caption{ The first four moments of the Monte Carlo weight,
    Eq.(\ref{appendix_W_G}) ({\it left}) and the first two moments 
    of multiplicity distributions ({\it right})
    , as described in the text.
   }  
  \label{weight_moments_mu}
\end{figure}

In Fig.(\ref{weight_moments_mu}) ({\it right}) we show, in open symbols,
the mean value $\langle N_1 \rangle$ and the variance 
$\langle \left( \Delta N_1 \right)^2 \rangle$ of the particle multiplicity 
distribution of the original GCE samples for different values 
of chemical potential $\mu_Q$. The closed symbols denote mean
value and variance of these samples after the
transformation Eq.(\ref{appendix_W_G}) was applied. Independent of the original
sample the result stays (within error bars) the same. However
the statistical error is lowest for a chemical potential close to:
\begin{equation}\label{GCE_mu}
\mu_Q ~=~ T ~ \sinh^{-1}\left(\frac{Q_g}{2V_g\psi} \right)~,
\end{equation}
i.e. when the initial sample is already similar 
(at least in terms of mean values) to the desired sample. This is reflected
in the moments of the Monte Carlo Weight factor, 
Fig.(\ref{weight_moments_mu}) ({\it left}). Higher moments have a strong
minimum around $\mu_Q=0.1896GeV$, i.e. the weights are most homogeneously
distributed amongst events, and most efficient used is made of them.


\begin{thebibliography}{100}
  
\bibitem{Fermi}
  E.~Fermi, Progr.Theor. Phys. {\bf 5} (1950) 570.

\bibitem{Hagedorn}
  R.~Hagedorn, Nucl. Phys. B {\bf 24}, 93 (1970).

\bibitem{GSIfits}
  J. Cleymans, D. Elliott, A. Keranen, E. Suhonen, Phys.Rev. C {\bf 57} (1998)
  3319;
  %
  J. Cleymans, H. Oeschler, K. Redlich, Phys.Rev. C {\bf 59} (1999) 1663;
  %
  R. Averbeck, R. Holzmann, V. Metag, R.S. Simon,
  Phys.Rev. C {\bf 67} (2003) 024903.

\bibitem{AGSfits}
  P.~Braun-Munzinger, J.~Stachel, J.~P.~Wessels and N.~Xu,   Phys.\ Lett.\  B
  {\bf 344}, 43 (1995).

\bibitem{SPSfits}
  P.~Braun-Munzinger, J.~Stachel, J.~P.~Wessels and N.~Xu, Phys.\ Lett.\  B {\bf
    365} (1996) 1;
  %
  P.~Braun-Munzinger, I.~Heppe and J.~Stachel,   Phys.\ Lett.\  B {\bf 465}, 15
  (1999);
  %
  F. Becattini, M. Ga\'zdzicki, A. Keranen, J. Manninen, R. Stock,
  Phys.Rev.C {\bf 69} 024905 (2004).

\bibitem{RHICfits}
  J.~Adams {\it et al.}  [STAR Collaboration], Nucl.\ Phys.\  A {\bf 757}, 102
  (2005).
\bibitem{FreezeOut}
  J. Cleymans, H. Oeschler, K. Redlich, and S. Wheaton, Phys. Rev. C {\bf 73},
  034905 (2006);
  J.Cleymans, and K.Redlich, Phys.Rev. C {\bf 60}, (1999) 054908;
  J.Cleymans, and K.Redlich, Phys.Rev.Lett. {\bf 81} (1998) 5284-5286;
  F. Becattini, J. Manninen, and M. Ga\'zdzicki,  Phys. Rev. C {\bf 73}, 044905
  (2006);
  A. Andronic, P. Braun-Munzinger, J. Stachel, Nucl. Phys. A {\bf 772}, 167
  (2006).

\bibitem{SHM_predictions_LHC}
  I.~Kraus, J.~Cleymans, H.~Oeschler, K.~Redlich and S.~Wheaton,
  arXiv:0707.1282 [hep-ph];
  %
  A.~Andronic, P.~Braun-Munzinger and J.~Stachel,
  arXiv:0707.4076 [nucl-th];
  %
  A.~Andronic, P.~Braun-Munzinger, K.~Redlich and J.~Stachel,
  arXiv:0707.4075 [nucl-th];
  %
  J.~Rafelski and J.~Letessier,
  J.\ Phys.\ G {\bf 35}, 044042 (2008);
  %
  J.~Rafelski and J.~Letessier,
  Eur.\ Phys.\ J.\  C {\bf 45}, 61 (2006);
  %
  F.~Becattini and J.~Manninen,
  J.\ Phys.\ G {\bf 35}, 104013 (2008).

\bibitem{SHM_predictions_FAIR}
  A.~Andronic, P.~Braun-Munzinger, K.~Redlich and J.~Stachel,
  J.\ Phys.\ G {\bf 35}, 104155 (2008).

\bibitem{QCD_pd}
  F.~Karsch, E.~Laermann and C.~Schmidt,
  Phys.\ Lett.\  B {\bf 520}, 41 (2001);
  %
  Z.~Fodor and S.~D.~Katz,
  JHEP {\bf 0203}, 014 (2002);
  %
  Z.~Fodor and S.~D.~Katz,
  JHEP {\bf 0404}, 050 (2004).

\bibitem{Model_pd}
  Y.~Hatta and T.~Ikeda,
  Phys.\ Rev.\  D {\bf 67}, 014028 (2003);
  %
  P.~de Forcrand and O.~Philipsen,
  Nucl.\ Phys.\  B {\bf 673}, 170 (2003);
  %
  B.~J.~Schaefer and J.~Wambach,
  Phys.\ Rev.\  D {\bf 75}, 085015 (2007);
  %
  K.~Fukushima,
  Phys.\ Rev.\  D {\bf 77}, 114028 (2008);
  %
  E.~S.~Bowman and J.~I.~Kapusta,
  Phys.\ Rev.\  C {\bf 79}, 015202 (2009).

\bibitem{OnsetOfDecon}
  M.~Ga\'zdzicki, M.~I.~Gorenstein and S.~Mrowczynski, Phys.\ Lett.\ B {\bf 585},
  115 (2004);
  M.~I.~Gorenstein, M.~Ga\'zdzicki and O.~S.~Zozulya, Phys.\ Lett.\ B {\bf 585},
  237 (2004).

\bibitem{PhaseTrans}
  I.N. Mishustin, Phys. Rev. Lett. {\bf 82}, 4779 (1999);
  Nucl. Phys. A {\bf 681}, 56-63 (2001);
  H. Heiselberg and A.D. Jackson, Phys. Rev. C {\bf 63}, 064904 (2001).

\bibitem{CriticalPoint}
  M.A.~Stephanov, K.~Rajagopal, and E.V.~Shuryak, Phys. Rev. Lett. {\bf 81},
  4816 (1998);
  Phys. Rev. D {\bf 60},114028 (1999);
  M.A.~Stephanov, Acta Phys.Polon.B {\bf 35} 2939 (2004);
  %
  M.~A.~Stephanov,
  Prog.\ Theor.\ Phys.\ Suppl.\  {\bf 153}, 139 (2004).

\bibitem{Koch}
  S.~Jeon and V.~Koch,
  arXiv:hep-ph/0304012;
  %
  V.~Koch,
  arXiv:0810.2520 [nucl-th].
  
\bibitem{Karsch_susc}
  M.~Cheng {\it et al.},
  Phys.\ Rev.\  D {\bf 79}, 074505 (2009)~.

\bibitem{Patriha}  
  R.~K.~Pathria, {\it Statistical Mechanics} (Butterworth Heinemann, Oxford, 1996), 2nd ed.  
  
\bibitem{Randrup}
  J.~Randrup, 
  Nucl.\ Phys.\ A {\bf 522}, 651  (1991);
  J.~Randrup
  Comput. Phys. Commun.\  {\bf 59}, 439 (1990).

\bibitem{Bec_MC}
  F.~Becattini, A.~Keranen, L.~Ferroni and T.~Gabbriellini,
  Phys.\ Rev.\  C {\bf 72}, 064904 (2005).

\bibitem{Bec_MCE}
  F.~Becattini and L.~Ferroni,
  Eur.\ Phys.\ J.\  C {\bf 35}, 243 (2004);
  %
  F.~Becattini and L.~Ferroni,
  Eur.\ Phys.\ J.\  C {\bf 38}, 225 (2004).

\bibitem{baseline}
  M.~Hauer, G.~Torrieri and S.~Wheaton,
  Phys.\ Rev.\  C {\bf 80}, 014907 (2009).

\bibitem{acc}
  M.~Hauer,
  Phys.\ Rev.\  C {\bf 77}, 034909 (2008).

\bibitem{feq}
  M.~I.~Gorenstein and M.~Hauer,
  Phys.\ Rev.\  C {\bf 78}, 041902 (2008).

\bibitem{THERMUS}
  S.~Wheaton, J.~Cleymans, and M.~Hauer,
  Comput.\ Phys.\ Commun.\  {\bf 180}, 84 (2009).

\bibitem{ROOT}
  R.~Brun and F.~Rademakers,
  Nucl.\ Instrum.\ Meth.\  A {\bf 389}, 81 (1997).

\bibitem{extgauss}  
  M.~S.~S.~Challa and J.~H.~Hetherington,  
  Phys.\ Rev.\  A {\bf 38}, 6324 (1988).  

\bibitem{clt}
  M.~Hauer, V.~V.~Begun and M.~I.~Gorenstein,
  Eur.\ Phys.\ J.\  C {\bf 58}, 83 (2008).

\bibitem{ResDecay}
  J.~Sollfrank, P.~Koch and U.~W.~Heinz,
  Phys.\ Lett.\  B {\bf 252}, 256 (1990);
  %
  J.~Sollfrank, P.~Koch and U.~W.~Heinz,
  Z.\ Phys.\  C {\bf 52}, 593 (1991);

\bibitem{THERMINATOR}
  A.~Kisiel, T.~Taluc, W.~Broniowski and W.~Florkowski,
  Comput.\ Phys.\ Commun.\  {\bf 174}, 669 (2006).

\bibitem{SolfrankHeinz}
  E.~Schnedermann, J.~Sollfrank and U.~W.~Heinz,
  Phys.\ Rev.\  C {\bf 48}, 2462 (1993).

\bibitem{BecCley}
  F.~Becattini and J.~Cleymans,
  J.\ Phys.\ G {\bf 34}, S959 (2007).

\bibitem{CEfirst}
  V.~V.~Begun, M.~Ga\'zdzicki, M.~I.~Gorenstein and O.~S.~Zozulya,
  Phys.\ Rev.\  C {\bf 70}, 034901 (2004).

\bibitem{MCEvsData}
  V.~V.~Begun, M.~Ga\'zdzicki, M.~I.~Gorenstein, M.~Hauer, V.~P.~Konchakovski and B.~Lungwitz,
  Phys.\ Rev.\  C {\bf 76}, 024902 (2007).

\bibitem{Res}
  V.~V.~Begun, M.~I.~Gorenstein, M.~Hauer, V.~P.~Konchakovski and O.~S.~Zozulya,
  Phys.\ Rev.\  C {\bf 74}, 044903 (2006).

\bibitem{beni_urqmd}
  B.~Lungwitz and M.~Bleicher,
  Phys.\ Rev.\  C {\bf 76}, 044904 (2007).
  
\bibitem{beni_data}
  C.~Alt {\it et al.}  [NA49 Collaboration],
  Phys.\ Rev.\  C {\bf 78}, 034914 (2008).

\bibitem{Turko}
  J.~Cleymans, K.~Redlich and L.~Turko,
  Phys.\ Rev.\  C {\bf 71}, 047902 (2005);
  J.~Cleymans, K.~Redlich and L.~Turko,
  J.\ Phys.\ G {\bf 31}, 1421 (2005).

\bibitem{CEsecond}
  V.~V.~Begun, M.~I.~Gorenstein and O.~S.~Zozulya,
  Phys.\ Rev.\  C {\bf 72}, 014902 (2005).

\bibitem{Abramowitz}
  M. Abramowitz and I.A. Stegun, {\it Handbook of Mathematical Functions with
  Formulas, Graphs, and Mathematical Tables}, New York, Dover (1965).

\end{thebibliography}
\end{document}